\documentclass[%
reprint,
superscriptaddress,
amsmath,amssymb,
aps,
twocolumn
]{revtex4-2}

\usepackage{graphicx}
\usepackage{amssymb}
\usepackage{pifont}
\usepackage{longtable}
\usepackage{rotating}
\usepackage{array}
\usepackage{multirow}
\usepackage{makecell}
\usepackage{booktabs}
\usepackage{xcolor}
\usepackage{ulem}
\usepackage{colortbl}
\usepackage{tikz}
\usepackage[flushleft]{threeparttable}

\usepackage{dcolumn}% Align table columns on decimal point
\usepackage{bm}% bold math
\usepackage{physics}
\usepackage{hyperref}% add hypertext capabilities
\newcommand{\cmark}{{\color[HTML]{FE0000} \ding{51}}}%
\newcommand{\xmark}{\ding{55}}%

\newcommand*{\citen}[1]{%
  \begingroup
    \romannumeral-`\x % remove space at the beginning of \setcitestyle
    \setcitestyle{numbers}%
    \cite{#1}%
  \endgroup   
}

\makeatletter
\def\maketitle{
\@author@finish
\title@column\titleblock@produce
\suppressfloats[t]}
\makeatother

\begin{document}

\preprint{APS/123-QED}

\title{Microscopic Origin of the Electric Dzyaloshinskii-Moriya Interaction}% Force line breaks with \\

\author{Peng Chen}
\affiliation{Physics Department and Institute for Nanoscience and Engineering, University of Arkansas, Fayetteville, Arkansas 72701, USA}
\author{Hong Jian Zhao}\email[]{physzhaohj@jlu.edu.cn}
\affiliation{International Center for Computational Method and Software (ICCMS), Jilin University, 2699, Qianjin Street, Changchun, 130012, China}
\affiliation{Key Laboratory of Physics and Technology for Advanced Batteries (Ministry of Education), College of Physics, Jilin University, Changchun, 130012, China}
\affiliation{Physics Department and Institute for Nanoscience and Engineering, University of Arkansas, Fayetteville, Arkansas 72701, USA}
\author{Sergey Prosandeev}
\affiliation{Physics Department and Institute for Nanoscience and Engineering, University of Arkansas, Fayetteville, Arkansas 72701, USA}
\author{Sergey Artyukhin}
\affiliation{Quantum Materials Theory, Istituto Italiano di Tecnologia, 16163 Genova, Italy.}
\author{Laurent Bellaiche}\email[]{laurent@uark.edu}
\affiliation{Physics Department and Institute for Nanoscience and Engineering, University of Arkansas, Fayetteville, Arkansas 72701, USA}

\date{\today}% It is always \today, today,

\begin{abstract}
The microscopic origin of the electric Dzyaloshinskii-Moriya interaction (eDMI) is unveiled and discussed by analytical analysis and first-principles based calculations. 
As similar to the magnetic Dzyaloshinskii-Moriya interaction (mDMI), eDMI also originates from electron-mediated effect and more specifically from certain electron hoppings that are being activated due to certain local inversion symmetry breaking.  However, the eDMI energy is found to be at least a third-order interaction in atomic displacements instead of bilinear in magnetic dipole moments for mDMI. 
Furthermore, the eDMI energy form is presented, and we find that novel electrical topological defects (namely, chiral electric bobbers) can arise from this eDMI.
Thus unraveling the microscopic origin of eDMI has the potential to lead to, and explain, the discovery of novel polar topological phases.
\end{abstract}

\maketitle

\section{Introduction}
The seminal works that addressed and explained magnetic Dzyaloshinskii-Moriya interaction (DMI) were done in 1957 and 1960 by I. E. Dzialoshinskii\cite{dzyaloshinskii1957,Dzyaloshinsky1958} and T. Moriya\cite{moriya1960}, and named after them.
Dzyaloshinskii firstly pointed out the existence of anti-symmetric form of interaction by symmetry analysis. 
Moriya then derived the microscopic origin of the magnetic Dzyaloshinskii-Moriya interaction (mDMI) by considering the spin-orbit coupling (SOC) and taking electron hopping as perturbation.
mDMI revolutionized  magnetism, since it, e.g., explains nontrivial non-collinear topological textures such as vortices\cite{Cooper1999,Donnelly2020}, skyrmions\cite{Skyrme1962,Aifer1996,Bogdanov2001,Rler2006,Muhlbauer2009,Yu2010,heinze11,Romming2013,Nagaosa2013,Dup2014,Hsu2016,wiesendanger16,bera19,Du2019}, and domain walls\cite{zang2018,Schoenherr2018} -- that is intriguing  both for fundamental theory and potential applications.

Magnetic effects normally have their electric analogue counterpart which is deep-rooted in the electromagnetic theory.
Strikingly, despite the fact that ferroelectric vortices and skyrmions have been recently reported in superlattices made of ferroelectric Pb(Zr,Ti)O$_3$ or PbTiO$_3$ sandwiched by either paraelectric SrTiO$_3$ dielectric layers \cite{Yadav2016,Zhang2017,das2019,Hsu2019,Bakaul2021} or SrRuO$_3$ metallic layers\cite{Rusu2022}, after having been predicted \cite{Naumov2004,Nahas2015,gonccalves2019}, it was not clear until recently if an electric analogue of mDMI exists. 
As a matter of fact, such exotic orders of electric dipoles were typically explained by electrostatic boundary conditions \cite{Hong2017} rather than by considering electric Dzyaloshinskii-Moriya interaction (eDMI). 
Recent symmetry analysis and ab-initio calculations by Zhao et. al.\cite{zhao2020} and the observation of helical textures of electric dipoles\cite{Khalyavin2020} in bulk perovskites (that are systems for which there is no depolarization field) have changed such perception. The discovery of eDMI not only should deepen our knowledge of electromagnetic phenomena (e.g. magnetic non-collinear spins versus electric non-collinear dipole patterns), but is also of technological importance. For example, mDMI generally plays an primordial role in generating magnetic topological phases ({\it e.g.}, helimagnets\cite{Uchida2006,Zhang2017scirep,Perreault2020}, skyrmions\cite{Skyrme1962,Aifer1996,Bogdanov2001,Rler2006,Muhlbauer2009,Yu2010,heinze11,Romming2013,Nagaosa2013,Dup2014,Hsu2016,wiesendanger16,bera19,Du2019}, merons\cite{pereiro14,bera19,gao2019}, etc.). These magnetic topological phases have potential applications in designing logical/storage devices based on magnetic fields or electric currents. However, challenges still exist for magnetic topological defects\cite{Scott2007,Luo2021}, that are (1) the size of  these defects, such as magnetic skyrmions, needs to be scaled down from micrometer (or 100 nm) to nanometers; (2) topological defects like magnetic skyrmions usually need an external magnetic field to assist their stability and the temperature at which they exist can be rather low; and (3) the response velocity of the magnetic topological phases under external field (especially electric field/current) needs to be improved. The discovery of possible electric topological phases ({\it e.g}., the electric counterpart of mDMI-based helical electric polar structures\cite{Khalyavin2020}, electric polar skyrmions\cite{gonccalves2019,das2019}, electric polar vorticies\cite{Yadav2016,Zhang2017,Hsu2019,Bakaul2021,Rusu2022}, etc.) has the potential to overcome these drawbacks, since (i) the observed electric topological phases are in nanometer scale and many have been reported at room temperature\cite{Khalyavin2020,das2019,Yadav2016,Zhang2017,Hsu2019,Bakaul2021,Rusu2022}; (ii) an electric field control of dipole textures would avoid the Joule heating (and thus leads to low-power devices)\cite{Scott2007}; and (iii) electric dipole textures normally have a fast response to the electric field. Moreover, from a fundamental physical knowledge viewpoint, the discussion of an intrinsic eDMI term that could stabilize the above-mentioned topological defects has long been overlooked when studying ferroelectric/polar materials, until recent {\it ab-initio} calculations\cite{zhao2020} clearly indicate its existence.

Let us recall that in Ref.\cite{zhao2020}, phenomenological models revealed a sequence of trilinear couplings, which showcase the existence of eDMI coming from oxygen octahedral rotations. However, many important ferroelectric materials (such as PbTiO$_3$, KNbO$_3$ etc.) do not have any oxygen octahedral tiltings and, as will be shown in this manuscript, they do possess eDMI. One may thus wonder if there is a general theory to explain the eDMI from a microscopic point of view and if the previously found oxygen-octahedral-tiling-mediated eDMI is only a special case. In this manuscript, we aim at addressing the microscopic origin of such a eDMI and answering several important open questions: (i) what is the microscopic origin of eDMI energy which gives rise to a cross-product form of ionic displacements  $\bm{u}_{i}$ and $\bm{u}_{j}$:
\begin{eqnarray}\label{eq:dmi}
\bm{\mathcal{D}}(i,j) \cdot (\bm{u}_{i}\times \bm{u}_{j})
\end{eqnarray}
where $\bm{\mathcal{D}}(i,j)$ is the eDMI vector? (ii) Is eDMI a classical or quantum effect? (iii) what is the energy scale of the eDMI?
Practically, bulk PbTiO$_3$ is selected as a study platform. We firstly show that eDMI can be obtained by extracting the antisymmetric part of the force constants.
Then, a Green's function perturbation method in a Tight-Binding (TB) electron Hamiltonian is adopted to study the origin of such antisymmetric feature.
By carefully looking into the forces coming from different orbital hopping channels, we then find that a similar mechanism that gives rise to mDMI also happens in ferroelectric materials and thus induces eDMI. 
More specifically, we discover that (1) certain local inversion symmetry breaking activates  forbidden electron hopping channels on adjacent atomic sites; and (2) it is the combination of the orbitals following certain selection rules that results in antisymmetric form of forces and thus eDMI. We also provide an analytical form of the eDMI energy in perovskite materials, as well as, compute  its coefficient which is estimated to be only one order smaller than that of  the typical energy that favors collinear polar texture. Consequently, noncollinear arrangments resulting from this eDMI have the potential to occur and be observed.\\

\section{Formalism}
\subsection{Antisymmetric force constants and eDMI.}
\begin{figure}[htp]
    \centering
    \includegraphics[width=0.9\linewidth]{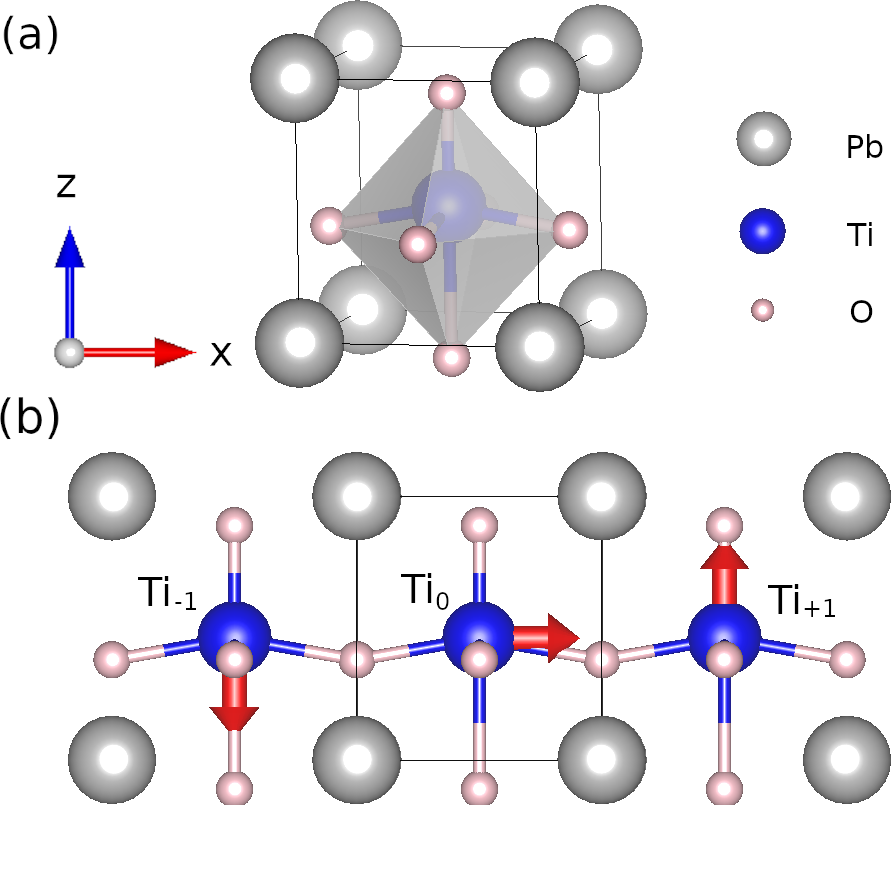}
    \caption{(a) Tetragonal phase of PbTiO$_3$, with titanium ions displaced along the $z$-direction; (b) Illustrative plot of some atomic displacements under a certain force constants $F_{\alpha\beta}(i,j)$ between titanium pairs with $i$=Ti$_{0}$ and $j$=Ti$_{+1}$ along the $x$-direction: $F_{xz}(\text{Ti}_{0},\text{Ti}_{+1})$ and $F_{zx}(\text{Ti}_{0},\text{Ti}_{+1})$; Note that $F_{zx}(\text{Ti}_{-1},\text{Ti}_{0})$ is equal to $F_{zx}(\text{Ti}_{0},\text{Ti}_{+1})$ due to the translational symmetry along the $x$-direction. The red vectors illustrate one possible set of the atomic displacement pattern that could result from a  negative $F_{xz}(\text{Ti}_{0},\text{Ti}_{+1})$ and positive $F_{zx}(\text{Ti}_{0},\text{Ti}_{+1})$.}\label{fig:pto}
\end{figure}
The ferroelectric polarization can be characterized by a collection of atomic displacements. Taking PbTiO$_3$ as in Fig.~\ref{fig:pto} (a) for example, the first-principles calculated normalized polar mode (that is soft in the cubic structure) consists of titanium and lead cations displaced by 0.78 and 0.31 \AA~ along the positive $z$-direction, respectively;  the oxygen anions that are on the side of  titanium ions within (001) planes being displaced by 0.38 \AA~ in the negative $z$-direction; and the oxygen anions that align with titanium ions along the z-axis being vertically displaced by 0.08 \AA~ in the positive $z$-direction.
Considering that the titanium cation has the largest displacement and is parallel to the total polarization, we use here the force constants\cite{born1955} between titanium sites to study qualitatively the force constants of polarization\footnote{Quantitatively, there exists a unitary transformation between the force constants of ions and the force constants of polarizations. Such transformation matrix can be obtained by taking the eigenvectors from diagonalizing the force constant matrix of the PbTiO$_3$ cubic structure.} (note that experiments also use atomic displacements, such as those of titanium and lead ions, to visualize non-collinear dipolar configurations\cite{Yadav2016,das2019,Hsu2019}, which is another reason we use the force constants of specific ionic sites).  
In the purpose of studying eDMI, we extract the antisymmetric part of the force constants matrix between different sites $i$ and $j$ via
\begin{align} \label{eq:antisymmetric}
\norm{F^{A}(i,j)}=\frac{1}{2}(\norm{F(i,j)}-\norm{F(i,j)}^{\text{T}}),
\end{align}
where T is the matrix transpose operation, $\norm{F(i,j)}$ is the force constants matrix between sites $i$ and $j$, and the achieved $\norm{F^{A}(i,j)}$ is antisymmetric as it satisfies $\norm{F^{A}(i,j)}=-\norm{F^{A}(i,j)}^{\text{T}}$.  
As detailed in Sec.~I of the Supplemental Material\cite{sm} (see, also, references \cite{landau1936,ginzburg1945,ginzburg1949,Feynman1939,Dyson1949,Adler1962,Wiser1963,hellmann2015,Born1928,Lannoo1979,Moraitis1984,Elstner1998,economou2006,Elstner1998,Szilva2013,Zhong1995,Baroni2001,Liechtenstein1987,Lounis2010,Szilva2013,He2021,linvariant,Campbell2006,Chen2008} therein), it is the antisymmetric part of force constants that gives rise the energy as in Eq.~(\ref{eq:dmi}) and defines the eDMI vector $\bm{\mathcal{D}}(i,j)$ via
\begin{align} \label{eq:DinFC}
\bm{\mathcal{D}}(i,j)=(F^{A}_{y,z},F^A_{z,x},F^{A}_{x,y}),
\end{align}
where $F^{A}_{y,z}=(F_{y,z}-F_{z,y})/2$, $F^{A}_{z,x}=(F_{z,x}-F_{x,z})/2$, $F^{A}_{x,y}=(F_{x,y}-F_{y,x})/2$ and the Cartesian directions of $x$, $y$, and $z$ are used here to indicate matrix entries of force constants matrix.
Note also that, by antisymmetric part of force constants, we mean the force constants matrix between a pair of ions rather than the overall force constants matrix that contains all ions, since such latter force constants matrix is always symmetric\cite{born1955}.

\subsection{orbital resolved force constants}
Normally, density functional perturbation theory (DFPT) takes the variation of electron density with respect to the ionic displacements and calculates the force constants in self-consistent processes. Though higher order corrections can be included, the physical insight is missing. For example, it does not reveal which orbital specifically contributes to the force constants and how it results in an antisymmetric character.
Alternatively, we thus decided to use a perturbation method derived from tight-binding (TB) Hamiltonian to calculate the antisymmetric part of the force constants, which allows us to analyse orbital-resolved force constants. The details of this method can be found in Refs.~\cite{Lannoo1979,Moraitis1984} and have also been summarized in Sec.~IV of the Supplemental Material\cite{sm}.
Interestingly, a similar formalisation has been previously used to calculate the magnetic exchange parameters\cite{Liechtenstein1987,Szilva2013,He2021} $J_{\alpha\beta}(i,j)$ which can be seen as a ``force constants'' of spins and whose antisymmetric part corresponds to mDMI.
The only difference between calculating $J_{\alpha\beta}(i,j)$ and $F_{\alpha\beta}(i,j)$ is that instead of taking ionic displacement as perturbation, $J_{\alpha\beta}(i,j)$ takes spin rotations.
Such formalism has also recently been used to explain mDMI in the tri-halides CrCl$_3$ and CrI$_3$\cite{Besbes2019}, Mn$_3$Sn\cite{Cardias2020_2}, and clusters of 3$d$ metals\cite{cardias2020}.

As detailed in Sec.~IV of the Supplemental Material\cite{sm}, the force constants $F_{\alpha\beta}(i,j)$ can be written as integration of force constants density\cite{Lannoo1979,Moraitis1984} $f_{\alpha,\beta}(\varepsilon,i,j)$ as:
\begin{align}\label{eq:fc1} 
F_{\alpha\beta}(i,j)=&\int_{-\infty}^{\varepsilon_{f}} f_{\alpha,\beta}(\varepsilon,i,j)\dd \varepsilon  
\end{align}
in which
\begin{align}\label{eq:fc2}
&f_{\alpha,\beta}(\varepsilon,i,j)=
\sum\limits_{m,n}\xi_{\alpha,\beta}^{m,n}(\varepsilon,i,j) 
\end{align}
where the orbital-resolved force constants density $\xi_{\alpha,\beta}^{m,n}$ can be further written as
\begin{align}\label{eq:fc3}
&\xi_{\alpha,\beta}^{m,n}(\varepsilon,i,j) = \nonumber \\ 
&-\frac{1}{2\pi}\text{Im}[\langle m,i | U_{i,\alpha}\hat{G}^{0}(\varepsilon)U_{j,\beta} | n,j \rangle \langle n,j | \hat{G}^{0}(\varepsilon) | m,i \rangle]
\end{align}
where U are effective perturbation potentials as defined in Sec.~IV of the Supplemental Material\cite{sm} and in Ref.~\cite{Lannoo1979}. Note that $U_{i,\alpha}$ is taken here as a short notation of $\frac{\partial U}{\partial \tau_{i,\alpha}}$, where $\bm{\tau}$ is the ionic position of site $i$. 
The Green's function operator $\hat{G}^{0}(\varepsilon)$ is defined as $\sum\limits_{p}\frac{\ket{p}\bra{p}}{\varepsilon-\varepsilon_{p}+i\eta}$, where $\varepsilon_{p}$ and $\ket{p}$ are the energy and wavefunction in the unperturbed system $H^{0}$; Im is the operation to take the imaginary part. 
Since the total force constants of Eq.~(\ref{eq:fc1}) is calculated by integrating $f_{\alpha,\beta}(\varepsilon,i,j)$, these latter are called force constants' density as a function of energy $\varepsilon$.  
Equation~(\ref{eq:fc2}) defines $\xi_{\alpha,\beta}^{m,n}(\varepsilon,i,j)$ as orbital-resolved force constants density from orbital $m$ on site $i$ and orbital $n$ on site $j$ and the summations of $\xi_{\alpha,\beta}^{m,n}(\varepsilon,i,j)$ over orbitals $m$ and $n$ gives rise to force constants density $f_{\alpha,\beta}(\varepsilon,i,j)$ in see Eq.~(\ref{eq:fc2}).
Employing Eq.~(\ref{eq:fc1}) in Eq.~(\ref{eq:DinFC}), the $\bm{\mathcal{D}}$ vector can be rewritten in force constants density expressions as 
\begin{align}
\bm{\mathcal{D}}(i,j)=&\bm{(}\mathcal{D}_{x}(i,j),\mathcal{D}_{y}(i,j),\mathcal{D}_{z}(i,j)\bm{)} \label{eq:edmivector}
\end{align}
where
\begin{align}
\mathcal{D}_{x}(i,j)=&\frac{1}{2}\int_{-\infty}^{\varepsilon_{f}}[f_{y,z}(\varepsilon,i,j)-f_{z,y}(\varepsilon,i,j)]\dd\varepsilon \label{eq:edmivectorx}\\
\mathcal{D}_{y}(i,j)=&\frac{1}{2}\int_{-\infty}^{\varepsilon_{f}}[f_{z,x}(\varepsilon,i,j)-f_{x,z}(\varepsilon,i,j)]\dd\varepsilon \label{eq:edmivectory}\\
\mathcal{D}_{z}(i,j)=&\frac{1}{2}\int_{-\infty}^{\varepsilon_{f}}[f_{x,y}(\varepsilon,i,j)-f_{y,x}(\varepsilon,i,j)]\dd\varepsilon \label{eq:edmivectorz}
\end{align}
Detailed derivations for reproducing the orbital-resolved force constants density as in Ref.~\cite{Lannoo1979,Moraitis1984} can be found in Sec. IV the Supplemental Material\cite{sm}.

\subsection{numerical details}
For the numerical calculation, we use the symmetry adapted Wannier basis\cite{Sakuma2013} for the TB Hamiltonian. 
The wannierization is performed using Wannier90\cite{pizzi2020} and Quantum Espresso\cite{Giannozzi2009,Giannozzi2017} to extract all the following orbitals of $\text{Ti}: 4s^{1},3p^{3},3d^{5}$, $\text{Pb}:6s^{1},6p^{3},5d^{5}$, and $\text{O}:2s^{1},2p^{3}$ orbitals, 30 Wannier functions in total. 
Note that each Wannier function is two-fold degenerated since we are working with spin non-polarized situation.
The core electrons are treated as tightly bond to the nucleus by optimized Norm-Conserving Vanderbilt (ONCV) pseudopotentials\cite{Hamann2013}.
The Green's function $G^{0}(\varepsilon,\bm{k})$ is calculated by numerically inverting $(\varepsilon+\varepsilon_{f})\text{I}-H^{0}(\bm{k})$, where $H^{0}$ is the TB Hamiltonian of the unperturbed structure, $\text{I}$ is an identity matrix, $\varepsilon_{f}$ is the Fermi energy level from the self-consistent first-principle calculation, and $\bm{k}$ is the Bloch vector defined in the first Brillouin zone. 
Fourier transformation can then be used to determine Green's function in real space  $G^{0}(\varepsilon,i,j)$.
The bare potential $V^{b}$, Hartree potential $V_{h}$, and exchange correlation potential $V_{xc}$ are extracted from self-consistent first-principle calculations in order to evaluate the $\tilde{U}$ and $U^{b}$ following the definitions in Sec. IV of the Supplemental Material\cite{sm}.
The perturbation is induced by shifting both the ionic potentials and Wannier functions by 0.15~\AA, numerically. 
Since the Wannier functions are predefined in real space on a coarse grid, the displacements of Wannier functions need to be performed in reciprocal space first and then transformed back.
Finite difference method is used to obtain the partial derivatives  $\frac{\partial U}{\partial \tau_{i,\alpha}}$.
Furthermore, the integration of the complex energy in Eq.~(\ref{eq:fc1}) is performed over a rectangular contour.
The $\eta$ in the Green's function is chosen to be 0.1 eV, such that the evaluation of $-\frac{1}{\pi}\text{Im}\text{Tr}[G^{0}(\varepsilon,0,0)]$ in the original crystal cell ($i=0$ and $j=0$) reproduces nicely the same density of states as calculated from first-principle calculations. The plane-wave energy cutoffs for wavefunction and electron density are 50 and 400 Ry, respectively, in  Quantum Espresso. A k-mesh of $9\cross9\cross8$ is used in the first-principle self-consistent calculations and a q-mesh of $4\cross4\cross4$ is used in the DFPT force constants calculations.

\section{the existence of eDMI} 
Note that eDMI was not thought to exist until the recent work\cite{zhao2020} by Zhao {\it et al.} that changed such perception by demonstrating that it is allowed by symmetry.
Note also that the phenomenological mechanism discussed by Zhao {\it et al.} includes oxygen octahedral tiltings and complex energy forms. 
However, the eDMI  should exist even in systems that do not have oxygen octahedral tiltings, such as PbTiO$_3$.
Such fact suggests that there could exist other and possibly simpler microscopic explanations of eDMI.

\subsection{intrinsic eDMI in bulk}
When performing DFPT calculations on PbTiO$_3$ tetragonal phase (as depicted in Fig.~\ref{fig:pto} (a) in which the polarization is along  the $z$-direction), we found that the antisymmetric part of the force constants between titanium sites being nearest neighbors along the $x$-direction is a matrix $\norm{F^{A}(i,j)}$ having the following elements:
\begin{eqnarray}\label{eq:fij}
\norm{F^{A}(i,j)}=\begin{pmatrix}
0 & 0 & -0.73 \\
0 & 0 & 0 \\
0.73 & 0 & 0
\end{pmatrix}
\end{eqnarray}
where the matrix entries go through the Cartesian $\{x,y,z\}$ directions, and $i$ and $j$ are used to indicate the nearest neighbor pair of the titanium sites along the $x$-direction, such as between Ti$_{-1}$ and Ti$_{0}$ or Ti$_{0}$ and Ti$_{+1}$ in Fig.~\ref{fig:pto} (b). 
Consequently, the $\bm{\mathcal{D}}(i,j)$ vector, according to Eq.~(\ref{eq:DinFC}), involving first-nearest neighbors along the x-axis of PbTiO$_3$ is equal to $(0,0.73,0)$ which favors titanium atoms to be displaced anticlock-wisely, as shown in Fig.~\ref{fig:pto} (b).
On the other hand, when calculating the $\norm{F^{A}(i,j)}$ in a case of  the polarization direction being reversed, the calculated $\bm{\mathcal{D}}(i,j)$ vector is found to reverse to $(0,-0.73,0)$, which favors titanium atoms to be displaced clock-wisely.
Note that the off-diagonal value 0.73 eV/\AA$^2$ is only one quarter of the largest component of the symmetric part of force constants $\norm{F^{S}(i,j)}_{x,x}=-2.84$ and even a little bit larger in the absolute value than $\norm{F^{S}(i,j)}_{y,y}=-0.64$ and $\norm{F^{S}(i,j)}_{z,z}=-0.66$, which suggests a strong competition between the collinear coupling decided by the symmetric part of the force constants and the noncollinear coupling decided by the antisymmetric part of the force constants.
Note also that in the famous bulk/interfacial mDMI pictures\cite{enwiki:1072177090}, the magnetic interaction between two neighboring ions and a single third ion (ligand) is the minimal model\cite{Keffer1962,Cheong2007,Luo2021} to discuss the mDMI. For example, in ABO3 perovskites, the mDMI is usually rooted in neighboring B-O-B pairs ({\it e.g.}, the mDMI of Fe-O-Fe pairs in BiFeO$_3$).
To understand the microscopic origin of eDMI in a way resembling the situation of the mDMI, we start from the B-B pairs ({\it e.g.} Ti-Ti pairs in our model system PbTiO$_3$). 
Noteworthy is that such an eDMI is not limited to Ti-Ti pairs, but is also valid for Pb-Pb pairs.
The calculated eDMI for the nearest neighboring Pb-Pb pair, given by the D(i,j) is (0,0.075,0) which favors A-site Pb to be displaced anticlock-wised looking from the positive y-direction.
Moreover, when calculating the $\bm{\mathcal{D}}(i,j)$ vector for polar modes that contain contributions from all atoms (Pb, Ti, and O), a unitary transformation was performed on the whole force constant matrix.\footnote{Such transformation matrix can be obtained by taking the eigenvectors from diagonalizing the force constant matrix of the PbTiO$_3$ cubic structure.}.

\subsection{coupling form and scale of eDMI}
In order to determine the energetic coupling form that can give rise to the  eDMI vector, we take the centrosymmetric $Pm\bar{3}m$ phase as reference and look for the energy invariants written in terms of the displacements $\bm{u}$.
The third order terms are found to be the lowest order that can give rise to antisymmetric force constants. 
Actually, the derived energy term associated with the polar modes on the nearest neighbor sites should always be in odd number of orders and the third order is therefore the minimal requirement.
This is because all $\bm{u}_i$, $\bm{u}_j$, and $\bm{e}_{ij}$ (vector that is pointing from site $i$ to $j$) reverse sign under inversion operation, which means that at least an extra odd order of $\bm{u}_i$ or $\bm{u}_j$ needs to be included to make the energy term invariant under inversion symmetry\cite{Zhao2021}. 
Thus a bilinear form as in Eq.~(\ref{eq:dmi}) with respect to polar modes is forbidden by symmetry, if $\bm{\mathcal{D}}(i,j)$ does not depend on the $\bm{u}$ displacements.
Assuming the eDMI to adopt the same form as the second-order mDMI\cite{Erb2020} may not be valid in some materials.
In fact, by symmetry analysis, we find there exists only one third-order eDMI energy and it can be written in compact form:
\begin{equation}\label{eq:uuu}
E_{dmi}=\mathcal{A}^{-}[(\bm{u}_i+\bm{u}_j)\cross \bm{e}_{ij}]\cdot (\bm{u}_i\cross \bm{u}_j)
\end{equation}
where $\mathcal{A}^{-}$ is constant and $\mathcal{A}^{-}(\bm{u}_i+\bm{u}_j)\cross \bm{e}_{ij}$ is eDMI $\bm{\mathcal{D}}(i,j)$ vector.
Note that we numerically found from DFT calculations on the ferroelectric tetragonal phase of PbTiO$_3$ that the magnitude of the eDMI vector $\mathcal{A}^{-}(\bm{u}_{i}+\bm{u}_{i})\cross\bm{e}_{ij}$ is $7.65\times10^6$ Nm$^2$/C$^2$. 
Note that Eq.~(\ref{eq:uuu}) is the pure chiral part (giving rise to antisymmetric forces) of the energy that is itself derived from two energy invariants (that can be found in Sec. VI of the Supplemental Material\cite{sm}).
An equivalent expression that is written in atomistic displacement basis is also presented in Secs. VI B and C of the Supplemental Material\cite{sm}, in order to see the roles from individual atoms.

It is also interesting to realize that the spin current model \cite{Katsura2005,Raeliarijaona2013} gives a mDMI for which the energy is proportional to $(\bm{u}_{i}\cross \bm{e}_{ij})\cdot(\bm{m}_{i}\cross\bm{m}_{j})$, and thus for which mDMI vector is proportional to $(\bm{u}_{i}\cross \bm{e}_{ij})$. Such latter vector  is very similar to eDMI vector $\mathcal{A}^{-}(\bm{u}_i+\bm{u}_j)\cross \bm{e}_{ij}$ (see Eq.~(\ref{eq:uuu})) -- which shows again an essential connection between magnetism and electricity.
Note that the spin current model assumes homogeneous dipole moments, thus $\bm{u}_{i}$ there is equal to $(\bm{u}_{i}+\bm{u}_{j})/2$. On the other hand, Eq.~(\ref{eq:uuu})
emphasizes that eDMI energy is third order with respect to the ionic displacements, while mDMI in the spin-current model is bilinear with respect to magnetic moments and linear with respect to the ionic displacement. 
%Note also that recent first-principle calculations find that magnetic materials can still present DMI though in absence of SOC\cite{cardias2020}, which has the same spirit of eDMI considering that no analogue of SOC in ferroelectrics exist {\bf LB: do we really need this sentence?}.

The calculated nonzero $\bm{\mathcal{D}}(i,j)$ vector in bulk PbTiO$_3$ suggests that, in addition to the depolarization field, another intrinsic mechanism involving now this kind of force constants can also contribute to the formation of polar vortices observed, e.g.,  in PbTiO$_3$/SrTiO$_3$ superlattices \cite{Yadav2016,das2019,Hsu2019} and especially in PbTiO$_3$/SrRuO$_3$ superlatices \cite{Rusu2022} (because the metallic SrRuO$_3$ layers can result in much weaker depolarization field).

\subsection{chiral electric bobbers from eDMI}
\begin{figure}[htp]
    \centering
    \includegraphics[width=0.9\linewidth]{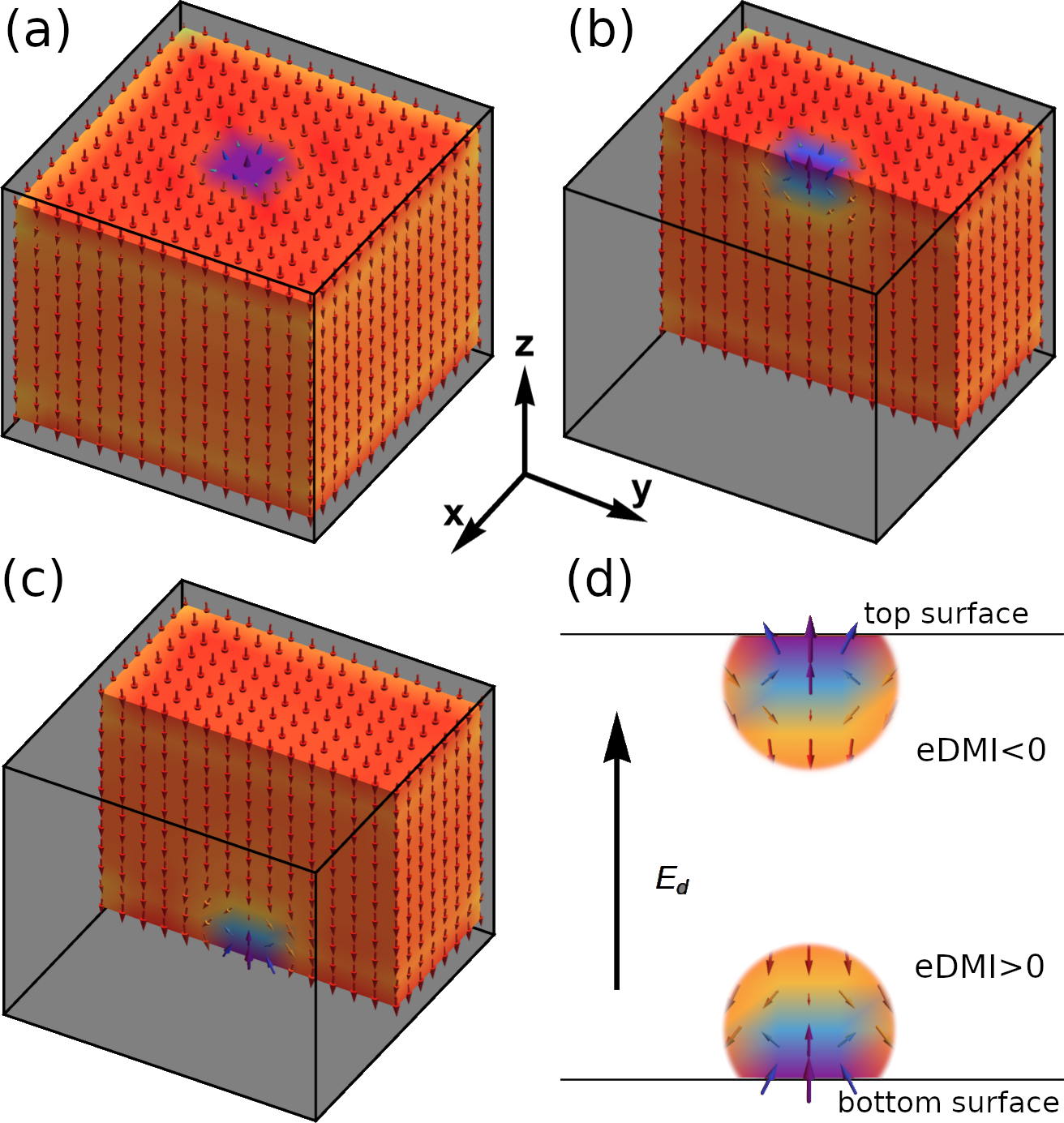}
    \caption{(a) top view of the surface bubble defect achieved with negative eDMI; (b) cross-section view of the (top) surface bubble defect from negative eDMI; (c) cross-section view of the (bottom) surface bubble defect from positive eDMI; (d) schematic illustration to show the charge and chirality of the surface bubble defects with respect to the sign of eDMI; the black arrow is to indicate the depolarization field ($E_{d}$) direction; the positive (negative) sign indicates tail-to-tail (head-to-head) dipolar configuration underneath the top (bottom) surface. In addition, the red domains have polarization along the negative z-direction, and the blue domains (defects) have polarization along the positive z-direction.}\label{fig:bobbers}
\end{figure}

It is interesting and important to know what type of dipolar structures one should expect from Eq.~(\ref{eq:uuu}). 
Thus we took the traditional effective Hamiltonian model for PbTiO$_3$ (see Supplemental Material\cite{sm} Sec.~VII for details) and additionally considered the eDMI energy in Eq.~(\ref{eq:uuu}) to explore such possible electric defects.
When we increase the magnitude of coefficient $\mathcal{A}^{-}_{nn}$ (nearest neighbor) and $\mathcal{A}^{-}_{nnn}$ (next-nearest neighbor) in front of Eq.~(\ref{eq:uuu}) from -0.001021 Hartree/Bohr$^{3}$ and -0.000353 Hartree/Bohr$^{3}$ (fitted for PTO) to values larger than -0.001813 Hartree/Bohr$^{3}$ and -0.000544 Hartree/Bohr$^{3}$ respectively, chiral electric bobbers (the electric counterpart of the ones in magnetic systems\cite{Rybakov2015,Ahmed2018,Zheng2018,Redies2019,Ran2021}) emerge, as depicted in Figs.~\ref{fig:bobbers} (a) and (b).  
The electric bobber survives only on the top surface of a monodomain with polarization pointing downwards, and its chirality is decided by the sign of the coefficient $\mathcal{A}^{-}$ of Eq.~(\ref{eq:uuu}). 
As can be seen from Fig.~\ref{fig:bobbers} (b), the up dipoles (on the edge of the blue region) rotate clockwise on the surface layer along the y-direction, due to a negative eDMI vector with respect to the x-axis.
When we rather use a positive $\mathcal{A}^{-}$ (i.e., reverse its sign), the electric bobber now only exists on the bottom surface of the monodomain.
Anticlockwise rotated down dipoles can be seen from Fig.~\ref{fig:bobbers} (c) (on the edge of the blue region) along the y-direction, due to a positive eDMI vector with respect to the x-axis.
As similar to the chiral magnetic bobbers, singularities (head-to-head and tail-to-tail dipoles) exist, however, chiral electric bobbers are rather small and are surface-dependent.
As illustrated in Fig.~\ref{fig:bobbers} (d), negative eDMI always gives a clockwise rotation of dipoles on the top surface, while positive eDMI favors the anticlockwise rotation of dipoles on the bottom surface; because of the existence of depolarization field in ferroelectric materials, the electric bobbers are surface-dependent, say bobbers with upwards dipoles will locate only on the top surface and bobbers with downwards dipoles will locate only on the bottom surface in a ferroelectric monodomain with polarization downwards.

Note the strength of eDMI needs to be increased by 78\% with respect to its ab-initio value to achieve such electric defects in PTO, but such resulting larger value may be found in other systems. We also find that such defects do not exist when the film is too thin, e.g., less than 10 unit cells. In addition, though the eDMI is necessary to stabilize electric bobbers, other intrinsic interactions, such as j5 and j7 are also found to be important to stabilize such a metastable phase on the surface. 
We have also found that eDMI is responsible for the formation of the mixed Ising-N{\'{e}}el type domain walls\cite{Lee2009} (see Supplemental Material\cite{sm} Fig. 8 for details).
Thus it can be seen that the effect of eDMI  can result in novel and/or complex textures and, due to the strong dipole-dipole interaction,  mainly survives near the surface and domain walls.

\section{Microscopic origin of eDMI.} 
As well known, mDMI originates from SOC\cite{moriya1960}. In contrast, the origin of the calculated eDMI (the antisymmetric force constants) is currently unknown and thus needs to be unsealed.
In the following, we will explain the origin of the antisymmetric feature of force constants as shown in Eq.~(\ref{eq:fij}) via electron hoppings, which will thus further explain the microscopic origin of eDMI vector $\bm{\mathcal{D}}_{ij}$ and its dependency on the polarization orientation as in Eq.~(\ref{eq:uuu}).

\subsection{eDMI as an electron-mediated quantum effect}
To determine the origin of eDMI, we decided to look in details at the microscopic full-Hamiltonian\cite{Baroni2001} (involving both electrons and ions) and derive the potential energy surface and its Hessian matrix. By following the textbook derivation as described in  sec.~III of Supplemental Material\cite{sm}, the force constants expression from the DFPT can be written as\cite{Baroni2001}
\begin{subequations}\label{eq:hessian}
\begin{align}
F_{\alpha\beta}(i,j)=&\frac{\partial^2 \Omega(\bm{\tau})}{\partial \tau_{i,\alpha}\partial \tau_{j,\beta}} \label{eq:hessiana} \\
=&\frac{\partial^{2}V_{ii}(\bm{\tau})}{\partial \tau_{i,\alpha}\partial \tau_{j,\beta}}+\int d\bm{r}\frac{\partial V_{ie}(\bm{r};\bm{\tau})}{\partial \tau_{i,\alpha}}\frac{\partial n(\bm{r};\bm{\tau})}{\partial \tau_{j,\beta}} \label{eq:hessianb}
\end{align}
\end{subequations}
It thus involves the second derivative of the potential energy surface $\Omega(\bm{\tau})$ with respect to ionic positions $\bm{\tau}_{i}$ and $\bm{\tau}_{j}$ $(i\neq j)$ (e.g., Ti$_{0}$ and Ti$_{+1}$ in Fig.~\ref{fig:pto}) along $\alpha$ and $\beta$ Cartesian directions, respectively.
Note that $n(\bm{r};\bm{\tau})$ is the electronic density, while the contributions to the force constant $F_{\alpha\beta}$ is divide into two parts: (1) the energy potential of ion-ion interaction $V_{ii}(\bm{\tau})$ that includes the nucleis and inner core electrons that rigidly follow the ionic displacements and (2) ion-electron interaction $V_{ie}(\bm{r};\bm{\tau})$ that includes both the ions (necleis combined with inner core electrons) and valence electrons $n(\bm{r};\bm{\tau})$.
Since $V_{ii}$ is a sum of repulsive ion-ion Coulomb interactions, the first term of the right side of Eq.~(\ref{eq:hessianb}) only contributes to the symmetric part of the force constants (see proofs in Sec~II of the Supplemental Material\cite{sm}).
Thus the anti-symmetric form in Eq.~(\ref{eq:fij}) has to come from the second term of the right side of Eq.~(\ref{eq:hessianb}), which in fact can be seen as the electron-density-mediated ion-ion indirect interaction.
More specifically, Eq.~(\ref{eq:hessianb}) indicates that (i) the change of the ionic position $\tau_{i,\alpha}$ induces a variation of electron-ion energy $\frac{\partial V_{ie}(\bm{r};\bm{\tau})}{\partial \tau_{i,\alpha}}$ at site $i$, (ii) which couples to the change of ionic position $\tau_{j,\beta}$ on site $j$ through the electron density fluctuation $\frac{\partial n(\bm{r};\bm{\tau})}{\partial \tau_{j,\beta}}$.
Thus though eDMI is mostly associated with ionic dipoles, Eq.~(\ref{eq:hessian}) tells that eDMI is not included in dipole-dipole interaction and, instead of treating electrons as point charges, treating electrons as wavefunctions are important to obtain eDMI. Thus eDMI originates from an electron-mediated quantum effect.
Note that the dipole-dipole interaction that takes consideration of the Born effective charges in a full tensor form should include the eDMI energy because the Born effective changes are associated with the electronic responses to the atomic displacements, which can be seen as another perspective to understand the electrons' role in the eDMI.

\begin{figure*}[!htp]
    \centering
    \includegraphics[width=1.0\linewidth]{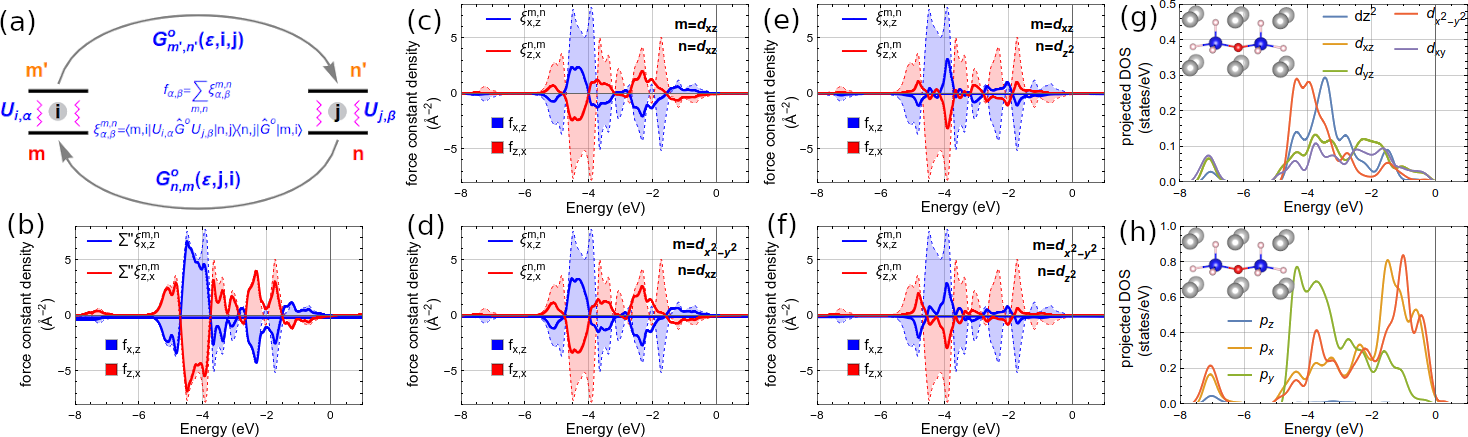}
    \caption{(a) Diagram plot of orbital-resolved force constants density $\xi_{\alpha,\beta}^{m,n}$ defined as $\sum\limits_{m^{\prime},n^{\prime}}\langle m,i|U_{i,\alpha}|m^{\prime},i\rangle$ $\langle m^{\prime},i | \hat{G}^{0}(\varepsilon) | n^{\prime},j \rangle$ $\langle n^{\prime},j | U_{j,\beta} | n,j \rangle$ $\langle n,j | \hat{G}^{0} | m,i \rangle$ (see Eq.~(\ref{eq:fc3} and Eq.~(32) in the Sec. IV of Supplemental Material\cite{sm}); Panels (b), (c), (d), and (e) are plots of orbital-resolved force constant density ($\xi_{x,z}^{m,n}$ in blue line and $\xi_{z,x}^{n,m}$ in red line) and total force constant density ($f_{x,z}(\varepsilon,i,j)$ in blue area and $f_{z,x}(\varepsilon,i,j)$ in red area), as defined in Eqs.~(\ref{eq:fc1}) and ((\ref{eq:fc2})).
    More specifically, Panel (c) contains $\xi_{x,z}^{m,n}$ and $\xi_{z,x}^{n,m}$ where $m=d_{xz}$ and $n=d_{xz}$; Panel (d) contains $\xi_{x,z}^{m,n}$ and $\xi_{z,x}^{n,m}$ where $m=d_{x^2-y^2}$ and $n=d_{xz}$; Panel (e) contains $\xi_{x,z}^{m,n}$ and $\xi_{z,x}^{n,m}$ where $m=d_{xz}$ and $n=d_{z^2}$; Panel (f) contains $\xi_{x,z}^{m,n}$ and $\xi_{z,x}^{n,m}$ where $m=d_{x^2-y^2}$ and $n=d_{z^2}$; and Panel (b) contains the sum of all the $\xi_{x,z}^{m,n}$ and $\xi_{z,x}^{n,m}$ plotted in Panels (c), (d), (e), and (f).
   Panels  (g) and (h) are density of states (DOS) projected on the titanium atom (blue ball in the subset) and oxygen atom (red ball in the inset); Note that the Fermi level is set at the zero of energy in all the plots.}\label{fig:loops}
\end{figure*}

In order to further understand the role of the electrons in the eDMI, the orbital-resolved force constants \cite{Lannoo1979,Moraitis1984} need to be calculated.
The second order perturbation is performed to calculate the orbital-resolved force constants density in Eq.~(\ref{eq:fc3}) and the perturbation process can be summarized as in Fig.~\ref{fig:loops} (a): (i) an atomic displacement $u_{i,\alpha}$ firstly induces an effective perturbation energy potential change $U_{i,\alpha}$ on site $i$ (left gray ball) and scatters state $\ket{m,i}$ (lower level on the left) to $\ket{m^{\prime},i}$ (top level on the left); (ii) such scattering is propagated by the Green's function $G^{0}_{m^{\prime},n^{\prime}}(\varepsilon,i,j)$ from site $i$ to site $j$ (gray arrow on the top); and (iii) couples to the atomic displacement $u_{j,\beta}$ on site $j$ (right gray ball) by alternating the effective perturbation potential $U_{j,\beta}$ and scattering state $\ket{n,j}$ (lower level on the right) to $\ket{n^{\prime},j}$ (top level on the right); (iv) another Green's function $G^{0}_{n,m}(\varepsilon,j,i)$ (gray arrow in the bottom) closes the ``loop'' by propagating the scattered state $n$ from site $j$ back to state $m$ on site $i$.
The specific mathematical expression of such loop can be found in Eq.~(32) of the Supplemental Material\cite{sm} and the more detailed mathematical definitions of $U_{i,\alpha}$, $U_{j,\beta}$, and the Green's function $G^{0}_{m^{\prime},n^{\prime}}(\varepsilon,j,i)$ and $G^{0}_{n,m}(\varepsilon,j,i)$ can be found in the Sec.~IV of the Supplemental Material\cite{sm}.
Note that the orbital scatterings in Fig.~\ref{fig:loops} (a) as well as Eq.~(\ref{eq:fc3}) show that 
the forces between atoms need a quantum treatment by the Hellman-Feynman theorem\cite{Feynman1939,hellmann2015}.
%calculating the forces on ions from electrons needs to involve the treatment of electrons in quantum descriptions. 
The eDMI should therefore be a quantum effect since the antisymmetric forces can only come from the interactions between (quantum) electrons and (classical) ions.
It is worth mentioning that such loop is physically equivalent to the DFPT process when calculating the force constants from Eq.~(\ref{eq:hessiana}).

Each loop as in Fig.~\ref{fig:loops} (a) is one contribution to the force constants from a set of orbitals $m,m^{\prime},n^{\prime}$, and $n$.
The summation of all the possible loops defined by $m,m^{\prime},n^{\prime}$, and $n$ orbitals gives rise to the force constants and can also be calculated by other methods such as DFPT.
In our calculations, there are in total 6561 (sum over $m,m^{\prime},n^{\prime}$, and $n$) loops that contribute to the force constants.
For the simplicity of further analysis, we define orbital-resolved force constants density $\xi_{\alpha,\beta}^{m,n}(\varepsilon,i,j)$ by summing out the $m^{\prime}$ and $n^{\prime}$ in the loops, see Eq.~(\ref{eq:fc3}), which represents the force constants density contribution from one orbital $m$ on site $i$ and another orbital $n$ on site $j$ to the force constants, where $\alpha$ and $\beta$ are elements of Cartesian $\{x,y,z\}$ directions, $m$ and $n$ range among all the orbitals on site $i$ and $j$ respectively, and $\varepsilon$ is the energy.
The sum of the orbital-resolved force constants density $\xi_{\alpha,\beta}^{m,n}(\varepsilon,i,j)$ over $m$ and $n$ is defined as the force constants density $f_{\alpha,\beta}(\varepsilon,i,j)$, as formulated in Eq.~(\ref{eq:fc2}).
Thus the total force constants, according to Eq.~(\ref{eq:fc1}), can be obtained by integrating  $f_{\alpha,\beta}(\varepsilon,i,j)$ from negative infinite to the Fermi energy
the highest occupied energy level.
$F_{x,z}$ and $F_{z,x}$ in Eq.~(\ref{eq:fij}) are calculated to be -0.58 and 0.58 eV/\AA$^2$ according to our TB model, respectively, which are comparable to the aforementioned DFPT results of -0.73 eV/\AA$^2$ and 0.73 eV/\AA$^2$. 
% The $F_{x,y}$, $F_{y,x}$, $F_{y,z}$, and $F_{z,y}$ are zeros just as the DFPT results.
We also calculated the next nearest neighbor $F_{x,z}$ and $F_{z,x}$ which are -0.16 eV/\AA$^2$ and 0.16 eV/\AA$^2$ according to our TB model, once again in good agreement with DFPT results of -0.14 eV/\AA$^2$ and 0.14 eV/\AA$^2$.
Note that the slight discrepancy between TB and DFPT results likely comes from the facts that our TB perturbation drops second- and higher-order electron density fluctuation and assumes a rigid Wannier orbital displacements, while the DFPT includes both the displacement of orbitals and the change of the orbital shapes during the self-consistent process of the electron density response.

So far three quantities are defined and will be used in the future analysis: the summation of (1) the orbital-resolved force constants density $\xi_{\alpha,\beta}^{m,n}(\varepsilon,i,j)$ (see Eq.~(\ref{eq:fc3})) over orbital $m$ on site $i$ and $n$ on site $j$ gives rise to (2) the force constants density $f_{\alpha,\beta}(\varepsilon,i,j)$ (see Eq.~(\ref{eq:fc2})), whose integration over energy $\varepsilon$ is (3) the force constants $F_{\alpha,\beta}(i,j)$ (see Eq.~(\ref{eq:fc1})) between sites $i$ and $j$.

\subsection{origin of the antisymmetric feature}
\begin{figure}[htp]
    \centering
    \includegraphics[width=1.0\linewidth]{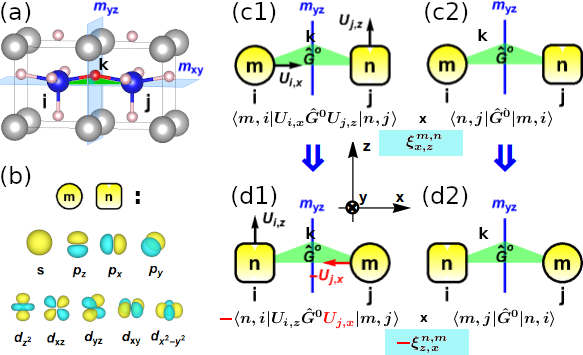}
    \caption{(a) Two PbTiO$_3$ ($P4mm$ phase) unit cells along the x-direction; The two blue balls on sites $i$ and $j$ are titanium atoms, and their intermediate oxygen atom is indicated by a red ball on site $k$; The vertical and horizontal blue planes represent mirrors $m_{yz}$ and $m_{xy}$ respectively; A green triangle is used to indicate the relative displacements between the two titanium atoms and their intermediate oxygen atom; 
    Panel (b) shows all the orbitals $m$ and $n$ in the orbital-resolved force constants expression; 
    Expression of $\xi_{x,z}^{m,n}$ is sketched by a multiplication between the expressions in (c1)  and (c2); The blue vertical lines in (c1) and (c2) are mirror $m_{yz}$ operations and transform the expressions in (c1) and (c2) to the expressions in (d1) and (d2), respectively, and whose multiplication gives rise to $-\xi_{z,x}^{n,m}$.
    In all the expression sketches, the yellow circles and squares represent orbital $m$ and $n$ respectively, the arrows are used to indicate effective perturbation potentials, and the green triangles (as also indicated in (a) among sites $i$, $j$ and $k$) are for Green's function $\hat{G}^{0}$.}\label{fig:rule2}
\end{figure}

The orbital-resolved force constants density $\xi_{\alpha,\beta}^{m,n}$ allows us to analyse the force constants contributions from different orbital combinations between sites $i$ and $j$.
As in Eq.~(\ref{eq:fc2}), the summation of the orbital-resolved force constants density over all possible $m$ and $n$ orbitals gives rise to the force constants density $f_{\alpha,\beta}(\varepsilon,i,j)$ which is plotted in Fig.~\ref{fig:loops} (b) represented by the colored areas with dashed outline (the same quantity is also plotted in Panels (c), (d), (e), and (f)). 
As can be seen, $f_{x,z}(\varepsilon,i,j)$ is equal to $-f_{z,x}(\varepsilon,i,j)$ at any given energy within the numerical round up of $0.01~\text{\AA}^{-2}$.
The y-component of eDMI vector $\bm{\mathcal{D}}(i,j)$ is thus derived from the non-zero antisymmetric force constants density $f_{z,x}-f_{x,z}\neq0$ (defined in Eq.~(\ref{eq:edmivectory})) that is corresponding to the numerical results in Eq.~(\ref{eq:fij}).
We are going to use the orbital-resolved force constants density $\xi_{\alpha,\beta}^{m,n}$ to understand the microscopic orbital origin of such eDMI vector.
More specifically, it is going to be seen that the commutation between orbitals $m$ and $n$ on sites $i$ and $j$ gives rise to inverse off-diagonal force constants density component, $\xi_{x,z}^{m,n}=-\xi_{z,x}^{n,m}$, which is symmetry protected. The relation of $f_{x,z}=-f_{z,x}$ in Fig.~\ref{fig:loops} (b) can thus be understood since $f_{x,z}=\sum\limits_{m,n}\xi_{x,z}^{m,n}$ and $f_{z,x}=\sum\limits_{n,m}\xi_{z,x}^{n,m}$ according to Eq.~(\ref{eq:fc2}), which explains the microscopic origin of the existence of eDMI vector along $y$ axis (see Eq.~(\ref{eq:edmivectory})).

Considering two PbTiO$_3$ unit cells as depicted in Fig.~\ref{fig:rule2} (a), the two titanium atoms (blue balls on sites $i$ and $j$) and their intermediate oxygen atom (red ball on site $k$) are displaced along negative and positive z-direction respectively. 
Orbitals $m$ and $n$ of the two titanium sites are chosen from the $4s^{1},3p^{3}$, and $3d^{5}$ orbitals as listed in Fig.~\ref{fig:rule2} (b) to calculate the orbital-resolved force constants density $\xi_{\alpha,\beta}^{m,n}$. 
(Note that the inner core electrons such as $1s,2s,3s$, and $2p$ orbitals are treated as tightly bond to the nucleus and thus only give rise to the symmetric part of the force constants.)
More specifically, according to Eq.~(\ref{eq:fc3}), the expression of $\xi_{x,z}^{m,n}$ can be written as $\langle m,i | U_{i,x}\hat{G}^{0}U_{j,z} | n,j \rangle$ (illustrated in Fig.~\ref{fig:rule2} (c1)) times $\langle n,j | \hat{G}^{0} | m,i \rangle$ (illustrated in Fig.~\ref{fig:rule2} (c2)).
In both Figs.~\ref{fig:rule2} (c1) and (c2), mirror $m_{yz}$ (vertical lines as also indicated in (a) as vertical blue plane) operations are performed.
Consequently, in Fig.~\ref{fig:rule2} (c1), the following functions are transformed: (1) orbital $m$ on site $i$ (left yellow circle) and orbital $n$ on site $j$ (right yellow square) are transformed to, in Fig.~\ref{fig:rule2} (d1), site $j$ on the right and site $i$ on the left, respectively; (2) effective perturbation potential $U_{i,x}$ on site $i$ (left black arrow) and $U_{j,z}$ on site $j$ (right black arrow) are transformed to, in Fig.~\ref{fig:rule2} (d1), $-U_{j,x}$  on site $j$ (right red arrow) and $U_{i,z}$ on site $i$ (left black arrow), respectively; (3) $G^{0}$ (green triangle) is unchanged since it is defined by the eigenfunctions of the unperturbed $H^{0}$ and follows the same crystalline symmetry, $P4mm$ as in the case of PbTiO$_3$ in Fig.~\ref{fig:rule2} (a).
Note that symmetry operations should never alternate the integration values, thus we have proved that $\langle m,i | U_{i,x}\hat{G}^{0}U_{j,z} | n,j \rangle$ (illustrated in Fig.~\ref{fig:rule2} (c1)) is equal to $-\langle n,i | U_{i,z}\hat{G}^{0}U_{j,x} | m,j \rangle$ (illustrated in Fig.~\ref{fig:rule2} (d1)).
Employing the same three transformation rules of the functions to (c2) and (d2), the following relations can also be proved: $\langle n,j | \hat{G}^{0} | m,i \rangle$ in Fig.~\ref{fig:rule2} (c2) is equal to $\langle m,j | \hat{G}^{0} | n,i \rangle$ in Fig.~\ref{fig:rule2} (d2).
Interestingly, the transformed expression, as illustrated in Fig.~\ref{fig:rule2} (d1) and (d2), is exactly the expression of $-\xi_{z,x}^{n,m}$, which means that $\xi_{x,z}^{m,n}=-\xi_{z,x}^{n,m}$ and $f_{x,z}=-f_{z,x}$ are constrained by the existence of the symmetry operation $m_{yz}$. 
One should notice that both orbitals $m$ and $n$ can be odd functions under the mirror operation $m_{yz}$ and give rise to minus signs, e.g. $p_{x}$ can be transformed to $-p_{x}$, $d_{xz}$ can be transformed to $-d_{xz}$, and $d_{xy}$ can be transformed to $-d_{xy}$. 
However, there are always two $m$ orbitals and two $n$ orbitals in the multiplication between $\langle m,i | U_{i,\alpha}\hat{G}^{0}U_{j,\beta} | n,j \rangle$ and $\langle n,j | \hat{G}^{0} | m,i \rangle$, which means no minus sign in total can be given to $\xi_{\alpha,\beta}^{m,n}$ due to the transformation of orbitals $m$ and $n$.
Moreover, not all the orbitals $m$ and $n$ can contribute to non-zero orbital-resolved force constants density $\xi_{\alpha,\beta}^{m,n}$ and eDMI vector. The symmetry of orbitals $m$ and $n$ decides if certain electron hopping channels are allowed to give rise to nonzero eDMI vector. Taking the structure in Fig.~\ref{fig:rule2} (a) for example, if one orbital is even (e.g. $m$ = $d_{x^2-y^2}$) and the other is odd (e.g. $n$ = $d_{xy}$) under the operation of mirror $m_{xz}$, the orbital-resolved force constants density will be zero ($\xi_{x,z}^{m,n}=\xi_{z,x}^{n,m}=0$) (see Sec. V of the Supplemental Material\cite{sm} for the proof from symmetry analysis and Fig.~1 of the Supplemental Material\cite{sm} for the numerical results).

The antisymmetric feature from the symmetry analysis is consistent with the numerical results as in Figs.~\ref{fig:loops} (c), (d), (e), and (f), where the red curve and blue curve are always in inverse sign. More specifically, Fig.~\ref{fig:loops} (c) shows $\xi_{x,z}^{d_{xz},d_{xz}}=-\xi_{z,x}^{d_{xz},d_{xz}}$, Fig.~\ref{fig:loops} (d) shows $\xi_{x,z}^{d_{x^2-y^2},d_{xz}}=-\xi_{z,x}^{d_{xz},d_{x^2-y^2}}$, Fig.~\ref{fig:loops} (e) shows $\xi_{x,z}^{d_{xz},d_{z^2}}=-\xi_{z,x}^{d_{z^2},d_{xz}}$, and Fig.~\ref{fig:loops} (f) shows $\xi_{x,z}^{d_{x^2-y^2},d_{z^2}}=-\xi_{z,x}^{d_{z^2},d_{x^2-y^2}}$. 
Figure~\ref{fig:loops} (b) sums up the orbital-resolved force constants density $\xi_{x,z}^{m,n}$ and $\xi_{z,x}^{n,m}$ that are in  Figs.~\ref{fig:loops} (c), (d), (e), and (f) (red and blue solid lines). We can see that the total force constants density $f_{x,z}$ and $f_{z,x}$ (colored area with dashed outline) are relatively well reproduced in Fig.~\ref{fig:loops}. (b). Contributions from other orbitals are relatively small or zeros and can be found in Fig.~1 of the Supplemental Material\cite{sm}.
Figure~\ref{fig:loops} (g) explains why the $d$ orbitals contributes most of the force constants. We can see that the DOS projected on titanium atom is in the same energy region as non-zero force constants density (colored areas in Fig.~\ref{fig:loops} (b)) and some peaks are also consistent, e.g. peaks at -3.5, -5.5, and -7.2 eV. 
The reason that the unoccupied titanium $d$ orbitals have contributed to DOS under the Fermi level is due to the $p$-$d$ interactions between titanium atoms (blue balls in the inset of (g)) and their intermediate oxygen atom (red ball in the inset of (g)), which can be seen in Fig.~\ref{fig:loops} (h) where the DOS projected on the intermediate oxygen atom roughly has the same shapes as the DOS projected on the titanium atom in Fig.~\ref{fig:loops} (g).

\subsection{Local inversion-symmetry-breaking and eDMI.} 
\begin{figure}[htp]
    \centering
    \includegraphics[width=1.0\linewidth]{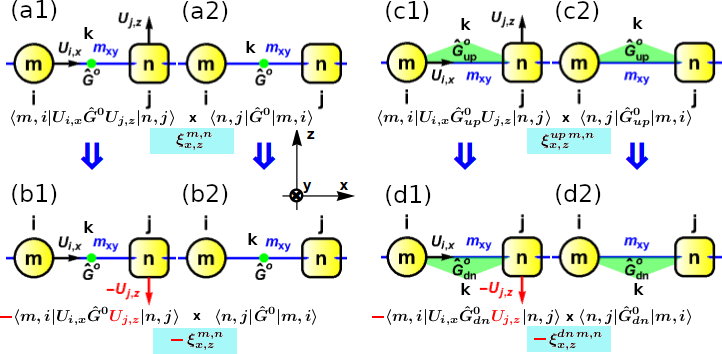}
    \caption{If sites $i$, $j$, and $k$ in Fig.~\ref{fig:rule2} (a) are collinear, expression of $\xi_{x,z}^{m,n}$ is sketched by a multiplication between expressions in (a1) and (a2); The blue horizontal lines in (a1) and (a2) are mirror $m_{xy}$ operations and transform the expressions in (a1) and (a2) to the expressions in (b1) and (b2), respectively, and whose multiplication gives rise to $-\xi_{x,z}^{m,n}$.
    When site $k$ is on top of sites $i$ and $j$ in the triangle, the expression of $\xi_{x,z}^{up,m,n}$ for such case is sketched by a multiplication between (c1) and (c2) which can be transformed by mirror $m_{xy}$ (blue horizontal lines) to the expressions in (d1) and (d2) whose multiplication gives rise to $-\xi_{x,z}^{dn,m,n}$ which is for the case that site $k$ is below sites $i$ and $j$. 
    In all the expression sketches, the yellow circles and squares represent orbital $m$ and $n$ respectively, the arrows are used to indicate effective perturbation potentials, and the green dots and triangles are for Green's function $\hat{G}^{0}$.}\label{fig:rule1}
\end{figure}
The intermediate oxygen plays an essential role in the indirect interaction between the nearest neighbor titanium atoms and so is to eDMI.
For instance, in Fig.~\ref{fig:rule1} (a), when the intermediate oxygen (site $k$) is collinear with the two nearest neighbor titanium atoms (sites $i$ and $j$), the orbital-resolved force constants density $\xi_{x,z}^{m,n}$ can be proved to be always zero.
Following a similar transformation rules as in Fig.~\ref{fig:rule2}, a mirror operation $m_{xy}$ (blue horizontal line in Figs.~\ref{fig:rule1} (a1) and (a2)) reverses only the sign in the front of the $U_{j,z}$ (black vertical arrow on site $j$) in (a1) and results in $-\langle m,i| U_{i,x}\hat{G}^{0}U_{j,z}|n,j \rangle$ in (b1), which indicates that $\xi_{x,z}^{m,n}=-\xi_{x,z}^{m,n}$ and $\xi_{x,z}^{m,n}$ has to be zero.
Thus we have proved that breaking such mirror symmetry (by forming a triangle among site $i$, $j$, and $k$) decides if eDMI exist, which is going to be another important rule in the ``orbital selection rules'' section.
Note that such mirror-symmetry-breaking (by forming a triangle) will result in a local-inversion-symmetry-breaking (inversion center on the middle point between sites $i$ and $j$) automatically. However, breaking such local-inversion-symmetry does not always give rise to mirror-symmetry-breaking. 
For example, site $k$ can be displaced along the line that connects sites $i$ and $j$ and away from the middle point between sites $i$ and $j$, but eDMI is still forbidden because of the existence of mirror $m_{xy}$ (see Fig.~\ref{fig:rule1}).

More importantly, Fig.~\ref{fig:rule1} has also proved that the direction of the displacement of the site $k$ with respect to sites $i$ and $j$ decides the direction of eDMI vector.
For instance, Figs.~\ref{fig:rule1} (c) and (d) prove that $\xi_{x,z}^{up,m,n}=-\xi_{x,z}^{dn,m,n}$, where $\xi_{x,z}^{up,m,n}$ is the orbital-resolved force constants density for the case that the site $k$ is displaced along the positive z-direction (green triangles in (c1) and (c2)) and $\xi_{x,z}^{dn,m,n}$ is the orbital-resolved force constants density for the case that the site $k$ is displaced along the negative z-direction (green triangles in (d1) and (d2)).
More specifically, in Figs.~\ref{fig:rule1} (c1) and (c2), the mirror $m_{xy}$ (horizontal blue lines) (1) transforms the $\hat{G}^{0}_{up}$ (corresponding the $P4mm$ phase of PbTiO$_3$ with polarization along negative z-direction) into the $\hat{G}^{0}_{dn}$ (corresponding to the $P4mm$ phase of PbTiO$_3$ with polarization along positive z-direction) in (d1) and (d2), (2) transforms the $U_{j,z}$  in (c1) (vertical black arrow on site $i$) into the $-U_{j,z}$ in (d1) (vertical red arrow on site $j$), and (3) changes no other functions.
Thus the $\xi_{x,z}^{up,m,n}$ defined by the multiplication between the expression in (c1) and (c2) is proved to be equal to $-\xi_{x,z}^{dn,m,n}$ defined by the multiplication between the expression in (d1) and (d2).
In addition to $\xi_{x,z}^{up,m,n}=-\xi_{x,z}^{dn,m,n}$, we have already known that (1) the $\bm{\mathcal{D}}(i,j)$ vector in the $P4mm$ phase of PbTiO$_3$ is proportional to $(0,-F_{xz},0)$ and (2)  $F_{xz}=\sum\limits_{m,n}\int\xi_{x,z}^{m,n}(\varepsilon,i,j)\dd\varepsilon$. 
Thus $\bm{\mathcal{D}}^{up}(i,j)=-\bm{\mathcal{D}}^{dn}(i,j)$ can be proved, where $\bm{\mathcal{D}}^{up}$ is eDMI vector in the case that the intermediate oxygen site displaced along the positive z-direction and $\bm{\mathcal{D}}^{dn}$ is eDMI vector in the case that intermediate oxygen site displaced along the negative z-direction. 
Such conclusion explains why the $\bm{\mathcal{D}}(i,j)$ vector is not homogeneous in some perovskite oxides\cite{zhao2020}, for example, with oxygen octahedral tilting where the oxygen atoms are displaced alternatively along positive and negative $z$-direction.

In Fig.~\ref{fig:rule2} (c1), we have also noticed that the inverse feature between the off-diagonal components of the force constants density (antisymmetric in force constants) $f_{x,z}=-f_{z,x}$ is due to the fact that under $m_{yz}$ operation one effective perturbation potential, $U_{i,x}$ (black arrow) that is perpendicular to $m_{yz}$ on site $i$ changes sign and the other $U_{j,z}$ (black arrow) that is parallel to $m_{yz}$ on site $j$ does not.
Such condition seems also satisfied by $U_{i,x}$ (vector perpendicular to $m_{yz}$) and $U_{j,y}$ (vector parallel to $m_{yz}$), which is corresponding to the the antisymmetric feature between force constants density $f_{x,y}=-f_{y,x}$.
However, $f_{x,y}=-f_{y,x}=0$ can be further proved by the symmetry operation of mirror $m_{xz}$ that goes through sites $i$, $j$, and $k$ (see Sec. V of the Supplemental Material\cite{sm}).
This is also consistent with the results of the force constants calculations in Eq.~(\ref{eq:fij}).
On the other hand, neither $U_{i,y}$ (vector parallel to $m_{yz}$) and $U_{j,z}$ (vector parallel to $m_{yz}$) nor $U_{i,z}$ (vector parallel to $m_{yz}$) and $U_{j,y}$ (vector parallel to $m_{yz}$) change sign under mirror $m_{yz}$, thus the off-diagonal components (in y- and z-directions) of the force constants density is only symmetric, $f_{y,z}=f_{z,y}$, and thus does not gives rise to any eDMI.
Taking consideration of $f_{yz}=f_{zy}=0$, $f_{zx}=-f_{xz}$, and $f_{xy}=f_{yx}$ into Eq.~(\ref{eq:edmivector}), we can see that the only non-zero eDMI vector component is along the y-axis which is perpendicular to the $(i,j,k)$ plane. This is going to be one important rule in the ``orbital selection rules'' section to be discussed below.

\subsection{eDMI with tiltings}
\begin{figure}[htp]
    \centering
    \includegraphics[width=0.85\linewidth]{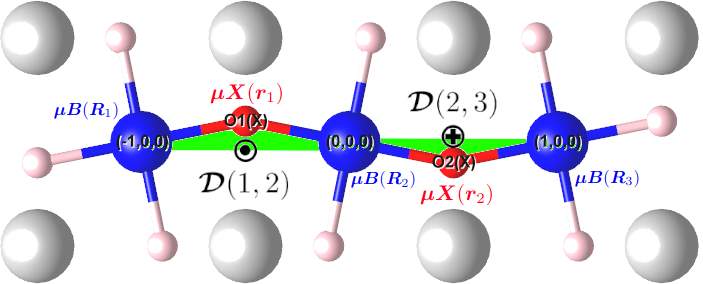}
    \caption{Illustration plot of the crystal structure with oxygen octahedral tiltings. Three 5-atom cells along [100] direction are plotted. The cell index are marked on the titanium atoms (blue ball) as $\bm{R}_{1}=(-1,0,0)$, $\bm{R}_{2}=(0,0,0)$, and $\bm{R}_{3}=(1,0,0)$, respectively. Note that the atomistic basis is defined in a unit cell in which the titanium atom locates at the origin (0,0,0) (see Fig.~5 (a) of the Supplemental Material\cite{sm}). Thus the oxygen atoms O1(X) and O2(X) site at positions $\bm{r}_{1}=\bm{R}_{1}+(0.5,0,0)$ and $\bm{r}_{2}=\bm{R}_{2}+(0.5,0,0)$ before the octahedra tilting displacements, respectively. The oxygen octahedra tiltings around titanium atoms in cell $\bm{R}_{1}$, $\bm{R}_{2}$, and $\bm{R}_{3}$ are anticlockwise, clockwise, and anticlockwise, respectively. Note that the atomic displacement variables for titanium atom ($\bm{\mu B}$) and oxygen atom along [100] directions ($\bm{\mu X}$) are labeled and their definition in unit cell can be found in Fig.~5 (a) of the Supplemental Material\cite{sm}. Note also that we use $\bm{R}$ for cell index and $\bm{r}$ for atom position in this manuscript.}\label{fig:tilting}
\end{figure}
So far we have noticed one important fact about eDMI: it comes from the indirect interaction between the nearest neighbor titanium sites through the displaced intermediate oxygen site in PbTiO$_3$, which suggests that eDMI is at least a three-body (indirect) interaction. 
As a matter of fact and as it can be seen from Fig.~\ref{fig:rule2} (a) and Fig.~\ref{fig:rule1}, the relative displacement of the intermediate site $k$ along the z-direction is critical to give non-zero antisymmetric force constants. 
It is consistent with the phenomenological model in Eq.~(\ref{eq:uuu}), in which the $\bm{u}_{i}+\bm{u}_j$ includes the relative displacement of the intermediate oxygen with respect to titanium sites $i$ and $j$. 
Such relative displacement can be seen from an equivalent expression in Eq.~(44) of the Supplemental Material\cite{sm} which is written in atomistic displacement basis which can be seen as expanding the polar mode $\bm{u}$ with the displacements ($\bm{\mu B}$) of titanium cations and their surrounding oxygen anions (O(X) in [100], O(Y) in [010], and O(Z) in [001] directions with respect to the titanium cations in Fig.~5 (b) of the Supplemental Material\cite{sm})
(As indicated in Fig. 1 (a), the polar mode $\bm{u}$ consists of titanium cations moving in one direction with respect to the oxygen anions that are on the side of titanium ions within (001) planes being moved towards the opposite direction.).
Equation~(44) of the Supplemental Material\cite{sm} tells us that the displacements of intermediate oxygen ($\bm{\mu X}$, $\bm{\mu Y}$, and $\bm{\mu Z}$ in Eq.~(44) of the Supplemental Material\cite{sm}) or equivalently polar modes $\bm{u}_{i}+\bm{u}_j$ (in Eq.~(\ref{eq:uuu}) of the main text) that are normal to $\bm{e}_{ij}$ will give rise to nonzero contribution to the eDMI energy.
On the other hand, if the local inversion symmetry breaking is from the intermediate oxygen that is displaced parallel to the line that connects sites $i$ and $j$ or equivalently $(\bm{u}_{i}+\bm{u}_j)\cross\bm{e}_{ij}=0$, the eDMI vector is always zero, as has been proven by Fig.~\ref{fig:rule1}.

Interestingly, such displacement of the intermediate oxygen site can also be associated with oxygen octahedral rotations in some perovskite materials, which is therefore consistent with a recent finding that eDMI vector can be related to such rotations \cite{zhao2020}.
More specifically, in a case with oxygen octahedra tilting pattern as in Fig.~\ref{fig:tilting}, the intermediate oxygen O1(X) that locates at position $\bm{r}_{1}=\bm{R}_{1}+(0.5,0,0)$ between titanium atoms in cells $\bm{R}_{1}=(-1,0,0)$ and $\bm{R}_{2}=(0,0,0)$ is displaced relatively upwards ($\bm{\mu X}(\bm{r}_{1})>0$) and the intermediate oxygen O2(X) that locates at position $\bm{r}_{2}=\bm{R}_{2}+(0.5,0,0)$ between titanium atoms in cells $\bm{R}_{2}=(0,0,0)$ and $\bm{R}_{3}=(1,0,0)$ is displaced relatively downwards ($\bm{\mu X}(\bm{r}_{2})<0$).
Thus two triangles (green in Fig.~\ref{fig:tilting}) are formed: (1) upward oriented triangle among titanium atom in cell $\bm{R}_{1}=(-1,0,0)$, titanium atom in $\bm{R}_{2}=(0,0,0)$, and oxygen atom O1(X) at position $\bm{r}_{1}=\bm{R}_{1}+(0.5,0,0)$ and (2) downward oriented triangle among titanium atom in cell $\bm{R}_{2}=(0,0,0)$, titanium atom in $\bm{R}_{3}=(1,0,0)$, and oxygen atom O2(X) at position $\bm{r}_{2}=\bm{R}_{2}+(0.5,0,0)$.
According to Eq.~(44) of the Supplemental Material\cite{sm} (that is the equivalent expression of Eq.~(\ref{eq:uuu}) of the main text but written in atomistic basis), the eDMI vector $\mathcal{D}(1,2)$ between $\bm{R}_{1}=(-1,0,0)$ and $\bm{R}_{2}=(0,0,0)$ (up triangle case) is proportional to $\bm{\mu X(\bm{r}_{1})}\cross\bm{e}_{ij}=(0, \mu X_{z}(\bm{r}_{1}), -\mu X_{y}(\bm{r}_{1}))$ and the eDMI vector $\mathcal{D}(2,3)$ between $\bm{R}_{2}=(0,0,0)$ and $\bm{R}_{3}=(1,0,0)$ (down triangle case) is proportional to $\bm{\mu X(\bm{r}_{2})}\cross\bm{e}_{ij}=(0, \mu X_{z}(\bm{r}_{2}), -\mu X_{y}(\bm{r}_{2}))$.
Thus $\mathcal{D}(1,2)$ is along [010] direction and $\mathcal{D}(2,3)$ is along [0-10] direction, considering that $\mu X_{z}(\bm{r}_{1})>0$, $\mu X_{z}(\bm{r}_{2})<0$, and $\mu X_{y}(\bm{r}_{1})=\mu X_{y}(\bm{r}_{2})=0$ in the tilting motions in Fig.~\ref{fig:tilting}), which is consistent with our discussion in Fig.~\ref{fig:rule1} (c) and (d) and, more importantly, reproduces the alternatively changed $\bm{\mathcal{D}}$ vectors due to oxygen octahedral tiltings as in Ref.~\cite{zhao2020}.

%Note that Eq.~(\ref{eq:uuu}) also reproduce such alternatively oriented eDMI vector, considering that, in Fig.~\ref{fig:tilting}, $\bm{u}_{1}+\bm{u}_{2}$ is along the negative vertical direction and $\bm{u}_{2}+\bm{u}_{3}$ is along the positive vertical direction and the eDMI vector is proportional to $(\bm{u}_{i}+\bm{u}_j)\cross\bm{e}_{ij}$. 
Thus our microscopic description presents a general explanation of the eDMI that is suitable to ferroelectric materials with or without oxygen octahedral tiltings.
Note that Eq.~(\ref{eq:uuu}) further implies that eDMI vector can be induced by other effects, such as polar motions. In fact, we perform DFT calculations that show that there can be noncollinear polar texture without the help of oxygen octahedral tiltings (see Sec. VII of the Supplemental Material\cite{sm}).

% \usepackage{multirow}
% \usepackage{graphicx}
\begin{table}[!htp]
\caption{\label{tab:rules} Orbital selection rules of the electric Dzyaloshinskii-Moriya Interaction in the case of three sites. The two rules are shown in the bottom of the table. In the sketch of rule 1, the $i$, $j$, and $k$ are used to indicate the three ionic sites, $m_{1}$ is a mirror that goes through $k$-site and is perpendicular to the line connecting $i$- and $j$-sites, and $m_{3}$ is a mirror that goes through all $i$-, $j$-, and $k$-sites. The table includes the results when such three rules are applied when both orbitals $m$ and $n$ range over all $s$, $p$, and $d$ orbitals. The check marks in the table indicate the non-zero orbital-resolved force constants density is in antisymmetric form $\xi_{x,z}^{m,n}=-\xi_{z,x}^{n,m}\neq 0$ and gives rise to the y-component of eDMI vector.}
\begin{center}
\resizebox{0.9\linewidth}{!}{
\begin{tabular}{|c|c|ccc|ccccc|}
\hline
              & $s$        & $p_z$      & $p_x$      & $p_y$      & $d_{z^2}$    & $d_{xz}$     & $d_{yz}$   & $d_{xy}$   & $d_{x^2-y^2}$ \\
\hline
$s$           & \checkmark & \checkmark & \checkmark &            & \checkmark   & \checkmark   &            &            & \checkmark    \\
\hline
$p_z$         & \checkmark & \checkmark & \checkmark &            & \checkmark   & \checkmark   &            &            & \checkmark    \\
$p_x$         & \checkmark & \checkmark & \checkmark &            & \checkmark   & \checkmark   &            &            & \checkmark    \\
$p_y$         &            &            &            & \checkmark &              &              &            & \checkmark &               \\
\hline
$d_{z^2}$     & \checkmark & \checkmark & \checkmark &            & \checkmark   & \checkmark   &            &            & \checkmark    \\
$d_{xz}$      & \checkmark & \checkmark & \checkmark &            &\checkmark$^e$&\checkmark$^c$&            &            & \checkmark    \\
$d_{yz}$      &            &            &            &            &              &              & \checkmark & \checkmark &               \\
$d_{xy}$      &            &            &            & \checkmark &              &              & \checkmark & \checkmark &               \\
$d_{x^2-y^2}$ & \checkmark & \checkmark & \checkmark &            &\checkmark$^f$&\checkmark$^d$&            &            & \checkmark    \\
\hline
\hline
\multicolumn{10}{|c|}{eDMI orbital selection rules:}                                            \\
\hline
\multicolumn{10}{|l|}{rule 1: non-collinear three orbitals}                                    \\
\multicolumn{10}{|l|}{\multirow{3}{*}{\includegraphics[width=0.24\textwidth]{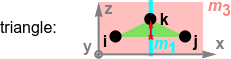}}} \\
\multicolumn{10}{|l|}{\hspace{4.5 cm},\hspace{0.2 cm}$\bm{\mathcal{D}}(i,j)\perp m_{3}$}                                                                        \\
\multicolumn{10}{|l|}{}                                                                        \\
\hline
\multicolumn{10}{|l|}{rule 2: mirror $m_3$}                                                 \\
\multicolumn{10}{|l|}{the orbitals on sites $i$ and $j$ should be either}                                                        \\
\multicolumn{10}{|l|}{both even or both odd with respect to}                                                        \\
\multicolumn{10}{|l|}{the operation of mirror $m_{3}$.}                                                        \\
\hline
\end{tabular}%
}
\end{center}
\begin{tablenotes}
\small
\item  ${}^{c}$ plotted in Fig.~\ref{fig:loops} (c); ${}^{d}$ plotted in Fig.~\ref{fig:loops} (d); 
\item  ${}^{e}$ plotted in Fig.~\ref{fig:loops} (e); ${}^{f}$ plotted in Fig.~\ref{fig:loops} (f).
\end{tablenotes}
\end{table}

\subsection{Orbital selection rules of eDMI} 
Based on the conclusions we made by the symmetry analysis as we derive the microscopic origin of eDMI in Figs.~\ref{fig:rule2} and \ref{fig:rule1}, selection rules can be summarized in order to determine what kind of orbital combinations can give rise to eDMI (see details in Sec. V of the Supplemental Material\cite{sm}).
With such orbital selection rules together with the DOS of the material, we can address which orbitals are allowed and and which orbitals possibly give rise to a quantitatively large eDMI, which makes engineering of eDMI possible.
Two rules are summarized in Tab.~\ref{tab:rules}, assuming a three-site model in which $i$ and $j$ characterize cation sites (e.g. titanium atoms) and $k$ is associated with the intermediate anion site (e.g. oxygen atom) located in the middle of sites $i$ and $j$:
(1) the local inversion symmetry should be broken by the $k$-site displacement that is off the line that goes through $i$ and $j$ sites, and thus a triangle is formed;
(2) the orbitals on sites $i$ and $j$ should be either both even or both odd with respect to the mirror that goes through all three sites.
Satisfying the two rules, eDMI $\bm{\mathcal{D}}$ vector will exist and has to be perpendicular to the triangle formed by sites $i$, $j$, and $k$;
In rule (1), not all the local inversion-symmetry breakings give rise to an eDMI vector. As a matter of fact, while there are two ways to break the local inversion symmetry, namely (i) site $k$ is displaced away from the line that connects sites $i$ and $j$, and (ii) site $k$ is displaced along the line that connects sites $i$ and $j$, the second way does not gives rise to eDMI because, as proven in Figs.~\ref{fig:rule1} (a) and (b), the orbital-resolved force constants density $\xi_{\alpha,\beta}^{m,n}$ vanishes due to the existence of mirror $m_{xy}$.
Extended tables that explain all the symmetry analysis (that are needed for summarizing Tab.~\ref{tab:rules}) are indicated in the appendix A of the Supplemental Material\cite{sm}.
Following eDMI orbital selection rules, all eDMI activated orbital combinations that involve $s$, $p$, and $d$ orbitals are derived and listed in Tab.~\ref{tab:rules}.
For instance, the main contribution to eDMI of PbTiO$_3$ from $\xi_{x,z}^{m,n}$ and $\xi_{z,x}^{n,m}$ as discussed in Figs.~\ref{fig:loops} (c), (d), (e), and (f) can also be found in Tab.~\ref{tab:rules}, check marks labeled by $c$, $d$, $e$, and $f$ respectively. 
% eDMI orbital selection rules should also work for other materials such as BiAlO$_3$. Considering that the valence shell electrons of the aluminium atom are $3s^2$ and $3p^1$ that interact with the valence shell electrons $2p^4$ of the oxygen atom which connects the two aluminium atoms, eDMI activated orbitals should be $m=s$ and $n=s$, $m=s$ and $n=p_z$, $m=s$ and $n=p_x$, $m=p_z$ and $n=s$, $m=p_z$ and $n=p_z$, $m=s$ and $n=p_x$, $m=p_x$ and $n=s$, $m=p_x$ and $n=p_z$, $m=p_x$ and $n=p_x$, and $m=p_y$ and $n=p_y$. Taking account that $m=p_{x}$ and $n=p_{x}$ orbitals overlap mostly with the intermediate oxygen orbitals, the main contribution to eDMI should come from $m=p_{x}$ on site $i$ and $n=p_{x}$ on site $j$. 
Note that Tab.~\ref{tab:rules} is also confirmed by our numerical results involving these $s$, $p$, and $d$ orbitals (see sections IV and V, and Fig.~1 of the Supplemental Material\cite{sm}).

\section{further remarks}
In this work, we discuss an overlooked intrinsic dipolar interaction (eDMI) that could give chiral polar structures; However, there are other energy terms and extrinsic conditions that compete with such kind of coupling term, which could result in that (1) noncollinear textures are metastable high energy phases, such as the Bloch component in the domain walls (high energy defects)\cite{Wojde2014,PereiraGonalves2019,Wang2014,Wu2018}; (2) only narrow region in the phase diagram (with respect to temperature, strain, and even external electric field) shows nontrivial topological polar texture, also similar to the magnetic situations\cite{Neubauer2009,Nagaosa2013}; and (3) hidden phases corresponds to such kind of coupling could exist under external probes\cite{Li2019}, such as laser pulses.

Based on the arguments provided in this manuscript, the eDMI should have no direct relation to the origin of the polar instability. Consequently, eDMI may also exist in different types of ferroelectrics ({\it e.g.} hyperferroelectrics, geometric/steric ferroelectrics, and even ferroelectric metals) if there are no other mechanisms/constraints forbidding it.

Note that the original effective Hamiltonian method\cite{Zhong1995} considers up to the second-order for polar-polar interaction between cells, which gives rise to only symmetric force constant matrix – unlike what the DFPT calculations tell us. 
The Ginzburg term in the phase field model\cite{Chen2008} (Ginzburg-Landau-Devonshire theory\cite{landau1936,ginzburg1945,ginzburg1949}) obviously cannot reproduce antisymmetric force constants either which can be easily seen by realizing that the Ginzburg term is only the power of the gradient of polar modes. 
%However, as calculated from DFPT, many important materials include (relatively large) antisymmetric parts of the force constants, such as PbTiO3, BaTiO3, KNbO3, etc. 
Consequently, our work should lead to revisiting currently used models.

It is also worth mentioning that our proposed microscopic theory for eDMI can be linked to several previous theoretical works. For example, the three-site model discussed in the present manuscript bears some analogy with the ``triple-dipole-interaction'' problem\cite{Axilrod1943,Axilrod1951,Barash1984} among three neutral atoms (van der Waals-type interaction). 
Interestingly, the proposed Axilrod-Teller potential for this ``triple-dipole-interaction'' can also give a nonzero antisymmetric part of the force constants, though the influence of the covalent bonding on the force constants was omitted (only polarization fluctuations were taken into account). 
In contrast, the study of vibronic instability\cite{Bersuker1966,bersuker1978,bersuker2012} of some crystals, discussed the effects of covalent bonding but did not pay attention to the antisymmetric part of the force constants. 
We are also aware of a simple three-site LCAO model developed by Prosandeev et. al.\cite{Sergey1989} discussing the correlation of the local atomic displacements in perovskites.
Such kind of correlations were earlier experimentally discovered from the diffuse scattering of neutrons in KNbO$_3$\cite{Coms1970,Currat1974}.

\section{Conclusion} 
In conclusion, the microscopic origin of the electric Dzyaloshinskii-Moriya interaction is unveiled and discussed thanks to analytical analysis of the orbital-resolved force constants calculations.
Our present study, therefore, emphasizes that eDMI exists and sheds some light into its origin, that  is eDMI is an electron-mediated quantum effect in which (1) the local inversion symmetry breaking activates previously forbidden electron hopping channels on adjacent atomic sites; and (2) combination of the orbitals with particular symmetry (following eDMI orbital selection rules detailed in Tab.~\ref{tab:rules}) results in the electric Dzyaloshinskii-Moriya interaction.  
Though both eDMI and mDMI need electron hopping channels and local-inversion-symmetry breaking to occur, mDMI needs spin-orbit coupling (soc) to connect spin up and down, unlike eDMI. Thus eDMI naturally exists in polar materials because of the general existence of local-inversion-symmetry breaking.
Moreover, eDMI energy is found to be at least third order in ionic displacements while mDMI energy ``only'' involves a second order in magnetic moments. 
Such differences may result in the formation of some exotic dipole textures that can differ from the extensively explored magnetic arrangements.

\begin{acknowledgements}
The work is supported by ONR under Grant No. N00014-17-1-2818 (P.C. and L.B.), the Vannevar Bush Faculty Fellowship (VBFF) grant no.
N00014-20-1-2834 from the Department of Defense (H.J.Z and L.B.) and the ARO Grant No. W911NF-21-1-0113. (L.B.). H.J.Z. and P.C. thank Prof. Wenhui Mi for the valuable discussions. The first-principle simulations and tight-binding calculations were done using the Arkansas High-Performance Computing Center.
\end{acknowledgements}

\setcounter{section}{0}
\setcounter{equation}{0}
\setcounter{figure}{0}
\setcounter{table}{0}

\title{~\\[0.4in]Supplemental Material: \\ Microscopic Origin of the Electric Dzyaloshinskii-Moriya Interaction}

\maketitle
\onecolumngrid
In this Supplemental Material (SM), we explain the relation between the antisymmetric part of force constants (FC) and electric Dzyaloshinskii-Moriya interaction (eDMI) in section I, and why two-body coulomb interaction only gives rise to symmetric FC in section II. We also give information about how to derive the FC from the potential energy surface in section III, and the formalism of the orbital-resolved FC in section IV; Then we describe how the orbital selection rules of eDMI are summarized in section V; Section VI shows and explains how to derive the third order form of eDMI in perovskite ferroelectrics; and, in the end, VII reports the effective Hamiltonian model parameters and results for domain walls.

\section{Antisymmetric part of force constant and eDMI}
The energy from the ionic displacements in harmonic approximation can be written in terms of the force constants (FC) as
\begin{equation}\label{eq:harmonic}
\begin{split}
E=\frac{1}{2}\sum\limits_{i,j}^{N}\sum\limits_{\alpha,\beta}^{x,y,z}F_{\alpha,\beta}(i,j)u_{i,\alpha}u_{j,\beta}
\end{split}
\end{equation}
where $i$ and $j$ are two ionic sites that run over all $N$ sites, $u_{i,\alpha}$ and $u_{j,\beta}$ are the ionic displacements on sites $i$ and $j$ along the $\alpha$ and $\beta$ Cartesian direction, respectively, and $F(i,j)$ is the total FC matrix.
The FC can be split into symmetric ($F^{S}$) and antisymmetric ($F^{A}$) parts as 
\begin{equation}\label{eq:fc}
\begin{split}
F(i,j)=F^{S}(i,j)+F^{A}(i,j),
\end{split}
\end{equation}
where 
\begin{align}
F^{S}(i,j)=\frac{1}{2}[F(i,j)+F^{\text{T}}(i,j)] \label{eq:fcs} \\ 
F^{A}(i,j)=\frac{1}{2}[F(i,j)-F^{\text{T}}(i,j)] \label{eq:fca}
\end{align}
in which T denotes the matrix transpose operation. 
Minding the energy from only the antisymmetric part of the FC, we get:
\begin{equation}\label{eq:Ea}
\begin{split}
E^{A}=&\frac{1}{2}\sum\limits_{i,j}[(u_{i,y}u_{j,z}-u_{i,z}u_{j,y})F^{A}_{yz}
+(u_{i,x}u_{j,z}-u_{i,z}u_{j,x})F^{A}_{xz}
+(u_{i,x}u_{j,y}-u_{i,y}u_{j,x})F^{A}_{xy}] \\
=&\frac{1}{2}\sum\limits_{i,j}\bm{\mathcal{D}}(i,j)\cdot(\bm{u}_{i}\cross\bm{u}_{j}) 
\end{split}
\end{equation}
where $\bm{\mathcal{D}}(i,j)=(F^{A}_{yz},F^{A}_{zx},F^{A}_{xy})$. The $\bm{\mathcal{D}}(i,j)$ vector that defines eDMI is thus directly related to the antisymmetric part of the FC. 

\section{two-body Coulomb interaction and symmetric FC}
The two-body Coulomb energy potential can be written as $V_{ion-ion}=\frac{1}{2}\sum\limits_{i\neq j}\frac{Z_{i}Z_{j}}{\abs{\bm{\tau}_{i}-\bm{\tau}_{j}}}$. 
Its antisymmetric part of the force constants is defined as $F_{\alpha,\beta}^{A}(i,j)=\frac{1}{2}(\frac{\partial^{2}V_{ion-ion}}{\partial \tau_{i,\alpha}\partial \tau_{j,\beta}}-\frac{\partial^{2}V_{ion-ion}^2}{\partial \tau_{i,\beta}\partial \tau_{j,\alpha}})$.
By taking the derivative of $V_{ion-ion}$ with respect to the ionic displacements $\tau_{i,\alpha}$ and $\tau_{j,\beta}$ one by one, we have
\begin{align}
\dfrac{\partial V_{ion-ion}}{\partial \tau_{i,\alpha}}=&-\sum\limits_{j}\dfrac{\tau_{i,\alpha}-\tau_{j,\alpha}}{\abs{\bm{\tau}_{i}-\bm{\tau}_{j}}^{3}} \label{eq:coulomb} \\
\dfrac{\partial^{2} V_{ion-ion}}{\partial\tau_{i,\alpha}\partial\tau_{j,\beta}}=&-\dfrac{(\tau_{i,\alpha}-\tau_{j,\alpha})(\tau_{i,\beta}-\tau_{j,\beta})}{\abs{\bm{\tau}_{i}-\bm{\tau}_{j}}^{5}} \label{eq:coulomb2} \\
\dfrac{\partial^{2} V_{ion-ion}}{\partial\tau_{i,\beta}\partial\tau_{j,\alpha}}=&-\dfrac{(\tau_{i,\beta}-\tau_{j,\beta})(\tau_{i,\alpha}-\tau_{j,\alpha})}{\abs{\bm{\tau}_{i}-\bm{\tau}_{j}}^{5}} \label{eq:coulomb3}
\end{align}
Looking at Eqs.~(\ref{eq:coulomb2}) and ~(\ref{eq:coulomb3}), the relation $\frac{\partial^{2}V_{ion-ion}}{\partial \tau_{i,\alpha}\partial \tau_{j,\beta}}=\frac{\partial^{2} V_{ion-ion}}{\partial \tau_{i,\beta}\partial \tau_{j,\alpha}}$ is always true. Thus the force constants from two body Coulomb interaction are always symmetric.  The $\bm{\mathcal{D}}(i,j)$ vector thus cannot be due to two-body Coulomb ion-ion interaction.

\section{potential energy surface and FC}
To determine origin of eDMI, we decided to look in details at the microscopic full-Hamiltonian\cite{Baroni2001} (involving both electrons and ions) and derive the potential energy surface and its Hessian matrix.
The full-Hamiltonian of solids involves both electrons' and ions' degree of freedoms and is written as
\begin{eqnarray}\label{eq:KS}
\hat{H}=\hat{T}_{e}+\hat{T}_{i}+\hat{V}_{ee}+\hat{V}_{ii}+\hat{V}_{ie}
\end{eqnarray}
where $\hat{T}_e$ and $\hat{T}_i$ are kinetic energies of electrons and ions, respectively; $\hat{V}_{ee}$ denotes the Coulomb interaction among electrons; $\hat{V}_{ii}$ is the Coulomb interaction among ions; and $\hat{V}_{ie}$ is the Coulomb interaction between electrons and ions.
To find the solutions (eigenfunction and eigenenergy) for both electrons and ions, the Schrodinger equation $\hat{H}\Psi(\bm{r},\bm{\tau})=\mathcal{E}\Psi(\bm{r},\bm{\tau})$ needs to be solved, where $\bm{r}$ and $\bm{\tau}$ are the electronic and ionic degree of freedoms, respectively.
Due to the large difference of the electron mass $m$ and ion mass $M$, the light electrons move much faster than the heavy ions and can instantaneously follow the motion of the ions. 
Thus the description of the electrons can be seen as small vibrations around the rest positions of the ions.
Such approximation is called Born–Oppenheimer approximation\cite{Born1928} or adiabatic approximation, within which the  total wavefunction can be written as a product of $\Psi(\bm{r},\bm{\tau})=\chi(\bm{\tau})\psi(\bm{r};\bm{\tau})$, where the electronic wavefunction depends parametrically on the ion coordinates. 
Then the ionic and electronic wavefunctions can be separately solved by
\begin{align}
(\hat{T}_{i}+\hat{V}_{ii}+E_{n}(\bm{\tau}))\chi(\bm{\tau})=\mathcal{E}(\bm{\tau})\chi(\bm{\tau}) \label{eq:ion}\\
(\hat{T}_{e}+\hat{V}_{ee}+\hat{V}_{ie}(\bm{\tau}))\psi_{n}(\bm{r};\bm{\tau})=E_{n}(\bm{\tau})\psi_{n}(\bm{r};\bm{\tau}) \label{eq:electron}
\end{align}
The  electronic system in Eq.~(\ref{eq:electron}) couples to the ionic system described by Eq.~(\ref{eq:ion}) via its n-th eigenstate with energy $E_n(\bm{\tau})$.
As can be seen from Eq.~(\ref{eq:ion}), the effective potential of ions in the adiabatic approximation can be written as\cite{Baroni2001}:
\begin{equation}\label{eq:omega}
\begin{split}
\Omega(\bm{\tau})=&V_{ii}(\bm{\tau}) \\
&+\sum\limits_{n}^{N_{occ}}\int\Psi_{n}^{*}(\bm{r};\bm{\tau})[\hat{T}_{e}+\hat{V}_{ee}+\hat{V}_{ie}]\Psi_{n}(\bm{r};\bm{\tau})\dd\bm{r}
\end{split}
\end{equation}
The first term on the right hand side of Eq.~(\ref{eq:omega}) is totally ionic in nature and the second term of Eq.~(\ref{eq:omega}) is the potential energy of ions interacting with the electronic ground state. Note that the electronic wave function $\Psi_{n}(\bm{r};\bm{\tau})$ implicitly depends upon $\bm{\tau}$ due to the electron-ion interaction, $\hat{V}_{ie}$.
According to the Hellmann-Feynman theorem\cite{Feynman1939,hellmann2015}, the force acting on the $i$-th ion from the electronic ground state can be written as 
$-\int d\bm{r}\Psi^{*}(\bm{r};\bm{\tau})\frac{\partial \hat{V}_{ie}(\bm{r};\bm{\tau})}{\partial\bm{\tau}_{i}}\Psi(\bm{r};\bm{\tau})=-\int \dd\bm{r}\frac{\partial \hat{V}_{ie}(\bm{r};\bm{\tau})}{\partial\bm{\tau}_{i}}n(\bm{r};\bm{\tau})$, where $n(\bm{r};\bm{\tau})$ is the electronic density,
considering that $\hat{T}_{e}$ and $\hat{V}_{ee}$ do not depend on the parameters $\bm{\tau}$. Then by taking second order of derivatives of the total energy with respect to ionic displacements, the force constants can be derived as:
\begin{equation}\label{smeq:hessian}
\begin{split}
F_{\alpha\beta}(i,j)=&\frac{\partial^2 \Omega(\bm{\tau})}{\partial \tau_{i,\alpha}\partial \tau_{j,\beta}} \\
=&\frac{\partial^{2}V_{ii}(\bm{\tau})}{\partial\tau_{i,\alpha}\partial \tau_{j,\beta}}+\int\dd\bm{r}\frac{\partial V_{ie}(\bm{r};\bm{\tau})}{\partial\tau_{i,\alpha}}\frac{\partial n(\bm{r};\bm{\tau})}{\partial\tau_{j,\beta}}.
\end{split}
\end{equation}
It thus involves the second derivative of the potential energy surface $\Omega(\bm{\tau})$ with respect to ionic displacements $\bm{\tau}_{i}$ and $\bm{\tau}_{j}$ $(i\neq j)$ (e.g., Ti$_{0}$ and Ti$_{+1}$ in Fig.~1 of the main text) along $\alpha$ and $\beta$ Cartesian directions, respectively.
Note that $n(\bm{r};\bm{\tau})$ is the electronic density, while $V_{ii}(\bm{\tau})$ and $V_{ie}(\bm{r};\bm{\tau})$ are the energy potential of ion-ion interaction and ion-electron interaction, respectively.
Since $V_{ii}$ is a sum of repulsive ion-ion Coulomb interactions, the first term of the right side of Eq.~(\ref{smeq:hessian}) only contributes to the symmetric part of the FC (see proofs in Sec~II of the SM).
Thus the anti-symmetric FC has to come from the second term of the right side of Eq.~(\ref{smeq:hessian}), which in fact can be seen as the electron-density-mediated ion-ion indirect interaction.
More specifically, the second part of Eq.~(\ref{smeq:hessian}) indicates that (i) the ionic displacement $\tau_{i,\alpha}$ induces a variation of electron-ion energy $\frac{\partial V_{ie}(\bm{r};\bm{\tau})}{\partial \tau_{i,\alpha}}$ at site $i$, (ii) which couples to the ionic displacement $\tau_{j,\beta}$ on site $j$ through the electron density fluctuation $\frac{\partial n(\bm{r};\bm{\tau})}{\partial \tau_{j,\beta}}$.
Thus though eDMI is mostly associated with ionic dipoles, Eq.~(\ref{smeq:hessian}) tells that eDMI is not an ionic dipole-dipole interaction but rather has to be an electron-mediated quantum effect.
 
\section{calculation of the orbital-resolved FC}
In this section, we provide the derivation of the orbital-resolved FC following the previous works of Refs.~\citen{Lannoo1979,Elstner1998,Szilva2013}. It is worth to mention that the perturbation with respect to the ionic displacements performed in this section gives rise to the energy variation as in Eq.~(\ref{eq:dEwgwg}) which is mathematically similar to the formalism of the perturbation with respect to the spin rotations as in Refs.~\citen{Liechtenstein1987,Lounis2010,Szilva2013,He2021} in which they call it magnetic force theorem because the ``force constants'' for spins were derived.

As discussed in both Sec. III of the SM and the main text, the antisymmetric part of the FC can be addressed from only the electronic system as in Eq.~(\ref{eq:electron}) in the adiabatic approximation.
Let us start with a perfect crystal Hamiltonian operator $\hat{H}^{0}$ for such electronic system.
Its matrix elements in Wannier basis can be written as $\langle m,i | \hat{H}^{0} | n,j \rangle$ where the first index in the each of the atomic state (Dirac bracket) denotes the atomic Wannier orbital and the second one the atomic site (Wannier center).
For the crystal with displaced atomic site, the Hamiltonian operator becomes $\hat{H}$  and the atomic state $|m^{\prime},i\rangle$ are rigidly translated from $|m,i\rangle$ according to the nuclei displacement.
The energy change can thus be calculated using standard perturbation theory with an screened effective perturbation potential\cite{Lannoo1979,economou2006} $\tilde{W}$ whose matrix elements are defined as $\mel{m,i}{\tilde{W}}{n,j}=\mel{m^{\prime},i}{\hat{H}}{n^{\prime},j}-\mel{m,i}{\hat{H}^{0}}{n,j}$. Note that the $\tilde{}$ symbol is to indicate the fact that the potential is screened by the change of the electron density due to the ionic displacements.
The best way to perform the perturbation is by Green's functions\cite{economou2006}.
We thus define $\hat{G}^{0}$ and $\hat{G}$ as the operators of the Green's functions for $\hat{H}^{0}$ and $\hat{H}$, respectively. According to the Dyson's equation\cite{Dyson1949,economou2006}, $\hat{G}$ can be related to $\tilde{W}$ via:
\begin{align}
\hat{G}=&~\hat{G}^{0}+\hat{G}^{0}\tilde{W}\hat{G} \nonumber \\
=&~\hat{G}^{0}+\hat{G}^{0}\tilde{W}\hat{G}^{0}+\hat{G}^{0}\tilde{W}\hat{G}^{0}\tilde{W}\hat{G}^{0}+\dotsb \label{eq:dyson}
\end{align}
This equation represents how the perturbation in $\tilde{W}$ from ionic displacements updates the Green's function, which will in the end alternates the energy of the system.
To write the total energy of the electronic system with respect to Green's function $G^{0}$ and the effective perturbation potential $\tilde{W}$,
we start with the total energy in the electron density\cite{Elstner1998}:
\begin{align}
E=2\int_{-\infty}^{\varepsilon_{f}}\varrho(\varepsilon)\varepsilon\dd\varepsilon-\frac{1}{2}\int n(\bm{r})V_{h}[n](\bm{r})\dd\bm{r}-\int n(\bm{r})V_{xc}[n](\bm{r})\dd\bm{r} \label{eq:Etot}
\end{align}
The first term represents the sum of one-electron Kohn-Sham energies (band energy) where $\varrho(\varepsilon)$ is the density of states (DOS) at energy $\varepsilon$, the integration being performed over the valence bands. 
Note that the factor 2 in front of the first term is due to the spin multiplicity.
The second term corrects for the double counting of the Hartree energy where $V_{h}[n](\bm{r})$ is the Hartree potential as a functional of electron density $n(\bm{r})$.
The third term corrects for the double counting of the exchange-correlation (XC) energy where $V_{xc}[n](\bm{r})$ is the XC potential, which is  also a functional of electron density.
The variation of the total energy with respect to the variation of electron density due to the ionic displacements can be derived from Eq.~(\ref{eq:Etot}) as:
\begin{subequations} \label{eq:dE}
\begin{align} 
    \delta E=&2\int_{-\infty}^{\varepsilon_{f}}\delta\varrho(\varepsilon)\varepsilon\dd\varepsilon \label{eq:dE1}\\
    &-\frac{1}{2}\int \delta n(\bm{r})W[n_{0}](\bm{r})\dd\bm{r} \label{eq:dE2} \\
    &-\frac{1}{2}\int n_{0}(\bm{r})\delta W[n](\bm{r})\dd\bm{r} \label{eq:dE3} \\
    &-\frac{1}{2}\int \delta n(\bm{r})\delta W[n](\bm{r})\dd\bm{r} \label{eq:dE4}
\end{align}
\end{subequations}
where the symbol $\delta$ means taking variation of functions, $n_{0}(\bm{r})$ is the unperturbed electronic density,  $n(\bm{r})=n_{0}(\bm{r})+\delta n(\bm{r})$, and $W$ is defined as $\mathrm{i}(V_{h}+2V_{xc})$ (the connection between $W$ and $\tilde{W}$ will show itself during the derivation later on). Note that $W$ is a functional of electron density, which is responsible to the energy change from the electronic density fluctuations.
Next, we will rewrite each term in Eq.~(\ref{eq:dE}) using Green's function and effective perturbation potential. 
\\
%\nointend{\bf To rewrite Eq.~(\ref{eq:dE1}):} 

{\bf\noindent Let us rewrite Eq.~(\ref{eq:dE1}):}\\
Using integration by parts, we have
\begin{align}
\int_{-\infty}^{\varepsilon_{f}}\delta\varrho(\varepsilon)\varepsilon\dd\varepsilon=-\int_{-\infty}^{\varepsilon_{f}}\delta\mathcal{N}(\varepsilon)\dd\varepsilon \label{eq:idos}
\end{align}
where $\mathcal{N}(\varepsilon)$ is the integrated DOS which denotes the total number of states having energy less than $\varepsilon$.
We can also write the DOS using Eq.~(\ref{eq:dyson})\cite{economou2006} in the second order approximation:
\begin{align}
\delta\varrho(\varepsilon)=&-\frac{1}{\pi}\text{Im}\text{Tr}(G-G^{0}) \nonumber \\
\simeq&-\frac{1}{\pi}\text{Im}\text{Tr}[G^{0}\tilde{W}G^{0}+G^{0}\tilde{W}G^{0}\tilde{W}G^{0}] \label{eq:dos}
\end{align}
where Im and Tr are operations to take imaginary part and trace, respectively. 
Using the fact that $(G^{0})^{2}$ is equal to $-\dd G^{0}/\dd\varepsilon$ and the cyclic property of the trace operation, Eq.~(\ref{eq:dos}) can be transformed into
\begin{align}
\delta\varrho(\varepsilon)=\frac{1}{\pi}\frac{\dd}{\dd\varepsilon}\text{Im}\text{Tr}[\tilde{W}G^{0}+1/2\tilde{W}G^{0}\tilde{W}G^{0}] \label{eq:dos2}
\end{align}
Considering $\varrho(\varepsilon)=\frac{\dd\mathcal{N}(\varepsilon)}{\dd\varepsilon}$ and substituting Eqs.~(\ref{eq:dos2}) and (\ref{eq:idos}), Eq.~(\ref{eq:dE1}) can finally be written as
\begin{align}
2\int_{-\infty}^{\varepsilon_{f}}\delta\varrho(\varepsilon)\varepsilon\dd\varepsilon=\int_{-\infty}^{\varepsilon_{f}}-2\frac{1}{\pi}\text{Im}\text{Tr}[\tilde{W}G^{0}+1/2\tilde{W}G^{0}\tilde{W}G^{0}]\dd\varepsilon \label{eq:dE1new}
\end{align}
\\
{\bf\noindent Let us now rewrite Eqs.~(\ref{eq:dE2}) and (\ref{eq:dE3}):}\\
Note that Hartree potential is written as $V_{h}[n](\bm{r})=\int\frac{n(\bm{r}^{\prime})}{\abs{\bm{r}-\bm{r}^{\prime}}}\dd\bm{r}^{\prime}$ which is linear with respect to electron density. Writing the change of the Hartree energy specifically gives $\int n_{0}(\bm{r})\delta V_{h}[n](\bm{r})\dd\bm{r}=\int\frac{n_{0}(\bm{r})\delta n(\bm{r}^{\prime})}{\abs{\bm{r}-\bm{r}^{\prime}}}\dd\bm{r}\dd\bm{r}^{\prime}$ and $\int\delta n(\bm{r})V_{h}[n_{0}](\bm{r})\dd\bm{r}=\int\frac{\delta n(\bm{r})n_{0}(\bm{r}^{\prime})}{\abs{\bm{r}-\bm{r}^{\prime}}}\dd\bm{r}\dd\bm{r}^{\prime}$, which gives:
\begin{align}
\int\delta n(\bm{r})V_{h}[n_{0}](\bm{r})\dd\bm{r}=\int n_{0}(\bm{r})\delta V_{h}[n](\bm{r})\dd\bm{r} \label{eq:vhar}
\end{align}
Similarly, in the linear response, $V_{xc}[n](\bm{r})$ can also be written as linear with respect to electron density $V_{xc}[n](\bm{r})\simeq V_{xc}[n_{0}](\bm{r})+(\frac{\delta V_{xc}}{\delta n}|_{n=n_{0}})(n(\bm{r})-n_{0}(\bm{r}))$.
Thus we can have $\int n_{0}(\bm{r})\delta V_{xc}[n](\bm{r})\dd\bm{r}\simeq\int n_{0}(\bm{r})\delta n(\bm{r})(\frac{\delta V_{xc}}{\delta n}|_{n=n_{0}})\dd\bm{r}$ and $\int \delta n(\bm{r}) V_{xc}[n](\bm{r})\dd\bm{r}\simeq\int\delta n(\bm{r})n_{0}(\bm{r})(\frac{\delta V_{xc}}{\delta n}|_{n=n_{0}})\dd\bm{r}+\int\delta n(\bm{r})\Pi[n_{0}](\bm{r})\dd\bm{r}$, where $\Pi[n_{0}](\bm{r})=V_{xc}[n_{0}](\bm{r})-n_{0}(\bm{r})(\frac{\delta V_{xc}}{\delta n}|_{n=n_{0}})$ and second and higher order electron density variation is dropped, which gives
\begin{align}
\int \delta n(\bm{r}) V_{xc}[n_{0}](\bm{r})\dd\bm{r}=\int n_{0}(\bm{r})\delta V_{xc}[n](\bm{r})\dd\bm{r}
+\int\delta n(\bm{r})\Pi[n_{0}](\bm{r})\dd\bm{r} \label{eq:vxc}
\end{align}
Substituting Eqs.~(\ref{eq:vhar}) and (\ref{eq:vxc}) into Eqs.~(\ref{eq:dE2}) and (\ref{eq:dE3}) and summing up gives:
\begin{align}
-\frac{1}{2}\int \delta n(\bm{r})W[n_{0}](\bm{r})\dd\bm{r}-\frac{1}{2}\int n_{0}(\bm{r})\delta W[n](\bm{r})\dd\bm{r}=-\int n_{0}(\bm{r})\delta W[n](\bm{r})\dd\bm{r}-\int\delta n(\bm{r})\Pi[n_{0}](\bm{r})\dd\bm{r} \label{eq:dndn}
\end{align}
Moreover, using the Cauchy's integral theorem and the completeness of the Wannier basis we can prove the following relation:
\begin{align}
\int_{-\infty}^{\varepsilon_{f}}\frac{2}{\pi}\text{Im}\text{Tr}[G^{0}\delta W]\dd\varepsilon = & 
\int_{-\infty}^{\varepsilon_{f}}\frac{2}{\pi}\text{Im}\sum\limits_{n}\sum\limits_{s}\sum\limits_{m}\frac{\bra{w_{n}}\ket{\phi^{0}_{m}}\bra{\phi^{0}_{m}}\ket{w_{s}}}{\varepsilon-\varepsilon_{m}+i\eta}\bra{w_{s}}\delta W\ket{w_{n}}\dd\varepsilon \nonumber \\
=&-2\sum\limits_{n}\sum\limits_{s}\sum\limits_{m}^{occ}\bra{w_{n}}\ket{\phi^{0}_{m}}\bra{\phi^{0}_{m}}\ket{w_{s}}\bra{w_{s}}\delta W\ket{w_{n}} \nonumber \\
=&-2\sum\limits_{m}^{occ}\bra{\phi^{0}_{m}}\delta W\ket{\phi^{0}_{m}}=-\int n_{0}(\bm{r})\delta W\dd\bm{r} \label{eq:cauchy1}
\end{align}
where $\ket{w_{n}}$ is the Wannier orbital (our TB basis), $\ket{\phi^{0}_n}$ is the $n^{th}$ wavefunction of $\hat{H}^{0}$, and ``occ'' means all occupied bands.
Following the same derivation as in Eq.~(\ref{eq:cauchy1}) and considering $\delta n(\bm{r})=n(\bm{r})-n_{0}(\bm{r})$, we can have
\begin{align}
\int_{-\infty}^{\varepsilon_{f}}\frac{2}{\pi}\text{Im}\text{Tr}[(G-G^{0})\Pi[n_{0}](\bm{r})]\dd\varepsilon=-\int \delta n(\bm{r})\Pi[n_{0}](\bm{r})\dd\bm{r} \label{eq:cauchy2}
\end{align}
Substituting Eqs.~(\ref{eq:cauchy1}) and (\ref{eq:cauchy2}) into Eq.~(\ref{eq:dndn}), the second and third terms of Eq.(\ref{eq:dE}) are finally written together as
\begin{align}
-\frac{1}{2}\int \delta n(\bm{r})W[n_{0}](\bm{r})\dd\bm{r}-\frac{1}{2}\int n_{0}(\bm{r})\delta W[n](\bm{r})\dd\bm{r}=
\int_{-\infty}^{\varepsilon_{f}}\frac{2}{\pi}\text{Im}\text{Tr}[G^{0}\delta W]\dd\varepsilon
+\int_{-\infty}^{\varepsilon_{f}}\frac{2}{\pi}\text{Im}\text{Tr}[(G-G^{0})\Pi[n_{0}](\bm{r})]\dd\varepsilon \label{eq:dE23final}
\end{align}
\\
{\bf\noindent Let us now rewrite Eq.~(\ref{eq:dE4}):}\\
Utilizing Eq.~(\ref{eq:cauchy1}) again and considering $\delta n(\bm{r})=n(\bm{r})-n_{0}(\bm{r})$,  Eq.~(\ref{eq:dE4}) can be rewritten as:
\begin{align}
-\frac{1}{2}\int\delta n(\bm{r})\delta W\dd\bm{r}=\int_{-\infty}^{\varepsilon_{f}}\frac{1}{\pi}\text{Im}\text{Tr}[(G-G^{0})\delta W]\dd\varepsilon \label{eq:dE4new}
\end{align}
Finally, collecting all the terms in Eqs.~(\ref{eq:dE1new}), (\ref{eq:dE23final}), and (\ref{eq:dE4new}), the variation of the total energy in Eq.~(\ref{eq:dE}) can be rewritten as\cite{Lannoo1979,Moraitis1984}
\begin{align}
\delta E=\int_{-\infty}^{\varepsilon_{f}}-\frac{2}{\pi}\text{Im}\text{Tr}[W^{b}G^{0}+1/2W^{b}G^{0}\tilde{W}G^{0}]\dd\varepsilon
-\int_{-\infty}^{\varepsilon_{f}}\frac{2}{\pi}\text{Im}\text{Tr}[\tilde{W}G^{0}\Pi[n_{0}]G^{0}]\dd\varepsilon
\label{eq:dEwgwg}
\end{align}
where the bare effective perturbation potential $W^{b}$ is defined as $\tilde{W}-\delta W$. Note that by bare, we mean that the effective perturbation potential comes from only the nuclei displacements, which is accomplished by subtracting $\delta W$ that is associated with the electron density fluctuations.
The screened effective perturbation\cite{Lannoo1979,economou2006} $\tilde{W}$ can also be related to $W^{b}$ via the inverse dielectric constant, that is $\tilde{W}=\epsilon^{-1}W^{b}$, where $\epsilon$ is the dielectric constant. 
Thus $\delta W=\tilde{W}-W^{b}=(1-\epsilon)\tilde{W}=\chi_{0}\tilde{W}$ can be derived, where $\chi_{0}$ is the charge susceptibility of the non-interacting Kohn-Sham system, which is consistent with the well-known Adler-Wiser form\cite{Adler1962,Wiser1963}.
The force constants $F_{\alpha\beta}(i,j)$ then can be derived by performing second derivatives of the energy with respect to $\bm{u}_{i}$ and $\bm{u}_{j}$:
\begin{align}
F_{\alpha,\beta}(i,j)=&\int_{-\infty}^{\varepsilon_{f}}-\frac{1}{2\pi}\text{Im}\text{Tr}[
W^{b}_{i,\alpha}G^{0}\tilde{W}_{j,\beta}G^{0}+\tilde{W}_{i,\alpha}G^{0}W^{b}_{j,\beta}G^{0}] \nonumber \\
=&\int_{-\infty}^{\varepsilon_{f}}-\frac{1}{2\pi}\text{Im}\text{Tr}[
W^{b}_{i,\alpha}G^{0}W^{b}_{j,\beta}G^{0}
+\tilde{W}_{i,\alpha}G^{0}\tilde{W}_{j,\beta}G^{0}
-(W^{b}_{i,\alpha}-\tilde{W}_{i,\alpha})G^{0}(W^{b}_{j,\beta}-\tilde{W}_{j,\beta})G^{0}
] \nonumber \\
=&\int_{-\infty}^{\varepsilon_{f}}-\frac{1}{2\pi}\text{Im}\text{Tr}[
W^{b}_{i,\alpha}G^{0}W^{b}_{j,\beta}G^{0}
+\tilde{W}_{i,\alpha}G^{0}\tilde{W}_{j,\beta}G^{0}
-W_{i,\alpha}G^{0}W_{j,\beta}G^{0}
] \nonumber \\
=&\sum\limits_{\gamma=1,3}\int_{-\infty}^{\varepsilon_{f}}-\frac{1}{2\pi}\text{Im}\text{Tr}[U^{\gamma}_{i,\alpha}G^{0}U^{\gamma}_{j,\beta}G^{0}] \label{eq:wgwg0}
\end{align}
where $U^{1}_{i,\alpha}=W^{b}_{i,\alpha}$, $U^{2}_{i,\alpha}=\tilde{W}_{i,\alpha}$, and $U^{3}_{i,\alpha}=\mathrm{i} W_{i,\alpha}$ (imaginary unit $\mathrm{i}$ is used for the convenience of making a compact form), $W_{i,\alpha}$ is taken here as a short notation of $\frac{\partial W}{\partial u_{i,\alpha}}$. Note that here only the first order electron density response in $W^{b}$ and $\tilde{W}$ are considered.
If we specifically write out the orbital summations in the trace of Eq.~(\ref{eq:wgwg0}):
\begin{align}
F_{\alpha\beta}(i,j)=&\int_{-\infty}^{\varepsilon_{f}} f_{\alpha,\beta}(\varepsilon,i,j)\dd \varepsilon \label{eq:wgwg1} \\
f_{\alpha\beta}(\varepsilon,i,j)=&
\sum\limits_{m}\sum\limits_{n}\xi^{m,n}_{\alpha\beta}(\varepsilon,i,j) 
\label{eq:wgwg2} 
\end{align}
in which the the orbital-resolved FC density is achieved as
\begin{align}
\xi^{m,n}_{\alpha\beta}(\varepsilon,i,j)=&\sum\limits_{m^{\prime}}\sum\limits_{n^{\prime}}\sum\limits_{\gamma=1,3}\zeta^{\gamma}_{\alpha\beta,mm^{\prime}n^{\prime}n}(\varepsilon,i,j) \label{eq:ofcd} \\
\zeta^{\gamma}_{\alpha\beta,mm^{\prime}n^{\prime}n}(\varepsilon,i,j)=&-\frac{1}{2\pi}\text{Im}[\langle m,i | U^{\gamma}_{i,\alpha} | m^{\prime},i\rangle 
\langle m^{\prime},i | \hat{G}^{0}(\varepsilon) | n^{\prime},j\rangle
\langle n^{\prime},j | U^{\gamma}_{j,\beta} | n,j\rangle
\langle n,j | \hat{G}^{0}(\varepsilon) | m,i\rangle] \label{eq:loops}
\end{align}
where $\xi^{m,n}_{\alpha\beta}(\varepsilon,i,j)$ is the orbital-resolved FC as used in the main text, $\zeta^{\gamma}_{\alpha\beta,mm^{\prime}n^{\prime}n}(\varepsilon,i,j)$ is the expression of what we call ``loop'' in the main text, $\alpha$ and $\beta$ goes through three Cartesian coordinates, and $m$, $m^{\prime}$ $n^{\prime}$, and $n$ are atomic orbitals.
Note that since all three $U_{i,\alpha}^{\gamma}$ follow the same symmetry as a vector along the $\alpha$ axis on site $i$, in the main text we are going to use $U_{i,\alpha}$ instead when discussing the symmetry operations.
The Green's function operator $\hat{G}^{0}(\varepsilon)$ is defined as $\sum\limits_{p}\frac{\ket{p}\bra{p}}{\varepsilon-\varepsilon_{p}+i\eta}$, where $\varepsilon_{p}$ and $\ket{p}$ are the energy and wavefunction $\ket{p}$ in the unperturbed system $H^{0}$.
Note that the trace operation of the matrix is replaced by the sum operation in order to write specifically the orbital-resolved FC density $\xi^{m,n}_{\alpha\beta}$ as in Eqs.~(\ref{eq:ofcd}) and Fig.~2 (a) of the main text  where we call it one ``loop'' composed by orbitals $m,m^{\prime},n^{\prime}$, and $n$.
Note also that Eq.~(\ref{eq:ofcd}) is identical to Eq.~(6) of the main text, where the trace is used instead of summing over $m,m^{\prime},n^{\prime}$, and $n$. 
\begin{figure}[htp]
    \centering
    \includegraphics[width=1.0\linewidth]{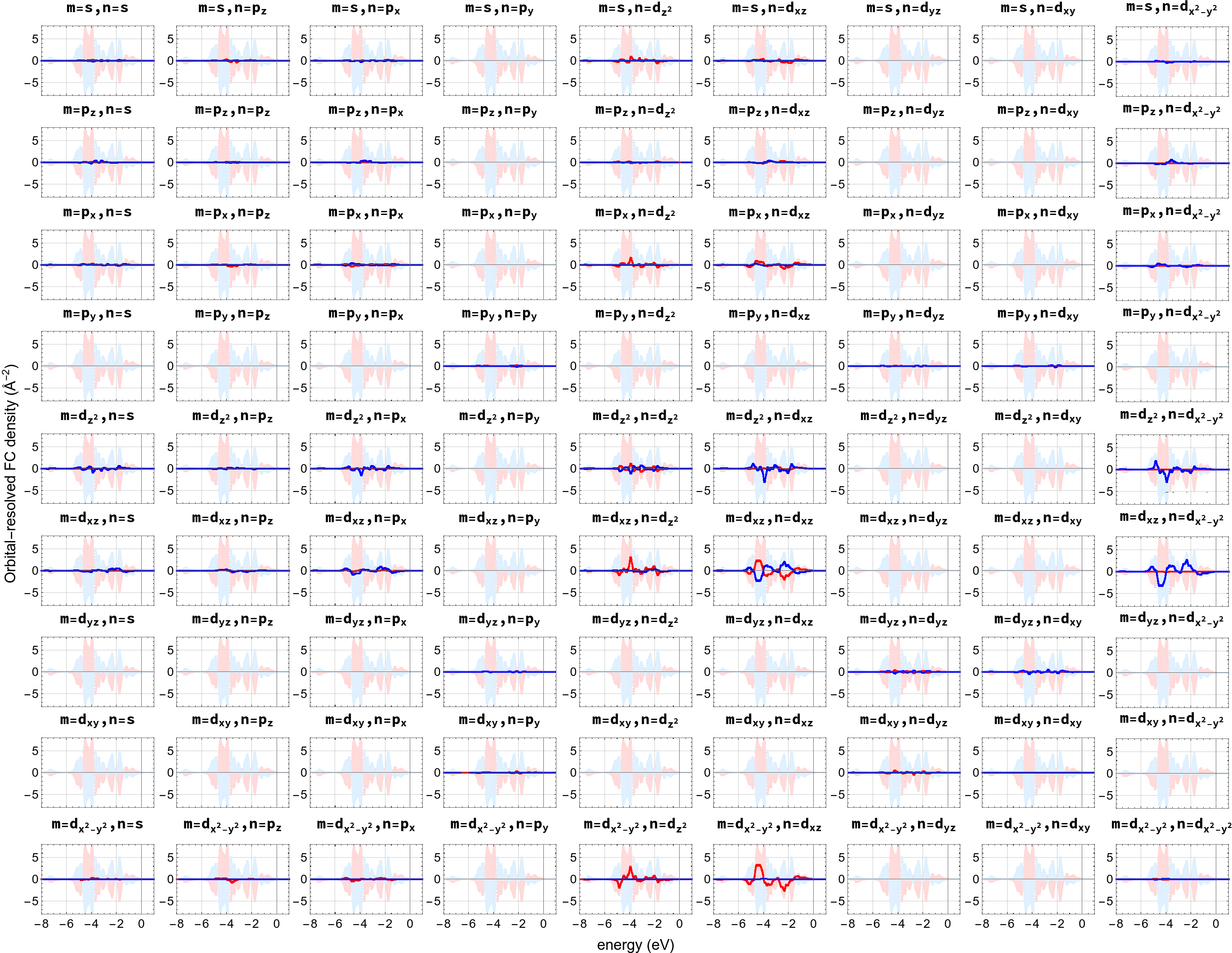}
    \caption{In each subplot, orbital-resolved FC density $\xi_{x,z}^{m,n}$ and $\xi_{z,x}^{m,n}$ are plotted in blue and red solid lines respectively and the total FC density $f_{x,z}$ and $f_{z,x}$ are also presented in blue and red shadow areas respectively. The Fermi energy is set to be zero eV. Note that the subplots without any solid lines are the eDMI forbidden cases for which the $\xi_{x,z}^{m,n}$ and $\xi_{z,x}^{m,n}$ are exactly zeros from the calculations when orbital $m$ is on site $i$ and orbital $n$ is on site $j$.}\label{fig:ximn}
\end{figure}

To calculate the $\xi_{\alpha,\beta}^{m,n}$ in PTO, we choose $4s^{1},3p^{3},3d^{5}$ as the valence electrons of the titanium atom and treat the others, $1s^{1},2s^{1},2p^{3},3s^{1}$, as core electrons that rigidly follows the nuclei. Thus the Wannier orbitals considered for titanium in our TB model allows us to verify the role of the orbital symmetry of all $s$, $p$, and $d$ orbitals in contributing to the non-zero eDMI. 
For instance, we have listed all the numerical results of orbital-resolved FC in Fig.~\ref{fig:ximn}, the $\xi_{x,z}^{m,n}$ (blue curves) and $\xi_{z,x}^{m,n}$ (red curves).
As can been seen in Fig.~\ref{fig:ximn}, some plots do not have have any solid line which means the numerical results for those orbitals $m$ and $n$ are exactly zeros, which are the cases for which the eDMI is forbidden by the symmetry.
The results in Fig.~\ref{fig:ximn}, with no exclusion, fulfills the relation of $\xi_{x,z}^{m,n}=\xi_{z,x}^{n,m}$ as long as the eDMI is allowed by symmetry.
Note that as have discussed in the main text Figs.~2 (g) and (h), $d$ orbitals contribute most of the orbital-resolved FC in PTO, but for other materials the other orbitals, e.g. both orbitals $m$ and $n$ are elements of $\{p_{x},p_{y},p_{z}\}$, could be the main contributor.

\section{orbital selection rules of eDMI}
In this section, we will show particular examples to explain how we summarize the orbital selection rules.
Noticing from  Eq.~(\ref{eq:ofcd}), the eDMI vector is decided by the multiplication between two integrations, one being $\langle m,i|U_{i,\alpha}\hat{G}^{0}U_{j,\beta} | n,j \rangle$ and the other being $\langle n,j | \hat{G}^{0} | m,i \rangle$, where the  $m$ and $n$ orbitals are located on sites $i$ and $j$, respectively.
More specifically, these two integrations should not be zero and their multiplication should give rise to antisymmetric feature.
In the following two subsections, we will explain firstly (in subsection A) the conditions to have antisymmetric orbital-resolved FC ($\xi_{\alpha,\beta}^{m,n}=-\xi_{\beta,\alpha}^{n,m}$) and secondly (in subsection B) the conditions to have nonzero orbital-resolved FC ($\xi_{\alpha,\beta}^{m,n}\neq 0$).

\subsection{orbital-resolved FC in antisymmetric form}
\begin{figure}[!htp]
    \centering
    \includegraphics[width=1.0\linewidth]{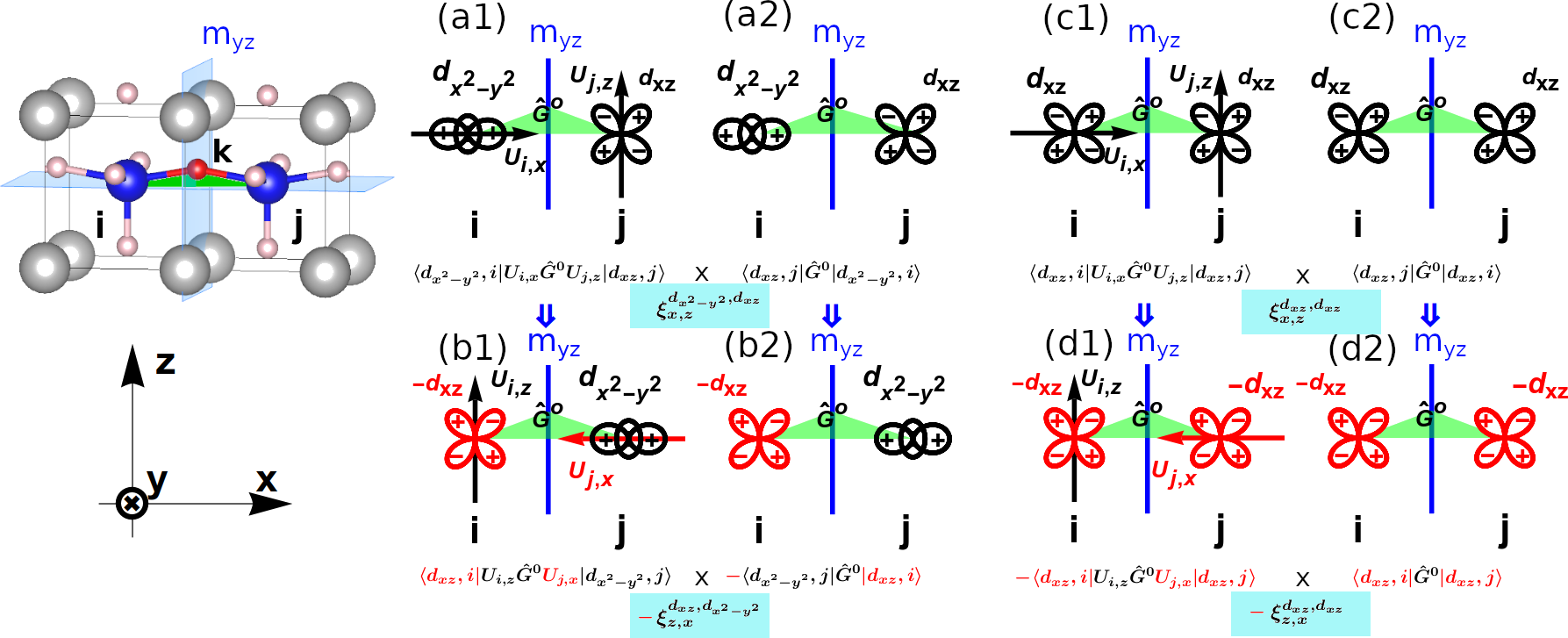}
    \caption{Illustration plots of the orbital resolved FC density. (a1) is for $\langle d_{x^2-y^2},i | U_{i,x}\hat{G}^{0}U_{j,z} | d_{xz},j \rangle$ and (a2) is for $\langle d_{x^2-y^2},i | \hat{G}^{0} | d_{xz},j \rangle$. The integrations in (a1) and (a2) times together gives rise to orbital resolved FC density $\xi_{x,z}^{m,n}$ where $m=d_{x^2-y^2}$ and $n=d_{xz}$. Blue lines in (a1) and (a2) are used to indicate operations of mirror $m_{yz}$, which transform (a1) to (b1) and (a2) to (b2) respectively. (b1) is the illustration plot for $-\langle d_{xz},i | \hat{G}^{0} | d_{x^2-y^2},j \rangle$ and (b2) is the illustration plot for $-\langle d_{xz},i | \hat{G}^{0} | d_{x^2-y^2},j \rangle$. The integrations in (b1) and (b2) times together gives rise to orbital resolved FC density $\xi_{z,x}^{n,m}$ where $m=d_{x^2-y^2}$ and $n=d_{xz}$. Similarly, Blue lines in (c1) and (c2) are used to indicate operations of mirror $m_{yz}$, which transform (c1) to (d1) and (c2) to (d2) respectively. The difference is that both orbitals $m$ and $n$ are $d_{xz}$ in (c1) $\langle d_{xz},i | U_{i,x}\hat{G}^{0}U_{j,z} | d_{xz},j \rangle$, (c2) $\langle d_{xz},i | \hat{G}^{0} | d_{xz},j \rangle$, (d1) $(-1)^{3}\langle d_{xz},i | U_{i,z}\hat{G}^{0}U_{j,x} | d_{xz},j \rangle$, and $(-1)^{3}\langle d_{xz},i | \hat{G}^{0} | d_{xz},j \rangle$. Thus the multiplication between (c1) and (c2) gives $\xi_{x,z}^{d_{xz},d_{xz}}$ and the multiplication between (d1) and (d2) gives $\xi_{z,x}^{d_{xz},d_{xz}}$. Note that the sites $i$, $j$, and $k$, and the mirror $m_{yz}$ in (a), (b), (c), and (d) are indicated on the left. The coordinate system is also indicated. If a orbital or $U$ function changes sign,it will be color coded by red in (b) and (d).}\label{fig:orbexamples1}
\end{figure}

The symmetry allowed orbitals $m$ and $n$ that contribute to eDMI should satisfy either the conditions of
\begin{align}
\langle n,j | \hat{G}^{0} | m,i \rangle &= -
\langle m,j | \hat{G}^{0} | n,i \rangle \label{eq:GA1-1}\\
\langle m,i | U_{i,\alpha}\hat{G}^{0}U_{j,\beta} | n,j \rangle &=
\langle n,i | U_{i,\beta}\hat{G}^{0}U_{j,\alpha} | m,j \rangle \label{eq:GA1-2}
\end{align}
or
\begin{align}
\langle n,j | \hat{G}^{0} | m,i \rangle &= \langle m,j | \hat{G}^{0} | n,i \rangle \label{eq:GA2-1}\\
\langle m,i | U_{i,\alpha}\hat{G}^{0}U_{j,\beta} | n,j \rangle &= -\langle n,i | U_{i,\beta}\hat{G}^{0}U_{j,\alpha} | m,j \rangle \label{eq:GA2-2}
\end{align}
which guarantee the antisymmetric feature of the orbital-resolved FC, $\xi_{\alpha,\beta}^{m,n}=-\xi_{\beta,\alpha}^{n,m}$.
Note that the left and right hand sides of both Eqs.~(\ref{eq:GA1-1}), (\ref{eq:GA1-2}), (\ref{eq:GA2-1}), and (\ref{eq:GA2-2}) can be related by the operation of a mirror that goes through the intermediate oxygen site $k$ and perpendicular to the line that connects the sites $i$ and $j$. 
This is exactly the symmetry analysis as depicted in Fig.~3 of the main text. Here, we give two examples in Fig.~\ref{fig:orbexamples1}: one as depicted in (a) and (b) that satisfies Eqs.~(\ref{eq:GA1-1}) and ((\ref{eq:GA1-2})) and the other as depicted in (c) and (d) that satisfies Eqs.~(\ref{eq:GA2-1}) and (\ref{eq:GA2-2}).

{\noindent\bf The first example ($m=d_{x^2-y^2}, n=d_{xz}$):} In both Fig.~\ref{fig:orbexamples1} (a1) and (a2), mirror $m_{yz}$ (vertical blue lines) operations are performed.
Consequently, in Fig.~\ref{fig:orbexamples1} (a1), the following functions are transformed: (1) orbital $d_{x^2-y^2}$ on site $i$ and orbital $d_{xz}$ on site $j$ are transformed to, in Fig.~\ref{fig:orbexamples1} (b1), $d_{x^2-y^2}$ on site $j$ and $-d_{xz}$ on site $i$, respectively; (2) effective perturbation potential $U_{i,x}$ on site $i$ (black arrow on the left) and $U_{j,z}$ on site $j$ (black arrow on the right) are transformed to, in Fig.~\ref{fig:orbexamples1} (b1), $-U_{j,x}$  on site $j$ (red arrow on the right) and $U_{i,z}$ on site $i$ (black arrow on the left), respectively; (3) $G^{0}$ (green triangle) is unchanged since it is defined by the eigenfunctions of the unperturbed $H^{0}$ and follows the same crystalline symmetry $P4mm$.
Note that symmetry operations should never alternate the integration values, thus we have proved that $\langle d_{x^2-y^2},i | U_{i,x}\hat{G}^{0}U_{j,z} | d_{xz},j \rangle$ (illustrated in Fig.~\ref{fig:orbexamples1} (a1)) is equal to $(-1)^{2}\langle d_{xz},i | U_{i,z}\hat{G}^{0}U_{j,x} | d_{x^2-y^2},j \rangle$ (illustrated in Fig.~\ref{fig:orbexamples1} (b1)).
While in Fig.~\ref{fig:orbexamples1} (a2), the following functions are transformed: (1) orbital $d_{x^2-y^2}$ on site $i$ and orbital $d_{xz}$ on site $j$ are transformed to, in Fig.~\ref{fig:orbexamples1} (b2), $d_{x^2-y^2}$ on site $j$ and $-d_{xz}$ on site $i$, respectively; (2) $G^{0}$ (green triangle) is unchanged for the same reason as explained in (a1).
Thus we can prove that $\langle d_{x^2-y^2},i | \hat{G}^{0} | d_{xz},j \rangle$ (illustrated in Fig.~\ref{fig:orbexamples1} (a2)) is equal to $-\langle d_{xz},i | \hat{G}^{0} | d_{x^2-y^2},j \rangle$ (illustrated in Fig.~\ref{fig:orbexamples1} (b2)).
In total, orbitals $d_{x^2-y^2}$ and $d_{xz}$ satisfies the first conditions as in Eqs.~(\ref{eq:GA1-1}) and (\ref{eq:GA1-2}).

{\noindent\bf The second example ($m=d_{xz}, n=d_{xz}$):} In both Figs.~\ref{fig:orbexamples1} (c1) and (c2), mirror $m_{yz}$ (vertical blue lines) operations are performed.
Consequently, in Fig.~\ref{fig:orbexamples1} (c1), the following functions are transformed: (1) orbital $d_{xz}$ on site $i$ and orbital $d_{xz}$ on site $j$ are transformed to, in Fig.~\ref{fig:orbexamples1} (d1), $-d_{xz}$ on site $j$ and $-d_{xz}$ on site $i$, respectively; (2) effective perturbation potential $U_{i,x}$ on site $i$ (black arrow on the left) and $U_{j,z}$ on site $j$ (black arrow on the right) are transformed to, in Fig.~\ref{fig:orbexamples1} (d1), $-U_{j,x}$  on site $j$ (red arrow on the right) and $U_{i,z}$ on site $i$ (black arrow on the left), respectively; (3) $G^{0}$ (green triangle) is unchanged since it is defined by the eigenfunctions of the unperturbed $H^{0}$ and follows the same crystalline symmetry $P4mm$.
Note that symmetry operations should never alternate the integration values, thus we have proved that $\langle d_{xz},i | U_{i,x}\hat{G}^{0}U_{j,z} | d_{xz},j \rangle$ (illustrated in Fig.~\ref{fig:orbexamples1} (c1)) is equal to $(-1)^{3}\langle d_{xz},i | U_{i,z}\hat{G}^{0}U_{j,x} | d_{xz},j \rangle$ (illustrated in Fig.~\ref{fig:orbexamples1} (d1)).
While in Fig.~\ref{fig:orbexamples1} (c2), the following functions are transformed: (1) orbital $d_{xz}$ on site $i$ and orbital $d_{xz}$ on site $j$ are transformed to, in Fig.~\ref{fig:orbexamples1} (d2), $-d_{xz}$ on site $j$ and $-d_{xz}$ on site $i$, respectively; (2) $G^{0}$ (green triangle) is unchanged for the same reason as explained in (a1).
Thus we can prove that $\langle d_{xz},i | \hat{G}^{0} | d_{xz},j \rangle$ (illustrated in Fig.~\ref{fig:orbexamples1} (c2)) is equal to $(-1)^{3}\langle d_{xz},i | \hat{G}^{0} | d_{xz},j \rangle$ (illustrated in Fig.~\ref{fig:orbexamples1} (d2)).
In total, orbitals $d_{xz}$ and $d_{xz}$ satisfies the second conditions as in Eqs.~(\ref{eq:GA2-1}) and (\ref{eq:GA2-2}).

Note that other orbital combinations that either satisfy Eqs.~(\ref{eq:GA1-1}) and (\ref{eq:GA1-2}) or Eqs.~(\ref{eq:GA2-1}) and (\ref{eq:GA2-2}) are summarized in the fifth and sixth columns of the Tab.~\ref{tab:dy}, \ref{tab:dz}, and \ref{tab:dx} in the appendix section \ref{appendix}, where ``-1'' means reverse sign and ``1'' means unchanged.
It is worth mentioning that, the results from the symmetry analysis in those tables are confirmed by the numerical result in Fig.~\ref{fig:ximn}, in which the $\xi_{x,z}^{m,n}$ (blue line) is always reversed sign compare to the $\xi_{z,x}^{n,m}$ (red line) as long as $\xi_{x,z}^{m,n}$ and $\xi_{z,x}^{n,m}$ exist.

\subsection{non-zero antisymmetric orbital-resolved FC}
\begin{figure}[!htp]
    \centering
    \includegraphics[width=0.8\linewidth]{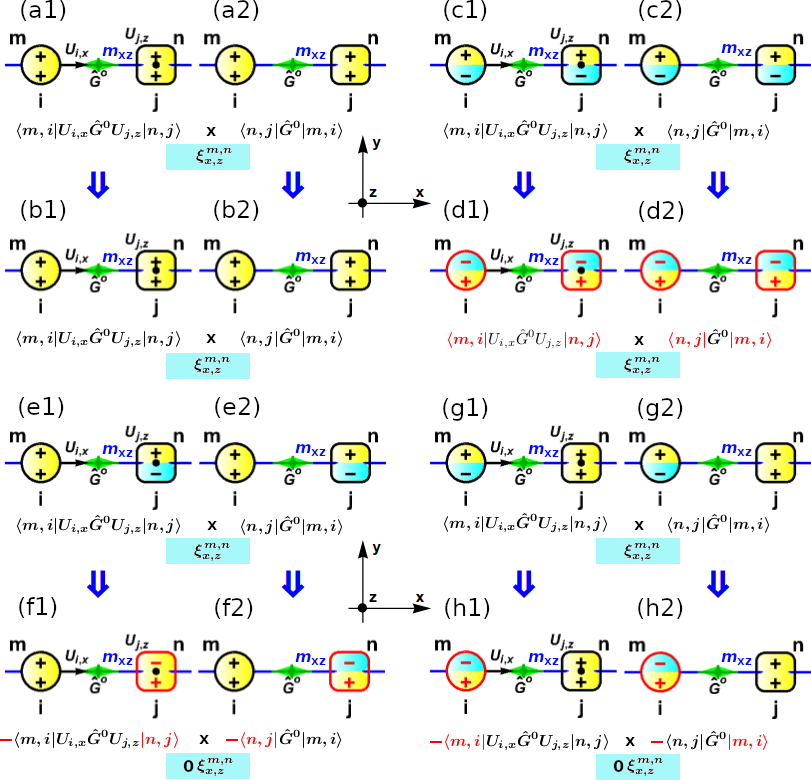}
    \caption{The orbital-resolved FC density $\xi_{x,z}^{m,n}$ of two PTO ($P4mm$ phase polarized along z-direction) unitcells along the x-direction are discussed.
    Expression of $\xi_{x,z}^{m,n}$ is sketched by a multiplication between the expressions in (a1)  and (a2); The blue horizontal lines in (a1) and (a2) are mirror $m_{xy}$ operations and transform the expressions in (a1) and (a2) to the expressions in (b1) and (b2), respectively, and whose multiplication also gives rise to $\xi_{x,z}^{m,n}$.
    Similarly, panel (c1) and (c2) are transformed to (d1) and (d2), panel (e1) and (e2) are transformed to (f1) and (f2), and panel (g1) and (g2) are transformed to (h1) and (h2) by mirror $m_{xy}$ operations.
    In all the expression sketches, the yellow circles and squares represent orbital $m$ and $n$ respectively, the plus and minus sign in the circles and squares is to show the symmetry of the orbitals; the arrows are used to indicate effective perturbation potentials, and the green triangles are for Green's function $\hat{G}^{0}$ which has the same symmetry as the $P4mm$ phase of PTO that is polarized along the z-direction.}\label{fig:mxy}
\end{figure}

Note that a trivial case in which $\xi_{\alpha,\beta}^{m,n}=0$ and $\xi_{\beta,\alpha}^{n,m}=0$ also satisfies the antisymmetric conditions $\xi_{\alpha,\beta}^{m,n}=-\xi_{\beta,\alpha}^{n,m}$, but should be ruled out. Thus neither $\langle m,i|U_{i,\alpha}\hat{G}^{0}U_{j,\beta} | n,j \rangle$ nor $\langle n,j | \hat{G}^{0} | m,i \rangle$ should be zero.
As summarized in Fig.~\ref{fig:mxy} where $m$ and $n$ are orbitals on site $i$ and $j$ respectively, $\xi_{\alpha,\beta}^{m,n}$ is nonzero when $m$ and $n$ are both even as in (a) or odd as in (c) under the crystalline symmetry operations, for example, the mirror $m_{xy}$ that goes through all the three sites $i$, $j$, and $k$ as in Fig.~\ref{fig:mxy}; $\xi_{\alpha,\beta}^{m,n}$ will be zero when one of $m$ and $n$ is odd under the crystalline symmetry operations, for example, under the mirror $m_{xy}$ that goes through all the three sites $i$, $j$, and $k$ as in Fig.~\ref{fig:mxy}, $m$ is even and $n$ is odd as in (e) or $m$ is odd and $n$ is even as in (g).

More specifically, in Figs.~\ref{fig:mxy} (a) and (b), the orbitals and functions are transformed by $m_{xy}$ according to the following rules: (1) both orbitals $m$ on site $i$ and $n$ on site $j$ are unchanged (even) from (a) to (b); (2) both $U_{i,x}$ on site $i$ and $U_{j,z}$ on site $j$ are unchanged from (a) to (b); (3) and $\hat{G}^{0}$ is also unchanged because $m_{xy}$ is one of the crystalline symmetry operations. While in Figs.~\ref{fig:mxy} (c) and (d), the orbitals and functions are transformed by $m_{xy}$ according to the following rules: (1) both orbitals $m$ on site $i$ and $n$ on site $j$ are reversed sign (odd) from (a) to (b); (2) both $U_{i,x}$ on site $i$ and $U_{j,z}$ on site $j$ are unchanged from (a) to (b); (3) and $\hat{G}^{0}$ is also unchanged because $m_{xy}$ is one of the crystalline symmetry operations. In both cases, $\xi_{x,z}^{m,n}$ is unchanged under the symmetry operation. Thus when $m$ and $n$ are both even as in (a) or odd as in (c) under the crystalline symmetry operations, there is no symmetry constrain to make FC density to be zero.
However, in Figs.~\ref{fig:mxy} (e) and (f), the orbitals and functions are transformed by $m_{xy}$ according to the following rules: (1) $m$ on site $i$ is unchanged (even) from (e) to (f); (2) $n$ on site $j$ reverse sign (odd) from (e) to (f); (3) both $U_{i,x}$ on site $i$ and $U_{j,z}$ on site $j$ are unchanged from (e) to (f); (4) and $\hat{G}^{0}$ is also unchanged because $m_{xy}$ is one of the crystalline symmetry operations. Thus that one orbital, $n$ on site $j$, is odd and the other, $m$ on site $i$, is even will result in $\xi_{x,z}^{m,n}=0$ because of $\langle m,i|U_{i,\alpha}\hat{G}^{0}U_{j,\beta} | n,j \rangle=-\langle m,i|U_{i,\alpha}\hat{G}^{0}U_{j,\beta} | n,j \rangle=0$ (transformation from (e1) to (f1)) and $\langle n,j | \hat{G}^{0} | m,i \rangle=-\langle n,j | \hat{G}^{0} | m,i \rangle=0$ (transformation from (e2) to (f2)). A similar situation happens in Figs.~\ref{fig:mxy} (g) and (h), where orbital $m$ on site $i$ is odd and orbital $n$ on site $j$ is even which also result in $\xi_{x,z}^{m,n}=0$ because of $\langle m,i|U_{i,\alpha}\hat{G}^{0}U_{j,\beta} | n,j \rangle=-\langle m,i|U_{i,\alpha}\hat{G}^{0}U_{j,\beta} | n,j \rangle=0$ (transformation from (g1) to (h1)) and $\langle n,j | \hat{G}^{0} | m,i \rangle=-\langle n,j | \hat{G}^{0} | m,i \rangle=0$ (transformation from (g2) to (h2)).

Note that such symmetry constrain on all orbital combinations are are summarized in the seventh and eighth columns of the Tab.~\ref{tab:dy}, \ref{tab:dz}, and \ref{tab:dx} in the appendix section \ref{appendix}, where ``0'' means the FC density is constrained to be zero by crystalline symmetry and ``-'' means no symmetry constrain.
It is worth mentioning that our symmetry analysis are consistent with the numerical results in Fig.~\ref{fig:ximn}, where orbitals $m$ and $n$ give rise zero orbital-resolved FC density ( $\xi_{x,z}^{m,n}=\xi_{z,x}^{n,m}=0$) when one of them is odd and the other is even.

\section{eDMI in third  order form}
\subsection{collective basis}
\begin{figure}[!htp]
    \centering
    \includegraphics[width=0.8\linewidth]{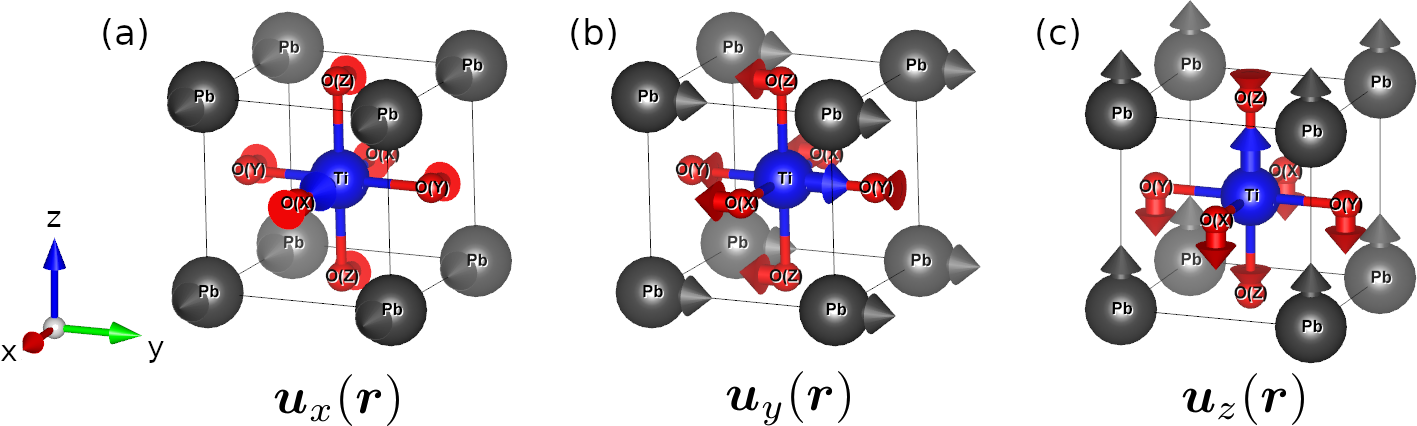}
    \caption{Collective basis for polar modes (a) $\bm{u}_{x}(\bm{r})$, (b) $\bm{u}_{y}(\bm{r})$, and (c) $\bm{u}_{z}(\bm{r})$, where $\bm{r}$ is the position index of an unit cell. Three oxygen atoms in one unit cell are labeled by O(X) (the oxygen located along positive x-direction with respect to titanium atom), O(Y) (the oxygen located along positive y-direction with respect to titanium atom), and O(Z) (the oxygen located along positive z-direction with respect to titanium atom). }\label{fig:atomistic}
\end{figure}
As discussed in the main text, the eDMI should be of odd order with respect to polar mode according to the symmetry arguments. We thus perform symmetry analysis to search for energy invariants with respect to cubic symmetry structure ($Pm\bar{3}m$), which is the high-temperature phase of most of the perovskite ferroelectrics. 
Note that we are looking for the lowest order of both polarization and gradient of polarization.
We use LINVARIANT\cite{linvariant} to perform such invariant generation, then a double check by ISOTROPY\cite{Campbell2006} is also conducted. 
We find there are only three energy terms that are of the first order of polarization gradient:
\begin{align}
E_{11}(\bm{R})=
+&u_{x}(\bm{R})^2 [u_{x}(\bm{R}+\bm{\Delta x}) - u_{x}(\bm{R}-\bm{\Delta x})] \nonumber \\
+&u_{y}(\bm{R})^2 [u_{y}(\bm{R}+\bm{\Delta y}) - u_{y}(\bm{R}-\bm{\Delta y})] \nonumber \\
+&u_{z}(\bm{R})^2 [u_{z}(\bm{R}+\bm{\Delta z}) - u_{z}(\bm{R}-\bm{\Delta z})] \label{eq:E11}\\
E_{12}(\bm{R})=
+&\frac{1}{2}u_{x}(\bm{R})^2 [u_{y}(\bm{R}+\bm{\Delta y}) - u_{y}(\bm{R}-\bm{\Delta y})] \nonumber\\
+&\frac{1}{2}u_{x}(\bm{R})^2 [u_{z}(\bm{R}+\bm{\Delta z}) - u_{z}(\bm{R}-\bm{\Delta z})] \nonumber\\
+&\frac{1}{2}u_{y}(\bm{R})^2 [u_{z}(\bm{R}+\bm{\Delta z}) - u_{z}(\bm{R}-\bm{\Delta z})] \nonumber\\
+&\frac{1}{2}u_{y}(\bm{R})^2 [u_{x}(\bm{R}+\bm{\Delta x}) - u_{x}(\bm{R}-\bm{\Delta x})] \nonumber\\
+&\frac{1}{2}u_{z}(\bm{R})^2 [u_{x}(\bm{R}+\bm{\Delta x}) - u_{x}(\bm{R}-\bm{\Delta x})] \nonumber\\
+&\frac{1}{2}u_{z}(\bm{R})^2 [u_{y}(\bm{R}+\bm{\Delta y}) - u_{y}(\bm{R}-\bm{\Delta y})] \label{eq:E12} \\
E_{44}(\bm{R})=
+&u_{x}(\bm{R})  u_{y}(\bm{R}) [u_{y}(\bm{R}+\bm{\Delta x}) - u_{y}(\bm{R}-\bm{\Delta x})] \nonumber\\
+&u_{x}(\bm{R})  u_{y}(\bm{R}) [u_{x}(\bm{R}+\bm{\Delta y}) - u_{x}(\bm{R}-\bm{\Delta y})] \nonumber \\
+&u_{y}(\bm{R}) u_{z}(\bm{R}) [u_{z}(\bm{R}+\bm{\Delta y}) - u_{z}(\bm{R}-\bm{\Delta y})] \nonumber\\
+&u_{y}(\bm{R})  u_{z}(\bm{R}) [u_{y}(\bm{R}+\bm{\Delta z}) - u_{y}(\bm{R}-\bm{\Delta z})] \nonumber\\
+&u_{x}(\bm{R})  u_{z}(\bm{R}) [u_{z}(\bm{R}+\bm{\Delta x}) - u_{z}(\bm{R}-\bm{\Delta x})] \nonumber\\
+&u_{x}(\bm{R})  u_{z}(\bm{R}) [u_{x}(\bm{R}+\bm{\Delta z}) - u_{x}(\bm{R}-\bm{\Delta z})] \label{eq:E44}
\end{align}
where $\bm{R}$ is the cell index, $\bm{\Delta x}$, $\bm{\Delta y}$, and $\bm{\Delta z}$ are one-unit-cell shifts vectors along x-, y-, and z-direction of the Cartesian coordinates, respectively.
$E_{11}(\bm{R})$, $E_{12}(\bm{R})$ and $E_{44}(\bm{R})$ can be seen as energies that come from the interaction of cell $\bm{R}$ and its surrounding cells that are reached by shifting $\bm{\Delta x}$, $\bm{\Delta y}$, and $\bm{\Delta z}$. 
Note that $E_{11}$, $E_{12}$, and $E_{44}$ in continuous form can be written as $E_{11}=u_{x}^{2}\frac{\partial u_{x}}{\partial x}+u_{y}^{2}\frac{\partial u_{y}}{\partial y}+u_{z}^{2}\frac{\partial u_{z}}{\partial z}$, $E_{12}=u_{x}^2\frac{\partial u_{y}}{\partial y}+u_{x}^2\frac{\partial u_{z}}{\partial z}+u_{y}^2\frac{\partial u_{x}}{\partial x}+u_{y}^2\frac{\partial u_{z}}{\partial z}+u_{z}^2\frac{\partial u_{x}}{\partial x}+u_{z}^2\frac{\partial u_{y}}{\partial y}$, and $E_{44}=u_{x}u_{y}\frac{\partial u_{x}}{\partial y}+u_{x}u_{y}\frac{\partial u_{y}}{\partial x}+u_{y}u_{z}\frac{\partial u_{y}}{\partial z}+u_{y}u_{z}\frac{\partial u_{z}}{\partial y}+u_{x}u_{z}\frac{\partial u_{x}}{\partial z}+u_{x}u_{z}\frac{\partial u_{z}}{\partial x}$,  as derived from ISOTROPY\cite{Campbell2006}.
% \begin{align}
%     E_{11}&=u_{x}^{2}\frac{\partial u_{x}}{\partial x}+u_{y}^{2}\frac{\partial u_{y}}{\partial y}+u_{z}^{2}\frac{\partial u_{z}}{\partial z},\\ E_{12}&=u_{x}^2\frac{\partial u_{y}}{\partial y}+u_{x}^2\frac{\partial u_{z}}{\partial z}+u_{y}^2\frac{\partial u_{x}}{\partial x}+u_{y}^2\frac{\partial u_{z}}{\partial z}+u_{z}^2\frac{\partial u_{x}}{\partial x}+u_{z}^2\frac{\partial u_{y}}{\partial y}, \\ E_{44}&=u_{x}u_{y}\frac{\partial u_{x}}{\partial y}+u_{x}u_{y}\frac{\partial u_{y}}{\partial x}+u_{y}u_{z}\frac{\partial u_{y}}{\partial z}+u_{y}u_{z}\frac{\partial u_{z}}{\partial y}+u_{x}u_{z}\frac{\partial u_{x}}{\partial z}+u_{x}u_{z}\frac{\partial u_{z}}{\partial x}
% \end{align}
According to group theory, linear combinations of $E_{11}$, $E_{12}$, and $E_{44}$ are still energy invariants.
Thus we can have equivalently three energy invariants $\mathcal{A}^{0}E_{11}$, $\mathcal{A}^{+}(E_{12}+E_{44})$, and $\mathcal{A}^{-}(E_{12}-E_{44})$, among which only $\mathcal{A}^{-}(E_{12}-E_{44})$ can give rise to antisymmetric force constants, and we will use it to define $E^{-}$ as
\begin{align}
E^{-}(\bm{R})=\mathcal{A}^{-}[E_{12}(\bm{R})- E_{44}(\bm{R})] \label{eq:Er}
\end{align}
where $\mathcal{A}^{0}$, $\mathcal{A}^{+}$, and $\mathcal{A}^{-}$ are coefficients.

$E^{-}(\bm{R})$ is an energy invariant with respect to the cubic symmetry origin at cell $\bm{R}$, which means $\hat{g}E^{-}(\bm{R})=E^{-}(\bm{R}), \forall\hat{g}\in\mathcal{G}^{0}$, where $\mathcal{G}^0$ is the quotient group $\mathcal{G}/\mathcal{T}$ with $\mathcal{G}$ being the space group $Pm\bar{3}m$ and $\mathcal{T}$ being the lattice translation group.
Thus the total energy that is invariant under space group $\mathcal{G}$ involves $E^{-}(\bm{R})$ can be written as $E^{tot}=\sum\limits_{\bm{R}}^{N}E^{-}(\bm{R})$, where $N$ indicates all the cells in the crystal.
\begin{table}[!htp]
\caption{The off-diagonal part of the force constants, $F_{\alpha,\beta}^{u}$, between polar modes and that comes from $E^{tot}=\sum\limits_{\bm{R}}^{N}E^{-}(\bm{R})$ are calculated.\label{tab:fc}}
%\resizebox{0.75\textwidth}{!}{%
\begin{tabular}{|c|c|c|cc|c|c|}
\hline
$\bm{R}_i$               & $\bm{R}_j$               & $\bm{\mathcal{D}}(i,j)$ & $\alpha$ & $\beta$ & $F_{\alpha\beta}^{u}(i,j)=\frac{\partial^2E^{tot}}{\partial u_{\alpha}(\bm{R}_{i}) \partial u_{\beta}(\bm{R}_{j})}$ & $F_{\beta\alpha}^{u}(i,j)=\frac{\partial^2E^{tot}}{\partial u_{\beta}(\bm{R}_{i}) \partial u_{\alpha}(\bm{R}_{j})}$ \\
\hline
\multirow{9}{*}{$\{0,0,0\}$} & \multirow{3}{*}{$\{1,0,0\}$} & $\mathcal{D}_x$ & $y$ & $z$ & 0                            & 0 \\
                             &                              & $\mathcal{D}_y$ & $z$ & $x$ & $ \mathcal{A}^{-}(u_{z}(0,0,0)+ u_{z}(1,0,0)$) & $-\mathcal{A}^{-}(u_{z}(0,0,0) + u_{z}(1,0,0))$ \\
                             &                              & $\mathcal{D}_z$ & $x$ & $y$ & $-\mathcal{A}^{-}(u_{y}(0,0,0)+
                             u_{y}(1,0,0))$ & $\mathcal{A}^{-}(u_{y}(0,0,0)+ u_{y}(1,0,0))$ \\
                             \cline{2-7}
                             & \multirow{3}{*}{$\{0,1,0\}$} & $\mathcal{D}_x$ & $y$ & $z$ & $-\mathcal{A}^{-}(u_{z}(0,0,0)+ u_{z}(0,1,0))$ & $\mathcal{A}^{-}(u_{z}(0,0,0)+ u_{z}(0,1,0))$ \\
                             &                              & $\mathcal{D}_y$ & $z$ & $x$ & 0                            & 0 \\
                             &                              & $\mathcal{D}_z$ & $x$ & $y$ & $ \mathcal{A}^{-}(u_{x}(0,0,0)+ u_{x}(0,1,0))$ & $-\mathcal{A}^{-}(u_{x}(0,0,0)+ u_{x}(0,1,0))$ \\
                             \cline{2-7}
                             & \multirow{3}{*}{$\{0,0,1\}$} & $\mathcal{D}_x$ & $y$ & $z$ & $ \mathcal{A}^{-}(u_{y}(0,0,0)+ u_{y}(0,0,1))$ & $-\mathcal{A}^{-}(u_{y}(0,0,0)+ u_{y}(0,0,1))$ \\
                             &                              & $\mathcal{D}_y$ & $z$ & $x$ & $-\mathcal{A}^{-}(u_{x}(0,0,0)+ u_{x}(0,0,1))$ & $\mathcal{A}^{-}(u_{x}(0,0,0)+ u_{x}(0,0,1))$ \\
                             &                              & $\mathcal{D}_z$ & $x$ & $y$ & 0                            & 0 \\
\hline
\end{tabular}%
%}
\end{table}
By performing second derivatives of $E^{tot}$ with respect to the polar modes $\bm{u}(\bm{R}_{i})$ and $\bm{u}(\bm{R}_{j})$, force constants for the polar modes $F_{\alpha,\beta}^{u}$ can be calculated. 
Such results are summarized in Tab.~\ref{tab:fc} of the SM, where site $i$ is chosen as $\bm{R}_i=(0,0,0)$ and site $j$ is chosen at the positions of  $(1,0,0)$, $(0,1,0)$, and $(0,0,1)$, respectively. 
More interestingly, the off-diagonal part of force constants in Tab.~\ref{tab:fc} of the SM is obviously antisymmetric and can be written as $\mathcal{A}^{-}(\bm{u}_i+\bm{u}_j)\cross \bm{e}_{ij}$, where $\bm{e}_{ij}=\bm{R}_{j}-\bm{R}_{i}$.
Thus the eDMI $\bm{\mathcal{D}}(i,j)$ vector is equal to $\mathcal{A}^{-}(\bm{u}_i+\bm{u}_j)\cross \bm{e}_{ij}$ and the corresponding energy can be written as:
\begin{equation}\label{eq:fdmi}
E_{dmi}=\mathcal{A}^{-}[(\bm{u}_i+\bm{u}_j)\cross \bm{e}_{ij}]\cdot (\bm{u}_i\cross\bm{u}_j)
\end{equation}
which is the eDMI term in third order.
To the best of our knowledge, energy terms $E_{dmi}$ as in Eq.~(\ref{eq:fdmi}), which are chiral, have never been considered and used in, e.g., effective Hamiltonian or phase-field simulations for ferroelectric materials.
Note that either energy terms $E^{-}$ as in Eq.~(\ref{eq:Er}) or $E_{dmi}$ as in Eq.~(\ref{eq:fdmi}) can be used in these simulations in order to take such eDMI into account. 
The difference is that $E_{dmi}$ is a pure eDMI energy term, while $E^{-}$ includes both eDMI energy ($E_{dmi}$) and other energy terms that give rise to the symmetric part of the force constants. 
A more general treatment to the first order gradient of polarization in order to automatically including eDMI in practice should use $\mathcal{A}^{0}E_{11}$, $\mathcal{A}^{+}(E_{12}+E_{44})$, and $\mathcal{A}^{-}(E_{12}-E_{44})$ and fit all three coefficients $\mathcal{A}^{0}$, $\mathcal{A}^{-}$, and $\mathcal{A}^{+}$ for the materials.

\begin{table}
\centering
\caption{Verification of the eDMI form in eq.~(eq:fdmi) along [110] (next-nearest neighbour) and [111] (next-next-neighbour) directions. The first column includes all the symmetry operations that makes $e_{ij}$ unchanged. In another word, for each $\bm{e}_{ij}$, the operations in the first column are a subgroup of the group $Pm\bar{3}m$.}
\label{tab:verification}
\begin{tabular}{|cc|ccc|c|}
\hline
Seitz Symbol  & triplets    & $\bm{u}_{i}$  & $\bm{u}_{j}$  & $\bm{e}_{ij}$ & $[(\bm{u}_{i}+\bm{u}_{j})\cross\bm{e}_{ij}]\cdot(\bm{u}_{i}\cross\bm{u}_{j})$     \\ 
\hline
$\{1|0\}$                    & ${x,y,z}$   & $(a,b,c)$     & $(d,e,f)$     & $(1,0,0)$     & $b^2 d + c^2 d - a b e + b d e - a e^2 - a c f + c d f - a f^2$ \\
$\{2_{100}|0\}$              & ${x,-y,-z}$ & $(a, -b, -c)$ & $(, -e, -f)$ & $(1,0,0)$     & $b^2 d + c^2 d - a b e + b d e - a e^2 - a c f + c d f - a f^2$ \\
$\{\text{m}_{010}|0\}$       & ${x,-y,z}$  & $(a, -b, c)$  & $(d, -e, f)$  & $(1,0,0)$     & $b^2 d + c^2 d - a b e + b d e - a e^2 - a c f + c d f - a f^2$ \\
$\{\text{m}_{001}|0\}$       & ${x,y,-z}$  & $(a, b, -c)$  & $(d, e, -f)$  & $(1,0,0)$     & $b^2 d + c^2 d - a b e + b d e - a e^2 - a c f + c d f - a f^2$ \\
$\{4^{+}_{100}|0\}$          & ${x,z,-y}$  & $(a, c, -b)$  & $(d, f, -e)$  & $(1,0,0)$     & $b^2 d + c^2 d - a b e + b d e - a e^2 - a c f + c d f - a f^2$ \\
$\{4^{-}_{100}|0\}$          & ${x,-z,y}$  & $(a, -c, b)$  & $(d, -f, e)$  & $(1,0,0)$     & $b^2 d + c^2 d - a b e + b d e - a e^2 - a c f + c d f - a f^2$ \\
$\{\text{m}_{011}|0\}$       & ${x,-z,-y}$ & $(a, -c, -b)$ & $(d, -f, -e)$ & $(1,0,0)$     & $b^2 d + c^2 d - a b e + b d e - a e^2 - a c f + c d f - a f^2$ \\
$\{\text{m}_{0\bar{1}1}|0\}$ & ${x,z,y}$   & $(a, c, b)$   & $(d, f, e)$   & $(1,0,0)$     & $b^2 d + c^2 d - a b e + b d e - a e^2 - a c f + c d f - a f^2$ \\ 
\hline
\multirow{2}{*}{$\{1|0\}$}                    & \multirow{2}{*}{${x,y,z}$}   & \multirow{2}{*}{$(a, b, c)$}   & \multirow{2}{*}{$(d, e, f)$}   & \multirow{2}{*}{$(1,1,0)$}     & $- a b d + b^2 d + c^2 d - b d^2 + a^2 e - a b e + c^2 e + a d e$  \\
                                              &                              &                                &                                &                                & $+ b d e - a e^2 - a c f - b c f + c d f + c e f - a f^2 - b f^2$ \\ \cline{6-6}
\multirow{2}{*}{$\{\text{m}_{001}|0\}$}       & \multirow{2}{*}{${x,y,-z}$}  & \multirow{2}{*}{$(a, b, -c)$}  & \multirow{2}{*}{$(d, e, -f)$}  & \multirow{2}{*}{$(1,1,0)$}     & $- a b d + b^2 d + c^2 d - b d^2 + a^2 e - a b e + c^2 e + a d e$  \\
                                              &                              &                                &                                &                                & $+ b d e - a e^2 - a c f - b c f + c d f + c e f - a f^2 - b f^2$  \\ \cline{6-6}
\multirow{2}{*}{$\{2_{110}|0\}$}              & \multirow{2}{*}{${y,x,-z}$}  & \multirow{2}{*}{$(b, a, -c)$}  & \multirow{2}{*}{$(e, d, -f)$}  & \multirow{2}{*}{$(1,1,0)$}     & $- a b d + b^2 d + c^2 d - b d^2 + a^2 e - a b e + c^2 e + a d e$  \\
                                              &                              &                                &                                &                                & $+ b d e - a e^2 - a c f - b c f + c d f + c e f - a f^2 - b f^2$  \\ \cline{6-6}
\multirow{2}{*}{$\{\text{m}_{1\bar{1}0}|0\}$} & \multirow{2}{*}{${y,x,z}$}   & \multirow{2}{*}{$(b, a, c)$}   & \multirow{2}{*}{$(e, d, f)$}   & \multirow{2}{*}{$(1,1,0)$}     & $- a b d + b^2 d + c^2 d - b d^2 + a^2 e - a b e + c^2 e + a d e$  \\ 
                                              &                              &                                &                                &                                & $+ b d e - a e^2 - a c f - b c f + c d f + c e f - a f^2 - b f^2$  \\ 
\hline
\multirow{3}{*}{$\{1|0\}$}                    & \multirow{3}{*}{${x,y,z}$}   & \multirow{3}{*}{$(a, b, c)$}   & \multirow{3}{*}{$(d, e, f)$}   & \multirow{3}{*}{$(1,1,1)$}     & $- a b d + b^2 d - a c d + c^2 d - b d^2 - c d^2 + a^2 e - a b e$  \\
                                              &                              &                                &                                &                                & $- b c e + c^2 e + a d e + b d e - a e^2 - c e^2 + a^2 f + b^2 f$  \\
                                              &                              &                                &                                &                                & $- a c f - b c f + a d f + c d f + b e f + c e f - a f^2 - b f^2$  \\ \cline{6-6}
\multirow{3}{*}{$\{\text{m}_{01\bar{1}}|0\}$} & \multirow{3}{*}{${x,z,y}$}   & \multirow{3}{*}{$(a, c, b)$}   & \multirow{3}{*}{$(d, f, e)$}   & \multirow{3}{*}{$(1,1,1)$}     & $- a b d + b^2 d - a c d + c^2 d - b d^2 - c d^2 + a^2 e - a b e$   \\
                                              &                              &                                &                                &                                & $- b c e + c^2 e + a d e + b d e - a e^2 - c e^2 + a^2 f + b^2 f$  \\
                                              &                              &                                &                                &                                & $- a c f - b c f + a d f + c d f + b e f + c e f - a f^2 - b f^2$  \\ \cline{6-6}
\multirow{3}{*}{$\{\text{m}_{1\bar{1}0}|0\}$} & \multirow{3}{*}{${y,x,z}$}   & \multirow{3}{*}{$(b, a, c)$}   & \multirow{3}{*}{$(e, d, f)$}   & \multirow{3}{*}{$(1,1,1)$}     & $- a b d + b^2 d - a c d + c^2 d - b d^2 - c d^2 + a^2 e - a b e$  \\
                                              &                              &                                &                                &                                & $- b c e + c^2 e + a d e + b d e - a e^2 - c e^2 + a^2 f + b^2 f$  \\
                                              &                              &                                &                                &                                & $- a c f - b c f + a d f + c d f + b e f + c e f - a f^2 - b f^2$  \\ \cline{6-6}
\multirow{3}{*}{$\{\text{m}_{10\bar{1}}|0\}$} & \multirow{3}{*}{${z,y,x}$}   & \multirow{3}{*}{$(c, b, a)$}   & \multirow{3}{*}{$(f, e, d)$}   & \multirow{3}{*}{$(1,1,1)$}     & $- a b d + b^2 d - a c d + c^2 d - b d^2 - c d^2 + a^2 e - a b e$  \\
                                              &                              &                                &                                &                                & $- b c e + c^2 e + a d e + b d e - a e^2 - c e^2 + a^2 f + b^2 f$  \\
                                              &                              &                                &                                &                                & $- a c f - b c f + a d f + c d f + b e f + c e f - a f^2 - b f^2$  \\ \cline{6-6}
\multirow{3}{*}{$\{3^{+}_{111}|0\}$}          & \multirow{3}{*}{${y,z,x}$}   & \multirow{3}{*}{$(b, c, a)$}   & \multirow{3}{*}{$(e, f, d)$}   & \multirow{3}{*}{$(1,1,1)$}     & $- a b d + b^2 d - a c d + c^2 d - b d^2 - c d^2 + a^2 e - a b e$  \\
                                              &                              &                                &                                &                                & $- b c e + c^2 e + a d e + b d e - a e^2 - c e^2 + a^2 f + b^2 f$  \\
                                              &                              &                                &                                &                                & $- a c f - b c f + a d f + c d f + b e f + c e f - a f^2 - b f^2$  \\ \cline{6-6}
\multirow{3}{*}{$\{3^{-}_{111}|0\}$}          & \multirow{3}{*}{${z,x,y}$}   & \multirow{3}{*}{$(c, a, b)$}   & \multirow{3}{*}{$(f, d, e)$}   & \multirow{3}{*}{$(1,1,1)$}     & $- a b d + b^2 d - a c d + c^2 d - b d^2 - c d^2 + a^2 e - a b e$ \\
                                              &                              &                                &                                &                                & $- b c e + c^2 e + a d e + b d e - a e^2 - c e^2 + a^2 f + b^2 f$  \\
                                              &                              &                                &                                &                                & $- a c f - b c f + a d f + c d f + b e f + c e f - a f^2 - b f^2$ \\
\hline
\end{tabular}
\end{table}

We also checked if the eDMI form that was derived from interactions between first-nearest neighbors is still valid between next-nearest neighbors ([110] direction) and next-next-nearest neighbors ([111] direction), by performing the symmetry operations on eq.~(\ref{eq:fdmi}). The results are summarized in Tab.~\ref{tab:verification}. Only the subgroup (that makes $\bm{e}_{ij}$ unchanged) symmetry operations are listed, because the other operations would generate the interactions between other equivalent $\bm{u}_{i}$ and $\bm{u}_{j}$ pairs. As can be seen from column sixth of the Tabl.~\ref{tab:verification}, the expression of eq.~(\ref{eq:fdmi}) gives unchanged energy contribution under all the symmetry operations in the subgroups, which means the eDMI form in eq.~(\ref{eq:fdmi}) is also correct for the next-nearest neighbors (all [110] equivalent directions) and next-next-nearest neighbors (all [111] equivalent directions). Note that for a general direction along which only the identity operation $\{1|0\}$ ($x,y,z$) is left, expression of eq.~(\ref{eq:fdmi}) is obviously unchanged.

\subsection{atomistic basis - case: Ti (B-site)}
\begin{figure}[!htp]
    \centering
    \includegraphics[width=0.8\linewidth]{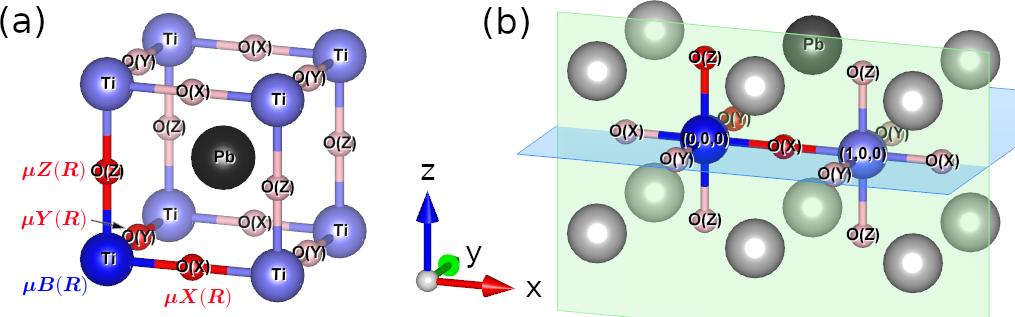}
    \caption{(a) Crystal structure to show the definition of the atomistic displacements (vectors in Cartesian coordinates) $\bm{\mu B}(\bm{R})$ for titanium atom, $\bm{\mu X}(\bm{R})$ for the oxygen located on the positive x-direction with respect to titanium atom (dark blue), $\bm{\mu Y}(\bm{R})$ for the oxygen located on the positive y-direction with respect to titanium atom (dark blue), and $\bm{\mu Z}(\bm{R})$ for the oxygen located on the positive z-direction with respect to titanium atom (dark blue) in one unit cell indexed as $\bm{R}$.
    Note that we use cell index $\bm{R}$ to label the oxygen displacements $\bm{\mu X}$, $\bm{\mu Y}$, and $\bm{\mu Z}$, instead of using the positions of O(X), O(Y), and O(Z) which should be $\bm{R}+(0.5,0,0)$, $\bm{R}+(0,0.5,0)$, and $\bm{R}+(0,0,0.5)$, respectively.
    (b) An illustration plot to show the structure of the nearest-neighbour titanium atoms pair, one titanium atom (dark blue) in cell $(0,0,0)$ and the other titanium atom (light blue) in cell $(1,0,0)$. Note that the oxygen atoms in red and the lead atom in black belong to the same cell $(0,0,0)$ as the titanium atom in blue. All other oxygen atoms that belong to other cells (such as O(X) in $(-1,0,0)$, O(Y) in $(0,-1,0)$, O(Z) in $(0,0,-1)$, O(X), O(Y), and O(Z) in $(1,0,0)$, O(Z) in $(1,0,-1)$, and O(Y) in $(1,-1,0)$) are marked in pink. The origin of each cell is chosen to be on a titanium atom, where the cell index is marked.}\label{fig:atomisticb}
\end{figure}
Equation~(\ref{eq:fdmi}) is an expression for the collective polar modes $\bm{u}$ that is a combination of the displacements of titanium, oxygen, and lead atoms. Such an expression using collective polar modes is convenient to be compared with magnetic DMI and useful for the implementation of effective Hamiltonians or phase field model\cite{Chen2008} (Ginzburg-Landau-Devonshire theory\cite{landau1936,ginzburg1945,ginzburg1949}), however, the role of the displacement of the intermediate oxygen atom on the eDMI is not clear as discussed in the main text Sec. III.B and Sec. IV.D.
Thus we generate an equivalent energy invariant but written in a basis of atomistic displacements, which includes, in each unit cell, the titanium atom displacement $\bm{\mu B}$, the displacements $\bm{\mu X}$, $\bm{\mu Y}$, and $\bm{\mu Z}$ (as indicated in Fig.~\ref{fig:atomisticb} (a)) for the oxygen atoms that are located on the positive x-, y-, and z-direction with respect to the titanium atom, respectively. By performing the same symmetry operation procedures, we only find one energy invariant that can give rise to antisymmetric force constants and is corresponding to the displacments of the nearest neighbour titanium atoms and their intermediate oxygen atoms:
\begin{align}
\mathcal{E}^{B}(\bm{R})=
&+\mu B_{x}(\bm{R}) \mu B_{y}(\bm{R}-\bm{\Delta x}) \mu X_{y}(\bm{R}-\bm{\Delta x}) \nonumber \\ 
&-\mu B_{x}(\bm{R}) \mu B_{y}(\bm{R}+\bm{\Delta x}) \mu X_{y}(\bm{R}) \nonumber \\ 
&+\mu B_{x}(\bm{R}) \mu B_{z}(\bm{R}-\bm{\Delta x}) \mu X_{z}(\bm{R}-\bm{\Delta x}) \nonumber \\ 
&-\mu B_{x}(\bm{R}) \mu B_{z}(\bm{R}+\bm{\Delta x}) \mu X_{z}(\bm{R}) \nonumber \\ 
&+\mu B_{x}(\bm{R}-\bm{\Delta y}) \mu B_{y}(\bm{R}) \mu Y_{x}(\bm{R}-\bm{\Delta y}) \nonumber \\ 
&-\mu B_{x}(\bm{R}+\bm{\Delta y}) \mu B_{y}(\bm{R}) \mu Y_{x}(\bm{R}) \nonumber \\ 
&+\mu B_{y}(\bm{R}) \mu B_{z}(\bm{R}-\bm{\Delta y}) \mu Y_{z}(\bm{R}-\bm{\Delta y}) \nonumber \\ 
&-\mu B_{y}(\bm{R}) \mu B_{z}(\bm{R}+\bm{\Delta y}) \mu Y_{z}(\bm{R}) \nonumber \\ 
&+\mu B_{x}(\bm{R}-\bm{\Delta z}) \mu B_{z}(\bm{R}) \mu Z_{x}(\bm{R}-\bm{\Delta z}) \nonumber \\ 
&-\mu B_{x}(\bm{R}+\bm{\Delta z}) \mu B_{z}(\bm{R}) \mu Z_{x}(\bm{R}) \nonumber \\
&+\mu B_{y}(\bm{R}-\bm{\Delta z}) \mu B_{z}(\bm{R}) \mu Z_{y}(\bm{R}-\bm{\Delta z}) \nonumber \\
&-\mu B_{y}(\bm{R}+\bm{\Delta z}) \mu B_{z}(\bm{R}) \mu Z_{y}(\bm{R})
 \label{eq:EB}
\end{align}
Note that the invariant generation considers all the O(X), O(Y), and O(Z) oxygen atoms within a distance equivalent to three cells ($\Delta x=(l,0,0)$, where $l\in\{-1,0,1\}$, $\Delta y=(0,l,0)$, where $l\in\{-1,0,1\}$, and $\Delta z=(0,0,l)$, where $l\in\{-1,0,1\}$) with respect to cell $\bm{R}=(0,0,0)$, which includes all the interactions until next-next-nearest-neighbour cells.
Following the same argument as when we derive the energy invariants for the collective polar modes,
$\mathcal{E}^{B}(\bm{R})$ is an energy invariant with respect to the cubic symmetry origin at cell $\bm{R}$, which means $\hat{g}\mathcal{E}^{B}(\bm{R})=\mathcal{E}^{B}(\bm{R}), \forall\hat{g}\in\mathcal{G}^{0}$, where $\mathcal{G}^0$ is the quotient group $\mathcal{G}/\mathcal{T}$ with $\mathcal{G}$ being the space group $Pm\bar{3}m$ and $\mathcal{T}$ being the lattice translation group.
Thus the total energy that is invariant under space group $\mathcal{G}$ involves $\mathcal{E}^{B}(\bm{R})$ can be written as $\mathcal{E}^{tot}=\sum\limits_{\bm{R}}^{N}\mathcal{E}^{B}(\bm{R})$, where $N$ indicates all the cells in the crystal.

\begin{table}[!htp]
\caption{The off-diagonal part of the force constants $F_{\alpha,\beta}^{B}$ between titanium atoms pairs ($\bm{\mu B}(\bm{R}_{i})$ and $\bm{\mu B}(\bm{R}_{j})$) that comes from  $\mathcal{E}^{tot}=\sum\limits_{\bm{R}}^{N}\mathcal{E}^{B}(\bm{R})$ are calculated.} \label{tab:fc2}
%\resizebox{0.75\textwidth}{!}{%
\begin{tabular}{|c|c|c|cc|c|c|}
\hline
$\bm{R}_i$               & $\bm{R}_j$               & $\bm{\mathcal{D}}(i,j)$ & $\alpha$ & $\beta$ & $F_{\alpha\beta}^{B}(i,j)=\frac{\partial^2\mathcal{E}^{tot}}{\partial \mu B_{\alpha}(\bm{R}_{i}) \partial \mu B_{\beta}(\bm{R}_{j})}$ & $F_{\beta\alpha}^{B}(i,j)=\frac{\partial^2\mathcal{E}^{tot}}{\partial \mu B_{\beta}(\bm{R}_{i}) \partial \mu B_{\alpha}(\bm{R}_{j})}$ \\
\hline
\multirow{9}{*}{$\{0,0,0\}$} & \multirow{3}{*}{$\{1,0,0\}$} & $\mathcal{D}_x$ & $y$ & $z$ & 0                            & 0 \\
                             &                              & $\mathcal{D}_y$ & $z$ & $x$ & $ \mu X_{z}(0,0,0)$ & $-\mu X_{z}(0,0,0)$ \\
                             &                              & $\mathcal{D}_z$ & $x$ & $y$ & $-\mu X_{y}(0,0,0)$ & $\mu X_{y}(0,0,0)$ \\
                             \cline{2-7}
                             & \multirow{3}{*}{$\{0,1,0\}$} & $\mathcal{D}_x$ & $y$ & $z$ & $-\mu Y_{z}(0,0,0)$ & $\mu Y_{z}(0,0,0)$ \\
                             &                              & $\mathcal{D}_y$ & $z$ & $x$ & 0                            & 0 \\
                             &                              & $\mathcal{D}_z$ & $x$ & $y$ & $ \mu Y_{x}(0,0,0)$ & $-\mu Y_{x}(0,0,0)$ \\
                             \cline{2-7}
                             & \multirow{3}{*}{$\{0,0,1\}$} & $\mathcal{D}_x$ & $y$ & $z$ & $ \mu Z_{y}(0,0,0)$ & $-\mu Z_{y}(0,0,0)$ \\
                             &                              & $\mathcal{D}_y$ & $z$ & $x$ & $-\mu Z_{x}(0,0,0)$ & $\mu Z_{x}(0,0,0)$ \\
                             &                              & $\mathcal{D}_z$ & $x$ & $y$ & 0                            & 0 \\
\hline
\end{tabular}%
%}
\end{table}

By performing second derivatives of $E^{tot}$ with respect to the atomistic displacements $\bm{\mu B}(\bm{R}_{i})$ and $\bm{\mu B}(\bm{R}_{j})$, force constants $F_{\alpha,\beta}^{B}$ between titanium atoms on sites i and j can be calculated. 
Such results are summarized in Tab.~\ref{tab:fc2} of the SM, where site $i$ is chosen as $\bm{R}_i=(0,0,0)$ and site $j$ is chosen at the positions of  $(1,0,0)$, $(0,1,0)$, and $(0,0,1)$, respectively. 
The off-diagonal part of force constants in Tab.~\ref{tab:fc2} of the SM is also antisymmetric (opposite sign between column 5 and 6.).
The eDMI can be summarized as:
\begin{equation}\label{eq:fdmib}
  E_{dmi}^{B}=
  \begin{cases}
    (\bm{\mu X}(\bm{R}_{i})\cross\bm{e}_{ij})\cdot (\bm{\mu B}(\bm{R}_{i})\cross\bm{\mu B}(\bm{R}_{j})), & \text{if $\bm{e}_{ij}=\bm{R}_{j}-\bm{R}_{i}=(1,0,0)$} \\
    (\bm{\mu Y}(\bm{R}_{i})\cross\bm{e}_{ij})\cdot (\bm{\mu B}(\bm{R}_{i})\cross\bm{\mu B}(\bm{R}_{j})), & \text{if $\bm{e}_{ij}=\bm{R}_{j}-\bm{R}_{i}=(0,1,0)$} \\
    (\bm{\mu Z}(\bm{R}_{i})\cross\bm{e}_{ij})\cdot (\bm{\mu B}(\bm{R}_{i})\cross\bm{\mu B}(\bm{R}_{j})), & \text{if $\bm{e}_{ij}=\bm{R}_{j}-\bm{R}_{i}=(0,0,1)$} 
  \end{cases}
\end{equation}

It is worth to notice that Eq.~(\ref{eq:fdmib}) share the similar form as Eq.~(\ref{eq:fdmi}) and the $\bm{\mathcal{D}}(i,j)$ vector between site i and j is proportional to the displacements of their intermediate oxygen site.
More specifically, according to Eq.~(\ref{eq:fdmib}), (1) along x-direction ($\bm{e}_{ij}=(1,0,0)$), $\bm{\mathcal{D}}(i,j)$ vector is equal to $(0,\mu X_{z},-\mu X_{y})$ which means that the displacement of intermediate oxygen O(X) ($\bm{\mu X}(\bm{R}_{i})$) along z-direction (y-direction) will result in y-component (z-component) of the eDMI vector and the x-component of $\bm{\mathcal{D}}(i,j)$ vector is always zero regardless of $\bm{\mu X}(\bm{R}_{i})$; (2) along y-direction ($\bm{e}_{ij}=(0,1,0)$), $\bm{\mathcal{D}}(i,j)$ vector is equal to $(-\mu Y_{z},0,\mu Y_{x})$ which means that the displacement of intermediate oxygen O(Y) ($\bm{\mu Y}(\bm{R}_{i})$) along z-direction (x-direction) will result in x-component (z-component) of the eDMI vector and the y-component of $\bm{\mathcal{D}}(i,j)$ vector is always zero regardless of $\bm{\mu Y}(\bm{R}_{i})$; (3) along z-direction ($\bm{e}_{ij}=(0,0,1)$), $\bm{\mathcal{D}}(i,j)$ vector is equal to $(\mu Z_{y},-\mu Z_{x},0)$ which means that the displacement of intermediate oxygen O(Z) ($\bm{\mu Z}(\bm{R}_{i})$) along y-direction (x-direction) will result in x-component (y-component) of the eDMI vector and the z-component of $\bm{\mathcal{D}}(i,j)$ vector is always zero regardless of $\bm{\mu Z}(\bm{R}_{i})$;.
It is also worth to mention that the calculated $\bm{\mathcal{D}}(i,j)$ vector in atomistic basis is connected with the $\bm{\mathcal{D}}(i,j)$ vector written in collective polar mode basis.
For instance, (1) along x-direction ($\bm{e}_{ij}=(1,0,0)$), the $\bm{\mathcal{D}}(i,j)$ vector is along y-direction (or z-direction) when the polar modes (that contains all ions' displacements) are along z-direction (or y-direction) (see $u_{z}(0,0,0)+u_{z}(1,0,0)$ or $u_{y}(0,0,0)+u_{y}(1,0,0)$ in Tab.~\ref{tab:fc}) which is consistent with the intermediate oxygen (O(X)) that is displaced towards also z-direction (or y-direction) (see $\mu X_{z}(0,0,0)$ or $\mu X_{y}(0,0,0)$ in Tab.~\ref{tab:fc2}); (2) along y-direction ($\bm{e}_{ij}=(0,1,0)$), the $\bm{\mathcal{D}}(i,j)$ vector is along x-direction (or z-direction) when the polar modes (that contains all ions' displacements) are along z-direction (or x-direction) (see $u_{z}(0,0,0)+u_{z}(0,1,0)$ or $u_{x}(0,0,0)+u_{x}(0,1,0)$ in Tab.~\ref{tab:fc}) which is consistent with the intermediate oxygen O(Y) that is displaced towards also z-direction (or x-direction) (see $\mu Y_{z}(0,0,0)$ or $\mu Y_{x}(0,0,0)$ in Tab.~\ref{tab:fc2}); (3) along z-direction ($\bm{e}_{ij}=(0,0,1)$), the $\bm{\mathcal{D}}(i,j)$ vector is along x-direction (or y-direction) when the polar modes (that contains all ions' displacements) are along y-direction (or x-direction) (see $u_{y}(0,0,0)+u_{y}(0,0,1)$ or $u_{x}(0,0,0)+u_{x}(0,0,1)$ in Tab.~\ref{tab:fc}) which is consistent with the intermediate oxygen O(X) that is displaced towards also y-direction (or x-direction) (see $\mu Z_{y}(0,0,0)$ or $\mu Z_{x}(0,0,0)$ in Tab.~\ref{tab:fc2}).

\subsection{atomistic basis - case: Pb (A-site)}
\begin{figure}[!htp]
    \centering
    \includegraphics[width=0.8\linewidth]{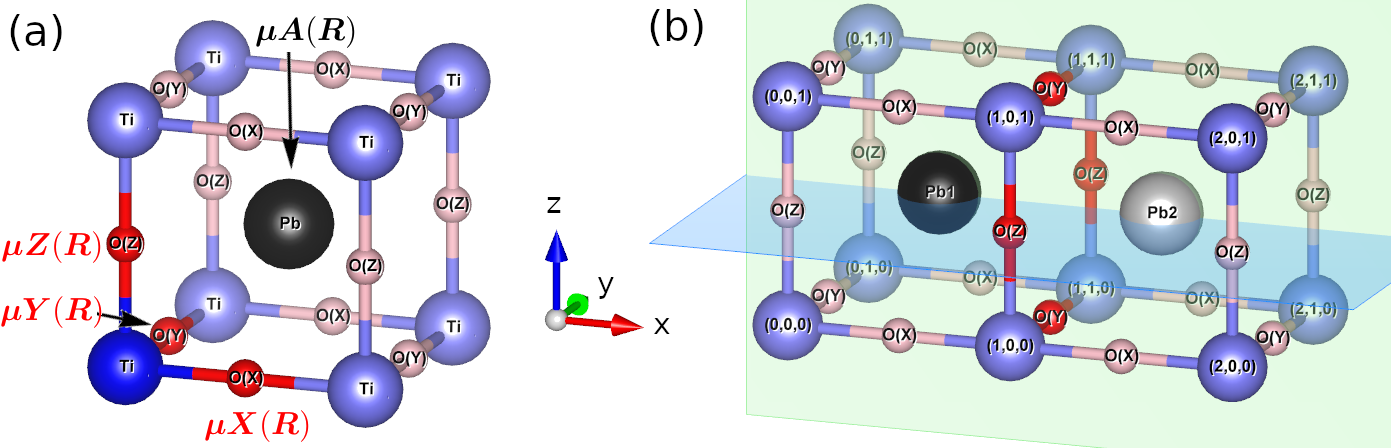}
    \caption{(a) Crystal structure to show the definition of the atomistic displacements (vectors in Cartesian coordinates) $\bm{\mu A}(\bm{R})$ for lead atom (black), $\bm{\mu X}(\bm{R})$ for the oxygen located on the positive x-direction with respect to titanium atom (dark blue), $\bm{\mu Y}(\bm{R})$ for the oxygen located on the positive y-direction with respect to titanium atom (dark blue), and $\bm{\mu Z}(\bm{R})$ for the oxygen located on the positive z-direction with respect to titanium atom (dark blue) in one unit cell at position $\bm{R}$. (b) An illustration plot to show the structure of the nearest-neighbor lead atoms pair, one lead atom (black) in cell (0,0,0) and the other lead atom (gray) in cell (100). Note that the oxygen atoms in red O(Y) (in cell (1,0,0)), O(Y) (in cell (1,0,1)), O(Z) (in cell (1,0,0)) and O(Z) (in cell (1,1,0)) are the intermediate ions between the lead atoms pair Pb1 (black) and Pb2 (gray) along x-direction. The origin of each cell is chosen to be on titanium atom, where the cell index is marked.}\label{fig:atomistica}
\end{figure}

We have also generated energy invariants that correspond to lead atom displacments $\bm{\mu A}$, oxygen atom displacements $\bm{\mu X}$, $\bm{\mu Y}$, and $\bm{\mu Z}$ (as indicated in Fig.~\ref{fig:atomistica} (a)). By performing the same symmetry operation procedures, we find two energy invariants that can give rise to antisymmetric force constants and is corresponding to the displacments of the nearest neighbour lead atoms and their intermediate oxygen atoms:
\begin{align}
\mathcal{E}_{1}^{A}(\bm{R})=
& + \mu A_{x}(\bm{R} - \Delta y) \mu A_{y}(\bm{R}) \mu X_{x}(\bm{R}) \nonumber \\
& - \mu A_{x}(\bm{R}) \mu A_{y}(\bm{R} - \Delta y) \mu X_{x}(\bm{R}) \nonumber \\
& + \mu A_{x}(\bm{R} - \Delta z) \mu A_{z}](\bm{R}) \mu X_{x}(\bm{R}) \nonumber \\
& - \mu A_{x}(\bm{R}) \mu A_{z}(\bm{R} - \Delta z) \mu X_{x}(\bm{R}) \nonumber \\ 
& + \mu A_{x}(\bm{R}) \mu A_{y}(\bm{R} + \Delta y) \mu X_{x}(\bm{R} + \Delta y) \nonumber \\
& + \mu A_{x}(\bm{R} - \Delta z) \mu A_{z}(\bm{R}) \mu X_{x}(\bm{R} + \Delta y) \nonumber \\
& - \mu A_{x}(\bm{R}) \mu A_{z}(\bm{R} - \Delta z) \mu X_{x}(\bm{R} + \Delta y) \nonumber \\
& + \mu A_{x}(\bm{R} - \Delta y) \mu A_{y}(\bm{R}) \mu X_{x}(\bm{R} + \Delta z) \nonumber \\
& - \mu A_{x}(\bm{R}) \mu A_{y}(\bm{R} - \Delta y) \mu X_{x}(\bm{R} + \Delta z) \nonumber \\
& + \mu A_{x}(\bm{R}) \mu A_{z}(\bm{R} + \Delta z) \mu X_{x}(\bm{R} + \Delta z) \nonumber \\
& + \mu A_{x}(\bm{R}) \mu A_{y}(\bm{R} + \Delta y) \mu X_{x}(\bm{R} + \Delta y + \Delta z) \nonumber \\
& + \mu A_{x}(\bm{R}) \mu A_{z}(\bm{R} + \Delta z) \mu X_{x}(\bm{R} + \Delta y + \Delta z) \nonumber \\
& - \mu A_{x}(\bm{R} - \Delta x) \mu A_{y}(\bm{R}) \mu Y_{y}(\bm{R}) \nonumber \\
& + \mu A_{x}(\bm{R}) \mu A_{y}(\bm{R} - \Delta x) \mu Y_{y}(\bm{R}) \nonumber \\ 
& + \mu A_{y}(\bm{R} - \Delta z) \mu A_{z}(\bm{R}) \mu Y_{y}(\bm{R}) \nonumber \\
& - \mu A_{y}(\bm{R}) \mu A_{z}(\bm{R} - \Delta z) \mu Y_{y}(\bm{R})  \nonumber \\
& + \mu A_{x}(\bm{R} + \Delta x) \mu A_{y}(\bm{R}) \mu Y_{y}(\bm{R} + \Delta x) \nonumber \\
& + \mu A_{y}(\bm{R} - \Delta z) \mu A_{z}(\bm{R}) \mu Y_{y}(\bm{R} + \Delta x) \nonumber \\
& - \mu A_{y}(\bm{R}) \mu A_{z}(\bm{R} - \Delta z) \mu Y_{y}(\bm{R} + \Delta x) \nonumber \\
& - \mu A_{x}(\bm{R} - \Delta x) \mu A_{y}(\bm{R}) \mu Y_{y}(\Delta z) \nonumber \\
& + \mu A_{x}(\bm{R}) \mu A_{y}(\bm{R} - \Delta x) \mu Y_{y}(\bm{R} + \Delta z)  \nonumber \\
& + \mu A_{y}(\bm{R}) \mu A_{z}(\bm{R} + \Delta z) \mu Y_{y}(\bm{R} + \Delta z) \nonumber \\
& + \mu A_{x}(\bm{R} + \Delta x) \mu A_{y}(\bm{R}) \mu Y_{y}(\bm{R} + \Delta x + \Delta z) \nonumber \\
& + \mu A_{y}(\bm{R}) \mu A_{z}(\bm{R} + \Delta z) \mu Y_{y}(\bm{R} + \Delta x + \Delta z) \nonumber \\
& - \mu A_{x}(\bm{R} - \Delta x) \mu A_{z}(\bm{R}) \mu Z_{z}(\bm{R}) \nonumber \\
& - \mu A_{y}(\bm{R} - \Delta y) \mu A_{z}(\bm{R}) \mu Z_{z}(\bm{R})  \nonumber \\
& + \mu A_{x}(\bm{R}) \mu A_{z}(\bm{R} - \Delta x) \mu Z_{z}(\bm{R}) \nonumber \\
& + \mu A_{y}(\bm{R}) \mu A_{z}(\bm{R} - \Delta y) \mu Z_{z}(\bm{R}) \nonumber \\
& + \mu A_{x}(\bm{R} + \Delta x) \mu A_{z}(\bm{R}) \mu Z_{z}(\bm{R} + \Delta x) \nonumber \\
& - \mu A_{y}(\bm{R} - \Delta y) \mu A_{z}(\bm{R}) \mu Z_{z}(\bm{R} + \Delta x) \nonumber \\
& + \mu A_{y}(\bm{R}) \mu A_{z}(\bm{R} - \Delta y) \mu Z_{z}(\bm{R} + \Delta x) \nonumber \\
& - \mu A_{x}(\bm{R} - \Delta x) \mu A_{z}(\bm{R}) \mu Z_{z}(\bm{R} + \Delta y) \nonumber \\
& + \mu A_{y}(\bm{R} + \Delta y) \mu A_{z}(\bm{R}) \mu Z_{z}(\bm{R} + \Delta y) \nonumber \\
& + \mu A_{x}(\bm{R}) \mu A_{z}(\bm{R} - \Delta x) \mu Z_{z}(\bm{R} + \Delta y) \nonumber \\
& + \mu A_{x}(\bm{R} + \Delta x) \mu A_{z}(\bm{R}) \mu Z_{z}(\bm{R} + \Delta x + \Delta y) \nonumber \\
& + \mu A_{y}(\bm{R} + \Delta y) \mu A_{z}(\bm{R}) \mu Z_{z}(\bm{R} + \Delta x + \Delta y) \label{eq:EA1}
\end{align}
and
\begin{align}
\mathcal{E}_{2}^{A}(\bm{R}) = &+ \mu A_{y}(\bm{R} - \Delta z) \mu A_{z}(\bm{R}) \mu X_{y})(\bm{R}) \nonumber \\
&- \mu A_{y}(\bm{R}) \mu A_{z}(\bm{R} - \Delta z) \mu X_{y})(\bm{R}) \nonumber \\
&+ \mu A_{y}(\bm{R} - \Delta z) \mu A_{z}(\bm{R}) \mu X_{y})(\bm{R} + \Delta y) \nonumber \\
&- \mu A_{y}(\bm{R}) \mu A_{z}(\bm{R} - \Delta z) \mu X_{y})(\bm{R} + \Delta y) \nonumber \\
&+ \mu A_{y}(\bm{R}) \mu A_{z}(\bm{R} + \Delta z) \mu X_{y})(\bm{R} + \Delta z) \nonumber \\
&+ \mu A_{y}(\bm{R}) \mu A_{z}(\bm{R} + \Delta z) \mu X_{y})(\bm{R} + \Delta y + \Delta z) \nonumber \\
&- \mu A_{y}(\bm{R} - \Delta y) \mu A_{z}(\bm{R}) \mu X_{z})(\bm{R}) \nonumber \\
&+ \mu A_{y}(\bm{R}) \mu A_{z}(\bm{R} - \Delta y) \mu X_{z})(\bm{R}) \nonumber \\
&+ \mu A_{y}(\bm{R} + \Delta y) \mu A_{z}(\bm{R}) \mu X_{z})(\bm{R} + \Delta y) \nonumber \\
&- \mu A_{y}(\bm{R} - \Delta y) \mu A_{z}(\bm{R}) \mu X_{z})(\bm{R} + \Delta z) \nonumber \\
&+ \mu A_{y}(\bm{R}) \mu A_{z}(\bm{R} - \Delta y) \mu X_{z})(\bm{R} + \Delta z) \nonumber \\
&+ \mu A_{y}(\bm{R} + \Delta y) \mu A_{z}(\bm{R}) \mu X_{z})(\bm{R} + \Delta y + \Delta z) \nonumber \\
&+ \mu A_{x}(\bm{R} - \Delta z) \mu A_{z}(\bm{R}) \mu Y_{x})(\bm{R}) \nonumber \\
&- \mu A_{x}(\bm{R}) \mu A_{z}(\bm{R} - \Delta z) \mu Y_{x})(\bm{R}) \nonumber \\
&+ \mu A_{x}(\bm{R} - \Delta z) \mu A_{z}(\bm{R}) \mu Y_{x})(\bm{R} + \Delta x) \nonumber \\
&- \mu A_{x}(\bm{R}) \mu A_{z}(\bm{R} - \Delta z) \mu Y_{x})(\bm{R} + \Delta x) \nonumber \\
&+ \mu A_{x}(\bm{R}) \mu A_{z}(\bm{R} + \Delta z) \mu Y_{x})(\bm{R} + \Delta z) \nonumber \\
&+ \mu A_{x}(\bm{R}) \mu A_{z}(\bm{R} + \Delta z) \mu Y_{x})(\bm{R} + \Delta x + \Delta z) \nonumber \\
&- \mu A_{x}(\bm{R} - \Delta x) \mu A_{z}(\bm{R}) \mu Y_{z})(\bm{R}) \nonumber \\
&+ \mu A_{x}(\bm{R}) \mu A_{z}(\bm{R} - \Delta x) \mu Y_{z})(\bm{R}) \nonumber \\
&+ \mu A_{x}(\bm{R} + \Delta x) \mu A_{z}(\bm{R}) \mu Y_{z})(\bm{R} + \Delta x) \nonumber \\
&- \mu A_{x}(\bm{R} - \Delta x) \mu A_{z}(\bm{R}) \mu Y_{z})(\bm{R} + \Delta z) \nonumber \\
&+ \mu A_{x}(\bm{R}) \mu A_{z}(\bm{R} - \Delta x) \mu Y_{z})(\bm{R} + \Delta z) \nonumber \\
&+ \mu A_{x}(\bm{R} + \Delta x) \mu A_{z}(\bm{R}) \mu Y_{z})(\bm{R} + \Delta x + \Delta z) \nonumber \\
&+ \mu A_{x}(\bm{R} - \Delta y) \mu A_{y}(\bm{R}) \mu Z_{x})(\bm{R}) \nonumber \\
&- \mu A_{x}(\bm{R}) \mu A_{y}(\bm{R} - \Delta y) \mu Z_{x})(\bm{R}) \nonumber \\
&+ \mu A_{x}(\bm{R} - \Delta y) \mu A_{y}(\bm{R}) \mu Z_{x})(\bm{R} + \Delta x) \nonumber \\
&- \mu A_{x}(\bm{R}) \mu A_{y}(\bm{R} - \Delta y) \mu Z_{x})(\bm{R} + \Delta x) \nonumber \\
&+ \mu A_{x}(\bm{R}) \mu A_{y}(\bm{R} + \Delta y) \mu Z_{x})(\bm{R} + \Delta y) \nonumber \\
&+ \mu A_{x}(\bm{R}) \mu A_{y}(\bm{R} + \Delta y) \mu Z_{x})(\bm{R} + \Delta x + \Delta y) \nonumber \\
&- \mu A_{x}(\bm{R} - \Delta x) \mu A_{y}(\bm{R}) \mu Z_{y})(\bm{R}) \nonumber \\
&+ \mu A_{x}(\bm{R}) \mu A_{y}(\bm{R} - \Delta x) \mu Z_{y})(\bm{R}) \nonumber \\
&+ \mu A_{x}(\bm{R} + \Delta x) \mu A_{y}(\bm{R}) \mu Z_{y})(\bm{R} + \Delta x) \nonumber \\
&- \mu A_{x}(\bm{R} - \Delta x) \mu A_{y}(\bm{R}) \mu Z_{y})(\bm{R} + \Delta y) \nonumber \\
&+ \mu A_{x}(\bm{R}) \mu A_{y}(\bm{R} - \Delta x) \mu Z_{y})(\bm{R} + \Delta y) \nonumber \\
&+ \mu A_{x}(\bm{R} + \Delta x) \mu A_{y}(\bm{R}) \mu Z_{y})(\bm{R} + \Delta x + \Delta y) \label{eq:EA2}
\end{align}

Note that the invariant generation considers all the O(X), O(Y), and O(Z) oxygen atoms within a distance equivalent to three cells ($\Delta x=(l,0,0)$, where $l\in\{-1,0,1\}$, $\Delta y=(0,l,0)$, where $l\in\{-1,0,1\}$, and $\Delta z=(0,0,l)$, where $l\in\{-1,0,1\}$) with respect to cell $\bm{R}=(0,0,0)$, which includes all the interactions until next-next-nearest-neighbour cells.
Energy $A(\mathcal{E}_{1}^{A}(\bm{R})+\mathcal{E}_{2}^{A}(\bm{R}))$  (A is a constant) is also an invariant with respect to the cubic symmetry origin at cell $\bm{R}$, since $\hat{g}\mathcal{E}_{1}^{A}(\bm{R})=\mathcal{E}_{1}^{A}(\bm{R}), \forall\hat{g}\in\mathcal{G}^{0}$ and $\hat{g}\mathcal{E}_{2}^{A}(\bm{R})=\mathcal{E}_{2}^{A}(\bm{R}), \forall\hat{g}\in\mathcal{G}^{0}$, where $\mathcal{G}^0$ is the quotient group $\mathcal{G}/\mathcal{T}$ with $\mathcal{G}$ being the space group $Pm\bar{3}m$ and $\mathcal{T}$ being the lattice translation group.
Thus the total energy that is invariant under space group $\mathcal{G}$ involves $A(\mathcal{E}_{1}^{A}(\bm{R})+\mathcal{E}_{2}^{A}(\bm{R}))$ can be written as $\mathcal{E}^{tot}=\sum\limits_{\bm{R}}^{N}A(\mathcal{E}_{1}^{A}(\bm{R})+\mathcal{E}_{2}^{A}(\bm{R}))$, where $N$ indicates all the cells in the crystal.

By performing second derivatives of $E^{tot}$ with respect to the atomistic displacements $\bm{\mu A}(\bm{R}_{i})$ and $\bm{\mu A}(\bm{R}_{j})$, force constants $F_{\alpha,\beta}^{A}$ between lead atoms on sites i and j can be calculated. 
Such results are summarized in Tab.~\ref{tab:fc3} of the SM, where site $i$ is chosen as $\bm{R}_i=(0,0,0)$ and site $j$ is chosen at the positions of  $(1,0,0)$, $(0,1,0)$, and $(0,0,1)$, respectively. 
\begin{table}[!htp]
\caption{The off-diagonal part of the force constants $F_{\alpha,\beta}^{A}$ between lead atoms pairs ($\bm{\mu A}(\bm{R}_{i})$ and $\bm{\mu A}(\bm{R}_{j})$) that comes from the $\mathcal{E}^{tot}=\sum\limits_{\bm{R}}^{N}A(\mathcal{E}_{1}^{A}(\bm{R})+\mathcal{E}_{2}^{A}(\bm{R}))$ are calculated.}\label{tab:fc3}
\resizebox{0.95\textwidth}{!}{%
\begin{tabular}{|c|c|c|cc|c|c|}
\hline
$\bm{R}_i$               & $\bm{R}_j$               & $\bm{\mathcal{D}}(i,j)$ & $\alpha$ & $\beta$ & $F_{\alpha\beta}^{A}(i,j)=\frac{\partial^2\mathcal{E}^{tot}}{\partial \mu A_{\alpha}(\bm{R}_{i}) \partial \mu A_{\beta}(\bm{R}_{j})}$ & $F_{\beta\alpha}^{A}(i,j)=\frac{\partial^2\mathcal{E}^{tot}}{\partial \mu A_{\beta}(\bm{R}_{i}) \partial \mu A_{\alpha}(\bm{R}_{j})}$ \\
\hline
\multirow{9}{*}{$\{0,0,0\}$} & \multirow{3}{*}{$\{1,0,0\}$} & $\mathcal{D}_x$ & $y$ & $z$ & 0                            & 0 \\
                             &                              & $\mathcal{D}_y$ & $z$ & $x$ & $\mu Y_{z}(1,0,0)+\mu Y_{z}(1,0,1)+\mu Z_{z}(1,0,0)+\mu Z_{z}(1,1,0)$ & $-\mu Y_{z}(1,0,0)-\mu Y_{z}(1,0,1)-\mu Z_{z}(1,0,0)-\mu Z_{z}(1,1,0)$ \\
                             &                              & $\mathcal{D}_z$ & $x$ & $y$ & $-\mu Y_{y}(1,0,0)-\mu Y_{y}(1,0,1)-\mu Z_{y}(1,0,0)-\mu Z_{y}(1,1,0)$ & $ \mu Y_{y}(1,0,0)+\mu Y_{y}(1,0,1)+\mu Z_{y}(1,0,0)+\mu Z_{y}(1,1,0)$ \\
                             \cline{2-7}
                             & \multirow{3}{*}{$\{0,1,0\}$} & $\mathcal{D}_x$ & $y$ & $z$ & $-\mu X_{z}(0,1,0)-\mu X_{z}(0,1,1)-\mu Z_{z}(0,1,0)-\mu Z_{z}(1,1,0)$ & $ \mu X_{z}(0,1,0)+\mu X_{z}(0,1,1)+\mu Z_{z}(0,1,0)+\mu Z_{z}(1,1,0)$ \\
                             &                              & $\mathcal{D}_y$ & $z$ & $x$ & 0                            & 0 \\
                             &                              & $\mathcal{D}_z$ & $x$ & $y$ & $\mu X_{x}(0,1,0)+\mu X_{x}(0,1,1)+\mu Z_{x}(0,1,0)+\mu Z_{x}(1,1,0)$ & $-\mu X_{x}(0,1,0)-\mu X_{x}(0,1,1)-\mu Z_{x}(0,1,0)-\mu Z_{x}(1,1,0)$ \\
                             \cline{2-7}
                             & \multirow{3}{*}{$\{0,0,1\}$} & $\mathcal{D}_x$ & $y$ & $z$ & $\mu X_{y}(0,0,1)+\mu X_{y}(0,1,1)+\mu Y_{y}(0,0,1)+\mu Y_{y}(1,0,1)$ & $-\mu X_{y}(0,0,1)-\mu X_{y}(0,1,1)-\mu Y_{y}(0,0,1)-\mu Y_{y}(1,0,1)$ \\
                             &                              & $\mathcal{D}_y$ & $z$ & $x$ & $-\mu X_{x}(0,0,1)-\mu X_{x}(0,1,1)-\mu Y_{x}(0,0,1)-\mu Y_{x}(1,0,1)$ & $ \mu X_{x}(0,0,1)+\mu X_{x}(0,1,1)+\mu Y_{x}(0,0,1)+\mu Y_{x}(1,0,1)$ \\
                             &                              & $\mathcal{D}_z$ & $x$ & $y$ & 0                            & 0 \\
\hline
\end{tabular}%
}
\end{table}

\begin{figure}[!htp]
    \centering
    \includegraphics[width=0.5\linewidth]{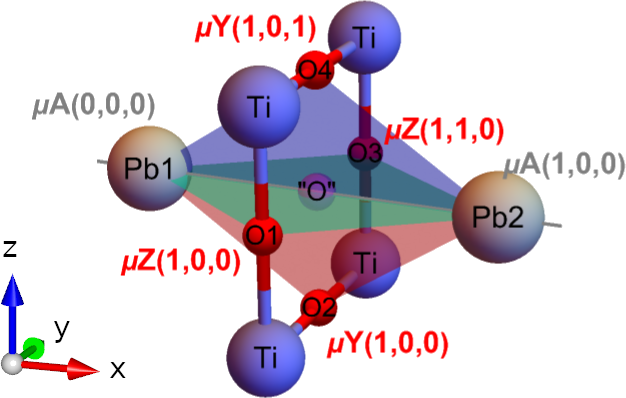}
    \caption{An illustration plot to show the structure of the nearest-neighbor lead atoms Pb1 in cell (0,0,0) and Pb2 in cell (100). The intermediate oxygen atoms are O1, O2, O3, and O4, corresponding to the O(Z) in cell (1,0,0), O(Y) in cell (1,0,0), O(Z) in cell (1,1,0), and O(Y) in cell (1,0,1), respectively. The magenta ball with label ``O'' is the average position of O1, O2, O3, and O4 and also the middle point between Pb1 and Pb2. It should not be treated as a real oxygen atom.}\label{fig:AA}
\end{figure}

The off-diagonal part of force constants $F_{\alpha,\beta}^{A}$ in Tab.~\ref{tab:fc3} of the SM is also antisymmetric (opposite sign between column 5 and 6.).
The eDMI for the case between lead atoms pairs are difficult to summarize, because there are four intermediate oxygen atoms as can be seen from supplementary fig.~\ref{fig:atomistica} (b) (red balls) and fig.~\ref{fig:AA} (O1, O2, O3, and O4).

However, the eDMI between lead atoms can be seen as assemble of four cation-ion-cation interactions (Pb1-O1-Pb2, Pb1-O2-Pb2, Pb1-O3-Pb2, and Pb1-O4-Pb2 as marked in fig.~\ref{fig:AA}) compared to the one Ti1-O-Ti2 interaction case in the previous subsection.
Taking for example the illustrated case in fig.~\ref{fig:AA} where $\bm{R}_{i}=(0,0,0)$, $\bm{R}_{j}=(1,0,0)$, and the eDMI vector $\mathcal{D}(i,j)=(0,\mu Y_{z}(1,0,0)+\mu Y_{z}(1,0,1)+\mu Z_{z}(1,0,0)+\mu Z_{z}(1,1,0),-\mu Y_{y}(1,0,0)-\mu Y_{y}(1,0,1)-\mu Z_{y}(1,0,0)-\mu Z_{y}(1,1,0))$, the energy coming from eDMI ($E_{dmi}^{A}$) can be split into two nonzero terms (by different components of the $\bm{\mathcal{D}}(i,j)$ vector):
\begin{subequations}
\begin{align}
E_{dmi}^{A}&=(0,\mu Y_{z}(1,0,0)+\mu Y_{z}(1,0,1)+\mu Z_{z}(1,0,0)+\mu Z_{z}(1,1,0),~0)\cdot(\bm{\mu A}(0,0,0)\cross\bm{\mu A}(1,0,0)) \label{eq:pbozpb}  \\
&-(0,~0,\mu Y_{y}(1,0,0)+\mu Y_{y}(1,0,1)+\mu Z_{y}(1,0,0)+\mu Z_{y}(1,1,0))\cdot(\bm{\mu A}(0,0,0)\cross\bm{\mu A}(1,0,0)) \label{eq:pboypb} 
\end{align}
\end{subequations}
In addition, to compare with the eDMI between titanium atoms (Ti-O-Ti), let's expand eq.~\ref{eq:fdmib} along [100] direction ($\bm{R}_{i}=(0,0,0)$, $\bm{R}_{j}=(1,0,0)$):
\begin{subequations}\label{eq:fdmibexpand}
\begin{align}
E_{dmi}^{B}&=(0,\mu X_{z}(0,0,0),~0)\cdot(\bm{\mu B}(0,0,0)\cross\bm{\mu B}(1,0,0)) \label{eq:tixzti} \\
&-(0,~0,\mu X_{y}(0,0,0))\cdot(\bm{\mu B}(0,0,0)\cross\bm{\mu B}(1,0,0)) \label{eq:tixyti}
\end{align}
\end{subequations}

Noticing that eq.~(\ref{eq:pbozpb}) and eq.~(\ref{eq:pboypb}) involve couplings between $\bm{\mu A}(0,0,0)$ (displacements of Pb1) and $\bm{\mu A}(1,0,0)$ (displacements of Pb2) through O1 (see Pb1-O1-Pb2 in fig.~\ref{fig:AA}), O2 (see Pb1-O2-Pb2 in fig.~\ref{fig:AA}), O3 (see Pb1-O3-Pb2 in fig.~\ref{fig:AA}), O4 (see Pb1-O4-Pb2 in fig.~\ref{fig:AA}).
More specifically, eq.~(\ref{eq:pbozpb}) is corresponding to coupling between $\bm{\mu A}(0,0,0)$ (displacements of Pb1) and $\bm{\mu A}(1,0,0)$ (displacements of Pb2) via the displacements of O1 ($\mu Z_{z}(1,0,0)$), O2 ($\mu Y_{z}(1,0,0)$), O3 ($\mu Z_{z}(1,1,0)$), and O4 ($\mu Y_{z}(1,0,1)$) along z-direction and eq.~(\ref{eq:pboypb}) is corresponding to coupling between $\bm{\mu A}(0,0,0)$ (displacements of Pb1) and $\bm{\mu A}(1,0,0)$ (displacements of Pb2) via the displacements of O1 ($\mu Z_{y}(1,0,0)$), O2 ($\mu Y_{y}(1,0,0)$), O3 ($\mu Z_{Y}(1,1,0)$), and O4 ($\mu Y_{y}(1,0,1)$) along y-direction.
Interestingly, the average position of O1, O2, O3, and O4 will be exactly on the middle point (``O'' in fig.~\ref{fig:AA}) between Pb1 and Pb2. 
Equation~(\ref{eq:pbozpb}) can also be interpreted as that the displacements along z-direction of all O1, O2, O3, and O4 result in (1) the displacements of (the average position) ``O'' along z-direction (away from the middle point between Pb1 and Pb2); (2) local inversion symmetry breaking by forming Pb1-``O''-Pb2 triangle with normal vector parallel to y-direction (compared to Ti1-O-Ti2 triangle); and (3) y-component of eDMI vector.
Thus, eq.~(\ref{eq:pbozpb}) is an analog of eq.~(\ref{eq:tixzti}) in which the oxygen O(X) that on the middle point of Ti1 and Ti2 has a displacement along z-direction and gives rise to the y-component of eDMI vector.
Similarly, eq.~(\ref{eq:pboypb}) can also be interpreted as that the displacements along y-direction of all O1, O2, O3, and O4 result in (1) the displacements of (the average position) ``O'' along y-direction (away from the middle point between Pb1 and Pb2); (2) local inversion symmetry breaking by forming Pb1-``O''-Pb2 triangle with normal vector parallel to z-direction (compared to Ti1-O-Ti2 triangle); and (3) z-component of eDMI vector.
Thus, eq.~(\ref{eq:pboypb}) is an analog of eq.~(\ref{eq:tixyti}) in which the oxygen O(X) that on the middle point of Ti1 and Ti2 has a displacement along y-direction and gives rise to the z-component of eDMI vector.

So far we have notice that the eDMI between lead atoms are more complicated than the eDMi between titanium atoms, because the former one has more than one intermediate oxygen atoms. 
However, they share the same physics as has been discussed in the main text, the local inversion breaking by the (average) displacements of the intermediate atoms gives rise to eDMI and the $\bm{\mathcal{D}}$ vector is parallel to the normal direction of the triangle that is formed by the two considered sites and their (average) displaced intermediate atoms.

One comment on the choice of the collective basis and atomistic basis is that the collective basis gives more concise in the formalism and is an direct expression when implementing into the effective Hamiltonian and phase field (Ginzburg-Landau-Devonshire model) simulations, while the atomistic basis gives complex expressions but can be more intuitive to understand the role of the symmetry and the intermediate atoms.

%\section{DFPT force constant results for some typical ferroelectric materials}

\section{Effective Hamiltonian model}
\begin{table}[!htp]
\begin{center}
\caption{Expansion parameters of the effective Hamiltonian for PbTiO$_3$. Atomic units are used here. The reference cubic lattice parameter
is 7.35 Bohr}\label{tab:parameters} 
\begin{tabular}{lcrcrcr} 
 \hline\hline
 Dipole & $Z^{*}$ & 8.329 & $\epsilon_{\infty}$ & 8.259 \\
 \hline
 $E^{onsite}(P)$ & $\kappa$ & 0.0444259 & $\alpha$ & 0.021716 & $\gamma$ & 0.0418993 \\
                      & $j_{1}$ & -0.012854 & $j_{2}$ & 0.012227 &    &  \\
 $E^{nn}$ & $j_{3}$ & -0.002757 & $j_{4}$ & -0.002761 & $j_{5}$ & -0.000583 \\
                      & $j_{6}$ & -0.001768 & $j_{7}$ & -0.000533 &    &  \\
$E^{dmi}$ & $\mathcal{A}^{-}_{nn}$ & -0.001021 & $\mathcal{A}^{-}_{nnn}$ & -0.000544\\
 $E^{int}(P,\eta)$ & $C_{1111}$ & -0.2186 & $C_{1122}$ & -0.051336 & $C_{1212}$ & -0.0111 \\
  $E^{elastic}$ & $B_{11}$ & 1.6246 & $B_{12}$ & 1.4521 & $B_{44}$ & 2.9658 \\
 \hline
 \hline
\end{tabular}
\end{center}
\end{table}
Here, we report the coefficients of the effective Hamiltonian ($H_{eff}$) for PbTiO$_3$ in Tab.~\ref{tab:parameters}.
The $H_{eff}$ has the following degrees of freedom:  vectors related to the ferroelectric soft mode $\mathbf{P}$ and inhomogeneous strain  ($\mathbf{u}$) in each 5-atom unit cell, as well as, the homogenous strain ($\mathbf{\eta}$). The P-mode local vectors in the
$H_{eff}$ are centered on titanium ions. 
The local vectors corresponding to the inhomogeneous strains are technically centered on lead ions.
The homogenous strain is defined with respect to cubic symmetry and has six independent components $\eta_{i}$, in Voigt notation.
The potential energy is formulated exactly as in Ref.~\citen{Zhong1995}, except for the newly added eDMI Eq.~\ref{eq:fdmi}. Such a model has a ferroelectric phase transition from cubic to tetragonal phase at 650 K; it reproduces the lowest energy phase, P4mm, with P-mode displaced by 0.329 \AA which is close enough to the DFT value 0.346 \AA.

\begin{figure}[!htp]
    \centering
    \includegraphics[width=0.8\linewidth]{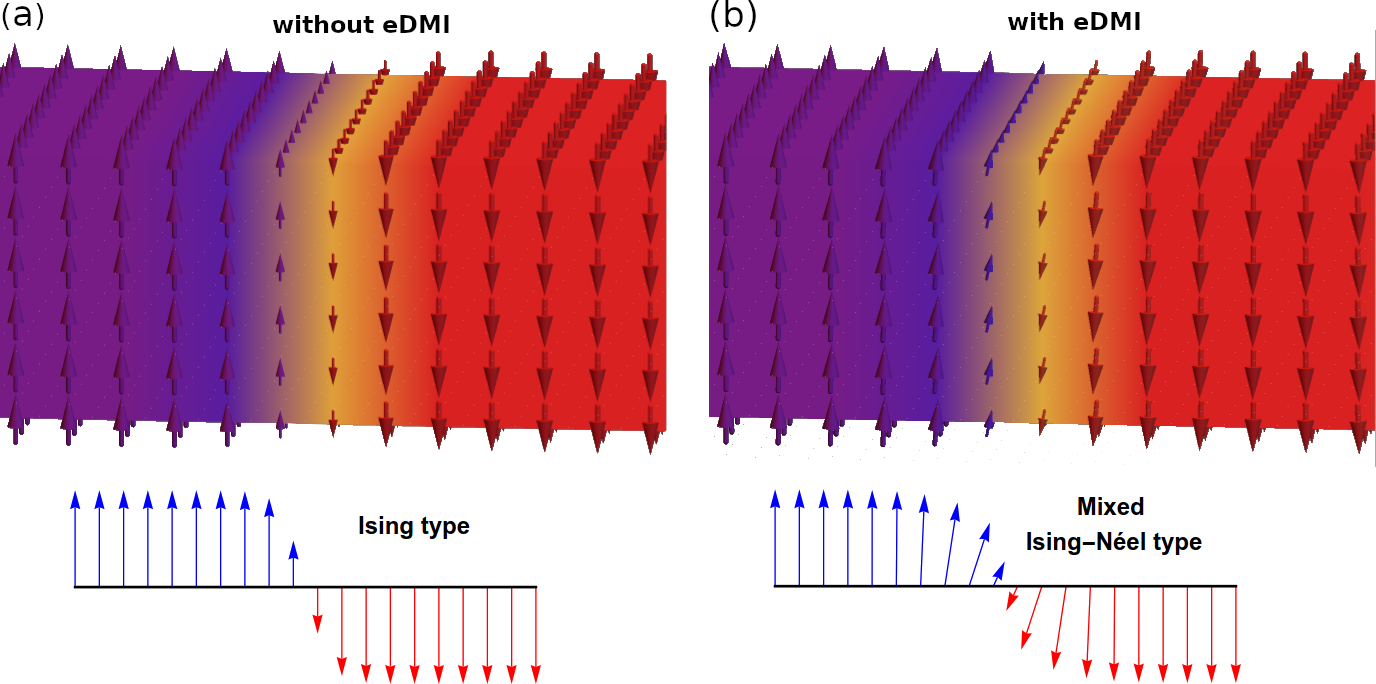}
    \caption{(a) Ising type domain wall achieved without eDMI and (b) mixed Ising-N{\'{e}}el type domain wall achieved with eDMI.}\label{fig:DW}
\end{figure}

We used 20x20x12 supercells that are all periodic along the [100] and [010] directions while finite along the z-direction. The thin film is mimicked to experience a 80\% of the polarization-induced surface charges. We use the Broyden–Fletcher–Goldfarb–Shanno (BFGS) algorithm for direct structural relaxation using the model potential from the effective Hamiltonian. To achieve electric bobbers, we simulated a thin film and used an initial configuration where all electric dipoles in a columnar nanodomain align strictly along the positive z-direction and are embedded in a big matrix of opposite polarization. When relaxing domain walls as in Fig.~\ref{fig:DW}, we used a bulk setup and initialized two adjacent domains with polarization aligned strictly along the positive and negative z-direction, respectively.

\appendix
\section{Extended tables of eDMI orbital selection rules\label{appendix}}
In this appendix, we present the results when orbital selection rules are applied to the $s$, $p$, and $d$ orbitals. Tables \ref{tab:dy}, \ref{tab:dz}, and \ref{tab:dx} list the three components $\mathcal{D}_{y}(i,j)$, $\mathcal{D}_{z}(i,j)$, and $\mathcal{D}_{x}(i,j)$ of the eDMI vector $\bm{\mathcal{D}}(i,j)$, respectively, from different orbitals combinations and mark if they are activated by symmetry.
The first two columns concern the orbitals $m$ and $n$ located on sites $i$ and $j$, respectively.
The third and fourth columns indicate if any of $\langle m,i|U_{i,\alpha}\hat{G}^{0}U_{j,\beta} | n,j \rangle$ and $\langle n,j | \hat{G}^{0} | m,i \rangle$ is zero (constrained by mirror $m_{xy}$) when sites $i$, $j$, and $k$ are collinearly aligned, which is against the orbital selection rule No. 1.
It can be seen that at least one of $\langle m,i|U_{i,\alpha}\hat{G}^{0}U_{j,\beta} | n,j \rangle$ and $\langle n,j | \hat{G}^{0} | m,i \rangle$ is zero through all the orbital combinations, which confirms the orbital selection rule No. 1.
The seventh and eighth columns indicate the antisymmetric feature that only one of $\langle m,i|U_{i,\alpha}\hat{G}^{0}U_{j,\beta} | n,j \rangle$ and $\langle n,j | \hat{G}^{0} | m,i \rangle$ changes sign when swapping orbitals $m$ and $n$ between sites $i$ and $j$, assuming that sites $i$, $j$, $k$ are not collinearly aligned.
The seventh and eighth columns summarize if $\langle m,i|U_{i,\alpha}\hat{G}^{0}U_{j,\beta} | n,j \rangle$ and $\langle n,j | \hat{G}^{0} | m,i \rangle$ are constrained to be zero by the operation of the mirror that goes through sites $i$, $j$, and $k$, which confirms the orbital selection rule No. 2.
The ninth column mark the component of the eDMI vector $\bm{\mathcal{D}}(i,j)$ that is  forbidden (via cross marks) or activated (via check marks).

% \usepackage{booktabs}
% \usepackage{multirow}
% \usepackage{graphicx}
\begin{table}[!htp]
\caption{The eDMI orbital selection rules for $\mathcal{D}_{y}(i,j)$ which is associated with $F_{x,z}$ and $F_{z,x}$.\label{tab:dy}}
\resizebox{\textwidth}{!}{%
\begin{tabular}{@{}|cc|cc|cc|cc|c|@{}}
\hline
\multicolumn{2}{|c|}{orbitals} & \multicolumn{2}{c|}{rule 1 (non-collinear)$^a$} & \multicolumn{2}{c|}{antisymmetric feature$^{b}$} & \multicolumn{2}{c|}{rule 2 (mirror $m_{3}$)$^{c}$} & \multirow{2}{*}{$\mathcal{D}_{y}(i,j)$} \\ 
m                          & n             & $\langle m,i | U_{i,x}\hat{G}^{0}U_{j,z} | n,j \rangle$ & $\langle n,j | \hat{G}^{0} | m,i \rangle$ & $\langle m,i | U_{i,x}\hat{G}^{0}U_{j,z} | n,j \rangle$ & $\langle n,j | \hat{G}^{0} | m,i \rangle$ & $\langle m,i | U_{i,x}\hat{G}^{0}U_{j,z} | n,j \rangle$ & $\langle n,j | \hat{G}^{0} | m,i \rangle$ &       \\
\hline
\multirow{9}{*}{$s$} & $s$       & 0 & - &-1 & 1 & - & - & \cmark \\
                     & $p_z$     & - & 0 &-1 & 1 & - & - & \cmark \\
                     & $p_x$     & 0 & - & 1 &-1 & - & - & \cmark \\
                     & $p_y$     & 0 & 0 &-1 & 1 & 0 & 0 & \xmark \\
                     & $d_{z^2}$ & 0 & - &-1 & 1 & - & - & \cmark \\
                     & $d_{xz}$  & - & 0 & 1 &-1 & - & - & \cmark \\
                     & $d_{yz}$  & - & 0 &-1 & 1 & 0 & 0 & \xmark \\
                     & $d_{xy}$  & 0 & 0 & 1 &-1 & 0 & 0 & \xmark \\
                & $d_{x^2-y^2}$  & 0 & - &-1 & 1 & - & - & \cmark \\
\hline
\multirow{8}{*}{$p_z$} & $p_z$     & 0 & - &-1 & 1 & - & - & \cmark \\
                       & $p_x$     & - & 0 & 1 &-1 & - & - & \cmark \\
                       & $p_y$     & 0 & 0 &-1 & 1 & 0 & 0 & \xmark \\
                       & $d_{z^2}$ & - & 0 &-1 & 1 & - & - & \cmark \\
                       & $d_{xz}$  & 0 & - & 1 &-1 & - & - & \cmark \\
                       & $d_{yz}$  & 0 & 0 &-1 & 1 & 0 & 0 & \xmark \\
                       & $d_{xy}$  & 0 & 0 & 1 &-1 & 0 & 0 & \xmark \\
                   & $d_{x^2-y^2}$ & - & 0 &-1 & 1 & - & - & \cmark \\
\hline
\multirow{7}{*}{$p_x$} & $p_x$     & 0 & - &-1 & 1 & - & - & \cmark \\
                       & $p_y$     & 0 & 0 & 1 &-1 & 0 & 0 & \xmark \\
                       & $d_{z^2}$ & 0 & - & 1 &-1 & - & - & \cmark \\
                       & $d_{xz}$  & - & 0 &-1 & 1 & - & - & \cmark \\
                       & $d_{yz}$  & 0 & 0 & 1 &-1 & 0 & 0 & \xmark \\
                       & $d_{xy}$  & 0 & 0 &-1 & 1 & 0 & 0 & \xmark \\
                   & $d_{x^2-y^2}$ & 0 & - & 1 &-1 & - & - & \cmark \\
\hline
\multirow{6}{*}{$p_y$} & $p_{y}$   & 0 & - &-1 & 1 & - & - & \cmark \\
                       & $d_{z^2}$ & 0 & 0 &-1 & 1 & 0 & 0 & \xmark \\
                       & $d_{xz}$  & 0 & 0 & 1 &-1 & 0 & 0 & \xmark \\
                       & $d_{yz}$  & - & 0 &-1 & 1 & - & - & \cmark \\
                       & $d_{xy}$  & 0 & - & 1 &-1 & - & - & \cmark \\
                   & $d_{x^2-y^2}$ & 0 & 0 &-1 & 1 & 0 & 0 & \xmark \\
\hline
\multirow{5}{*}{$d_{z^2}$} & $d_{z^2}$ & 0 & - &-1 & 1 & - & - & \cmark \\
                           & $d_{xz}$  & - & 0 & 1 &-1 & - & - & \cmark \\
                           & $d_{yz}$  & 0 & 0 &-1 & 1 & 0 & 0 & \xmark \\
                           & $d_{xy}$  & 0 & 0 & 1 &-1 & 0 & 0 & \xmark \\
                      & $d_{x^2-y^2}$  & 0 & - &-1 & 1 & - & - & \cmark \\
\hline
\multirow{4}{*}{$d_{xz}$} & $d_{xz}$ & 0 & - &-1 & 1 & - & - & \cmark \\
                          & $d_{yz}$ & 0 & 0 & 1 &-1 & 0 & 0 & \xmark \\
                          & $d_{xy}$ & 0 & 0 &-1 & 1 & 0 & 0 & \xmark \\
                     & $d_{x^2-y^2}$ & - & 0 & 1 &-1 & - & - & \cmark \\
\hline
\multirow{3}{*}{$d_{yz}$} & $d_{yz}$ & 0 & - &-1 & 1 & - & - & \cmark \\
                          & $d_{xy}$ & - & 0 & 1 &-1 & - & - & \cmark \\
                     & $d_{x^2-y^2}$ & 0 & 0 &-1 & 1 & 0 & 0 & \xmark \\
\hline
\multirow{2}{*}{$d_{xy}$} & $d_{xy}$ & 0 & - &-1 & 1 & - & - & \cmark \\
                     & $d_{x^2-y^2}$ & 0 & 0 & 1 &-1 & 0 & 0 & \xmark \\
\hline
$d_{x^2-y^2}$ & $d_{x^2-y^2}$ & 0 & - &-1 & 1 & - & - & \cmark \\
\hline
\end{tabular}%
}
\begin{tablenotes}
\small
\item  ${}^{a}$ from operation of mirror $m_{xy}$;
\item  ${}^{b}$ from operation that swaps orbitals $m$ and $n$ between sites $i$ and $j$, e.g. mirror $m_{yz}$;
\item  ${}^{c}$ from operation of mirror $m_{xz}$ that goes through sites $i$, $j$, and $k$.
\end{tablenotes}
\end{table}

% \usepackage{booktabs}
% \usepackage{multirow}
% \usepackage{graphicx}
\begin{table}[!htp]
\caption{The eDMI orbital selection rules for $\mathcal{D}_{z}(i,j)$ which is associated with $F_{x,y}$ and $F_{y,x}$.\label{tab:dz}}
\resizebox{\textwidth}{!}{%
\begin{tabular}{@{}|cc|cc|cc|cc|c|c|c|c|c|@{}}
\hline
\multicolumn{2}{|c|}{orbitals} & \multicolumn{2}{c|}{rule 1 (non-collinear)$^a$} & \multicolumn{2}{c|}{antisymmetric feature$^b$} & \multicolumn{2}{c|}{rule 3 (mirror $m_{3}$)$^c$} & \multirow{2}{*}{$\mathcal{D}_{z}(i,j)$} \\ 
m                          & n             & $\langle m,i | U_{i,x}\hat{G}^{0}U_{j,y} | n,j \rangle$ & $\langle n,j | \hat{G}^{0} | m,i \rangle$ & $\langle m,i | U_{i,x}\hat{G}^{0}U_{j,y} | n,j \rangle$ & $\langle n,j | \hat{G}^{0} | m,i \rangle$ & $\langle m,i | U_{i,x}\hat{G}^{0}U_{j,y} | n,j \rangle$ & $\langle n,j | \hat{G}^{0} | m,i \rangle$ &      \\
\hline
\multirow{9}{*}{$s$} & $s$       & 0 & - &-1 & 1 & 0 & - & \xmark \\
                     & $p_z$     & 0 & 0 &-1 & 1 & 0 & - & \xmark \\
                     & $p_x$     & 0 & - & 1 &-1 & 0 & - & \xmark \\
                     & $p_y$     & - & 0 &-1 & 1 & - & 0 & \xmark \\
                     & $d_{z^2}$ & 0 & - &-1 & 1 & 0 & - & \xmark \\
                     & $d_{xz}$  & 0 & 0 & 1 &-1 & 0 & - & \xmark \\
                     & $d_{yz}$  & 0 & 0 &-1 & 1 & - & 0 & \xmark \\
                     & $d_{xy}$  & - & 0 & 1 &-1 & - & 0 & \xmark \\
                & $d_{x^2-y^2}$  & 0 & - &-1 & 1 & 0 & - & \xmark \\
\hline
\multirow{8}{*}{$p_z$} & $p_z$     & 0 & - &-1 & 1 & 0 & - & \xmark \\
                       & $p_x$     & 0 & 0 & 1 &-1 & 0 & - & \xmark \\
                       & $p_y$     & 0 & 0 &-1 & 1 & - & 0 & \xmark \\
                       & $d_{z^2}$ & 0 & 0 &-1 & 1 & 0 & - & \xmark \\
                       & $d_{xz}$  & 0 & - & 1 &-1 & 0 & - & \xmark \\
                       & $d_{yz}$  & - & 0 &-1 & 1 & - & 0 & \xmark \\
                       & $d_{xy}$  & 0 & 0 & 1 &-1 & - & 0 & \xmark \\
                   & $d_{x^2-y^2}$ & 0 & 0 &-1 & 1 & 0 & - & \xmark \\
\hline
\multirow{7}{*}{$p_x$} & $p_x$     & 0 & - &-1 & 1 & 0 & - & \xmark \\
                       & $p_y$     & - & 0 & 1 &-1 & - & 0 & \xmark \\
                       & $d_{z^2}$ & 0 & - & 1 &-1 & 0 & - & \xmark \\
                       & $d_{xz}$  & 0 & 0 &-1 & 1 & 0 & - & \xmark \\
                       & $d_{yz}$  & 0 & 0 & 1 &-1 & - & 0 & \xmark \\
                       & $d_{xy}$  & - & 0 &-1 & 1 & - & 0 & \xmark \\
                   & $d_{x^2-y^2}$ & 0 & - & 1 &-1 & 0 & - & \xmark \\
\hline
\multirow{6}{*}{$p_y$} & $p_{y}$   & 0 & - &-1 & 1 & 0 & - & \xmark \\
                       & $d_{z^2}$ & - & 0 &-1 & 1 & - & 0 & \xmark \\
                       & $d_{xz}$  & 0 & 0 & 1 &-1 & - & 0 & \xmark \\
                       & $d_{yz}$  & 0 & 0 &-1 & 1 & 0 & - & \xmark \\
                       & $d_{xy}$  & 0 & - & 1 &-1 & 0 & - & \xmark \\
                   & $d_{x^2-y^2}$ & - & 0 &-1 & 1 & - & 0 & \xmark \\
\hline
\multirow{5}{*}{$d_{z^2}$} & $d_{z^2}$ & 0 & - &-1 & 1 & 0 & - & \xmark \\
                           & $d_{xz}$  & 0 & 0 & 1 &-1 & 0 & - & \xmark \\
                           & $d_{yz}$  & 0 & 0 &-1 & 1 & - & 0 & \xmark \\
                           & $d_{xy}$  & - & 0 & 1 &-1 & - & 0 & \xmark \\
                      & $d_{x^2-y^2}$  & 0 & - &-1 & 1 & 0 & - & \xmark \\
\hline
\multirow{4}{*}{$d_{xz}$} & $d_{xz}$ & 0 & - &-1 & 1 & 0 & - & \xmark \\
                          & $d_{yz}$ & - & 0 & 1 &-1 & - & 0 & \xmark \\
                          & $d_{xy}$ & 0 & 0 &-1 & 1 & - & 0 & \xmark \\
                     & $d_{x^2-y^2}$ & 0 & 0 & 1 &-1 & 0 & - & \xmark \\
\hline
\multirow{3}{*}{$d_{yz}$} & $d_{yz}$ & 0 & - &-1 & 1 & 0 & - & \xmark \\
                          & $d_{xy}$ & 0 & 0 & 1 &-1 & 0 & - & \xmark \\
                     & $d_{x^2-y^2}$ & 0 & 0 &-1 & 1 & - & 0 & \xmark \\
\hline
\multirow{2}{*}{$d_{xy}$} & $d_{xy}$ & 0 & - &-1 & 1 & 0 & - & \xmark \\
                     & $d_{x^2-y^2}$ & - & 0 & 1 &-1 & - & 0 & \xmark \\
\hline
$d_{x^2-y^2}$ & $d_{x^2-y^2}$ & 0 & - &-1 & 1 & 0 & - & \xmark \\
\hline
\end{tabular}%
}
\begin{tablenotes}
\small
\item  ${}^{a}$ from operation of mirror $m_{xy}$;
\item  ${}^{b}$ from operation that swaps orbitals $m$ and $n$ between sites $i$ and $j$, e.g. mirror $m_{yz}$;
\item  ${}^{c}$ from operation of mirror $m_{xz}$ that goes through sites $i$, $j$, and $k$.
\end{tablenotes}
\end{table}

% \usepackage{booktabs}
% \usepackage{multirow}
% \usepackage{graphicx}
\begin{table}[!htp]
\caption{The eDMI orbital selection rules for $\mathcal{D}_{x}(i,j)$ which is associated with $F_{y,z}$ and $F_{z,y}$.\label{tab:dx}}
\resizebox{\textwidth}{!}{%
\begin{tabular}{@{}|cc|cc|cc|cc|c|@{}}
\hline
\multicolumn{2}{|c|}{orbitals} & \multicolumn{2}{c|}{rule 1 (non-collinear)$^a$} & \multicolumn{2}{c|}{antisymmetric feature$^b$} & \multicolumn{2}{c|}{rule 3 (mirror$m_{3}$)$^c$} & \multirow{2}{*}{$\mathcal{D}_{x}(i,j)$}  \\ 
m                          & n             & $\langle m,i | U_{i,y}\hat{G}^{0}U_{j,z} | n,j \rangle$ & $\langle n,j | \hat{G}^{0} | m,i \rangle$ & $\langle m,i | U_{i,y}\hat{G}^{0}U_{j,z} | n,j \rangle$ & $\langle n,j | \hat{G}^{0} | m,i \rangle$ & $\langle m,i | U_{i,y}\hat{G}^{0}U_{j,z} | n,j \rangle$ & $\langle n,j | \hat{G}^{0} | m,i \rangle$ &  \\
\hline
\multirow{9}{*}{$s$} & $s$       & 0 & - & 1 & 1 & 0 & - & \xmark \\
                     & $p_z$     & 0 & 0 & 1 & 1 & 0 & - & \xmark \\
                     & $p_x$     & 0 & - &-1 &-1 & 0 & - & \xmark \\
                     & $p_y$     & 0 & 0 & 1 & 1 & - & 0 & \xmark \\
                     & $d_{z^2}$ & 0 & - & 1 & 1 & 0 & - & \xmark \\
                     & $d_{xz}$  & 0 & 0 &-1 &-1 & 0 & - & \xmark \\
                     & $d_{yz}$  & - & 0 & 1 & 1 & - & 0 & \xmark \\
                     & $d_{xy}$  & 0 & 0 &-1 &-1 & - & 0 & \xmark \\
                & $d_{x^2-y^2}$  & 0 & - & 1 & 1 & 0 & - & \xmark \\
\hline
\multirow{8}{*}{$p_z$} & $p_z$     & 0 & - & 1 & 1 & 0 & - & \xmark \\
                       & $p_x$     & 0 & 0 &-1 &-1 & 0 & - & \xmark \\
                       & $p_y$     & - & 0 & 1 & 1 & - & 0 & \xmark \\
                       & $d_{z^2}$ & 0 & 0 & 1 & 1 & 0 & - & \xmark \\
                       & $d_{xz}$  & 0 & - &-1 &-1 & 0 & - & \xmark \\
                       & $d_{yz}$  & 0 & 0 & 1 & 1 & - & 0 & \xmark \\
                       & $d_{xy}$  & 0 & 0 &-1 &-1 & - & 0 & \xmark \\
                   & $d_{x^2-y^2}$ & 0 & 0 & 1 & 1 & 0 & - & \xmark \\
\hline
\multirow{7}{*}{$p_x$} & $p_x$     & 0 & - & 1 & 1 & 0 & - & \xmark \\
                       & $p_y$     & 0 & 0 &-1 &-1 & - & 0 & \xmark \\
                       & $d_{z^2}$ & 0 & - &-1 &-1 & 0 & - & \xmark \\
                       & $d_{xz}$  & 0 & 0 & 1 & 1 & 0 & - & \xmark \\
                       & $d_{yz}$  & - & 0 &-1 &-1 & - & 0 & \xmark \\
                       & $d_{xy}$  & 0 & 0 & 1 & 1 & - & 0 & \xmark \\
                   & $d_{x^2-y^2}$ & 0 & - &-1 &-1 & 0 & - & \xmark \\
\hline
\multirow{6}{*}{$p_y$} & $p_{y}$   & 0 & - & 1 & 1 & 0 & - & \xmark \\
                       & $d_{z^2}$ & 0 & 0 & 1 & 1 & - & 0 & \xmark \\
                       & $d_{xz}$  & - & 0 &-1 &-1 & - & 0 & \xmark \\
                       & $d_{yz}$  & 0 & 0 & 1 & 1 & 0 & - & \xmark \\
                       & $d_{xy}$  & 0 & - &-1 &-1 & 0 & - & \xmark \\
                   & $d_{x^2-y^2}$ & 0 & 0 & 1 & 1 & - & 0 & \xmark \\
\hline
\multirow{5}{*}{$d_{z^2}$} & $d_{z^2}$ & 0 & - & 1 & 1 & 0 & - & \xmark \\
                           & $d_{xz}$  & 0 & 0 &-1 &-1 & 0 & - & \xmark \\
                           & $d_{yz}$  & - & 0 & 1 & 1 & - & 0 & \xmark \\
                           & $d_{xy}$  & 0 & 0 &-1 &-1 & - & 0 & \xmark \\
                      & $d_{x^2-y^2}$  & 0 & - & 1 & 1 & 0 & - & \xmark \\
\hline
\multirow{4}{*}{$d_{xz}$} & $d_{xz}$ & 0 & - & 1 & 1 & 0 & - & \xmark \\
                          & $d_{yz}$ & 0 & 0 &-1 &-1 & - & 0 & \xmark \\
                          & $d_{xy}$ & 0 & 0 & 1 & 1 & - & 0 & \xmark \\
                     & $d_{x^2-y^2}$ & 0 & 0 &-1 &-1 & 0 & - & \xmark \\
\hline
\multirow{3}{*}{$d_{yz}$} & $d_{yz}$ & 0 & - & 1 & 1 & 0 & - & \xmark \\
                          & $d_{xy}$ & 0 & 0 &-1 &-1 & 0 & - & \xmark \\
                     & $d_{x^2-y^2}$ & - & 0 & 1 & 1 & - & 0 & \xmark  \\
\hline
\multirow{2}{*}{$d_{xy}$} & $d_{xy}$ & 0 & - & 1 & 1 & 0 & - & \xmark \\
                     & $d_{x^2-y^2}$ & 0 & 0 &-1 &-1 & - & 0 & \xmark \\
\hline
$d_{x^2-y^2}$ & $d_{x^2-y^2}$ & 0 & - & -1 & 1 & 0 & - & \xmark \\
\hline
\end{tabular}%
}
\begin{tablenotes}
\small
\item  ${}^{a}$ from operation of mirror $m_{xy}$;
\item  ${}^{b}$ from operation that swaps orbitals $m$ and $n$ between sites $i$ and $j$, e.g. mirror $m_{yz}$;
\item  ${}^{c}$ from operation of mirror $m_{xz}$ that goes through sites $i$, $j$, and $k$.
\end{tablenotes}
\end{table}

%\noindent{\bf References}\\
\bibliography{ref}% Produces the bibliography via BibTeX.
\end{document}